\documentclass[aps,prx,reprint,superscriptaddress,nofootinbib,longbibliography,floatfix,amsmath,amssymb]{revtex4-2}
\usepackage{prx-format}
\usepackage{placeins}


\definecolor{NNCommentColor}{HTML}{A00070}

\newif\ifArxivSupplementary
\newcommand{\ArxivSkipContentsline}[4]{}
\newcommand{\ArxivTocMain}{\let\contentsline\ArxivSkipContentsline}
\newcommand{\ArxivTocSupplement}{\let\contentsline\ArxivRealContentsline}

\begin{document}

\title{Quantum geometry drives orbital paramagnetism in superconducting fluctuations}
\affiliation{RIKEN Center for Emergent Matter Science (CEMS), Wako, Saitama 351-0198, Japan}
\affiliation{Fundamental Quantum Science Program (FQSP), TRIP Headquarters, RIKEN, Wako 351-0198, Japan}
\affiliation{Institute of Condensed Matter Physics, School of Physics and Astronomy, Shanghai Jiao Tong University, Shanghai 200240, China}
\affiliation{Department of Physics and Astronomy, University of Kentucky, Lexington, Kentucky 40506-0055, USA}
\noaffiliation
\author{Zi-Ting Sun}
\affiliation{RIKEN Center for Emergent Matter Science (CEMS), Wako, Saitama 351-0198, Japan}
\author{Ying-Ming Xie}
\affiliation{Institute of Condensed Matter Physics, School of Physics and Astronomy, Shanghai Jiao Tong University, Shanghai 200240, China}
\author{Chunli Huang}
\affiliation{Department of Physics and Astronomy, University of Kentucky, Lexington, Kentucky 40506-0055, USA}
\author{Naoto Nagaosa}
\email{nagaosa@riken.jp}
\affiliation{RIKEN Center for Emergent Matter Science (CEMS), Wako, Saitama 351-0198, Japan}
\affiliation{Fundamental Quantum Science Program (FQSP), TRIP Headquarters, RIKEN, Wako 351-0198, Japan}
\date[]{}

\begin{abstract}
Superconductors expel magnetic fields through the Meissner effect.
Even slightly above the superconducting transition temperature, electrons
can briefly form Cooper pairs, leading to an orbital diamagnetic susceptibility that diverges as the transition is approached. In this work, we use Gaussian fluctuation theory to study a generic multicomponent superconductor above its transition temperature and show that fluctuations can instead produce a paramagnetic response. We identify two competing contributions: the usual diamagnetic contribution from the center-of-mass motion of Cooper pairs and a paramagnetic contribution from interference between different components of the superconducting order parameter.
The latter is related to pair quantum geometry, which describes how the internal components of the Cooper pair change with momentum. 
We then apply our theory to a model of a kagome superconductor with two chiral pairing modes. When the parent state preserves time-reversal symmetry, the paramagnetic contribution can outweigh the conventional diamagnetic contribution, producing net fluctuation paramagnetism.  However, when the orbital time-reversal symmetry is broken, e.g.,~by forming a weak loop-current order, the degeneracy between the two chiral modes is then lifted. Upon cooling toward the superconducting transition, the lower-energy mode dominates the fluctuations, and diamagnetism wins as single-component physics takes over.
Our results provide a mechanism for fluctuation paramagnetism in superconductors and show that the orbital magnetic response can probe the quantum geometry of Cooper pairs.
%Our results show that superconducting fluctuations need not always oppose magnetic fields. Their orbital magnetic response can reveal the quantum geometry of collective Cooper pairs, extending the observable consequences of quantum geometry beyond those of individual electrons in a crystal.
\end{abstract}
\maketitle
\addtocontents{toc}{\protect\ArxivTocMain}
\raggedbottom

\section{Introduction}

Fluctuating Cooper pairs above the superconducting transition temperature
$T_c$ produce singular thermodynamic and transport corrections
\cite{LarkinVarlamov2005}.
Among these effects, fluctuation orbital magnetism is a representative
phenomenon \cite{schmid1969diamagnetic,LeePayne1971,LeePayne1972}: the fluctuation contribution to
the orbital magnetic susceptibility is diamagnetic and diverges as
$T\to T_c^+$ \cite{prange1970diamagnetic}.
This response provides a precursor to the perfect diamagnetism of the
superconducting Meissner effect \cite{Tinkham2004}, and has been widely
observed in conventional superconductors \cite{GollubEtAl1969Diamagnetism}
and cuprates \cite{LiEtAl2010Diamagnetism}.
This classical result follows from Gaussian fluctuations of the
Ginzburg--Landau (GL) theory for uniform superconductors with a single-component
order parameter \cite{LarkinVarlamov2005} or several pairing modes that
contribute independently \cite{AdachiIkeda2016,KoshelevVarlamovVinokur2005}.
For a general superconductor with multiple coupled components of the
order parameter, it remains unclear whether the fluctuation orbital
magnetic response must stay diamagnetic or can become paramagnetic
instead.

\begin{figure}[tbp]
 \centering
 \includegraphics[width=\columnwidth]{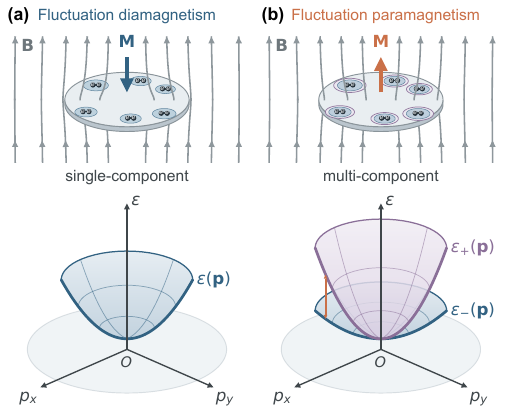}
 \caption[Single-component versus multi-component fluctuation response]{%
 Orbital magnetic response of Gaussian superconducting fluctuations above $T_c$.
 Upper panels show short-lived Cooper pairs in a thin film under a
 perpendicular field $\mathbf B$, with induced magnetization $\mathbf M$ satisfying $M_z = \chi^{\text{fl}}_{zz} B_z$ in linear response. Here $\chi^{\text{fl}}_{zz}$ denotes the out-of-plane fluctuation orbital magnetic susceptibility. Lower panels show the eigenvalue dispersions of the inverse pair propagator.
 Their curvatures near the minima characterize the mode stiffnesses,
 while their positive minima indicate $T>T_c$.
 (a) A single-component superconductor has one branch of eigenvalue $\varepsilon(\mathbf p)$ and a diamagnetic response.
 (b) A multi-component superconductor has two branches with eigenvalues
 $\varepsilon_-(\mathbf p)$ and $\varepsilon_+(\mathbf p)$.
 Intermode orbital coupling associated with pair quantum geometry can
 result in a paramagnetic orbital magnetic response.}
 \label{fig:mechanism}
\end{figure}

In this work, we reveal a counterintuitive possibility: intercomponent
coherence in a multicomponent superconductor can reverse the orbital magnetic
response from diamagnetism to paramagnetism in the Gaussian fluctuation regime.
For a single-component order parameter, the fluctuating Cooper pairs
have no internal order-parameter degrees of freedom, and the magnetic field acts only on
their center-of-mass motion, producing diamagnetic screening
[Fig.~\ref{fig:mechanism}(a)].
However, for a multicomponent superconductor, the intercomponent coherence
mixes different order-parameter components at finite pair momentum, and
the orbital magnetic field can couple to this internal motion of Cooper
pairs [Fig.~\ref{fig:mechanism}(b)].
This coupling lowers the fluctuation free energy in the field and can
outweigh the diamagnetic screening, giving rise to net fluctuation
paramagnetism.

A more concrete description of the effects of intercomponent coherence
is provided by nontrivial pair quantum geometry, which is defined by
variations of the eigenstates of the matrix-valued pair propagator with
Cooper-pair center-of-mass momentum
\cite{sun2026quantum,keskiner2026superconductivity,KalkanIskin2026CollectivePairing},
and has been shown to be directly relevant to the fluctuation transport
of multicomponent superconductors \cite{sun2026quantum}.
We find that if we express the orbital magnetic susceptibility of multicomponent
superconducting fluctuations in the spectral representation of the
inverse pair propagator, the effects of intercomponent coherence
are naturally exposed as contributions from the pair quantum metric
and Berry curvature.
The conventional relation between susceptibility and coherence
length~\cite{LarkinVarlamov2005,schmid1969diamagnetic} can be modified
at leading order by pair Berry curvature in a multicomponent superconductor,
even when only one of its pairing modes becomes soft.
These pair quantum geometric contributions give rise to a significant
paramagnetic fluctuation response and vanish in single-component
cases.

Finally, we demonstrate the mechanism in a nearest-neighbor tight-binding model
on a kagome lattice with onsite spin-singlet pairing
\cite{NagashimaEtAl2024,HolbaekFischer2025}.
Near the van Hove \mbox{filling}, the model favors a chiral $d$-wave
pairing doublet with net fluctuation paramagnetism.
Besides, we show that a weak parent loop-current order~\cite{FengEtAl2021,DennerThomaleNeupert2021} lifts the degeneracy
between the $d\pm id$ pairing branches and drives a crossover back to
diamagnetism as $T$ approaches $T_c$.
This system provides a concrete realization of fluctuation
orbital paramagnetism and can be related to the recently discovered
kagome superconductors $A\mathrm{V}_3\mathrm{Sb}_5$
($A=\mathrm{K},\mathrm{Rb},\mathrm{Cs}$)~\mbox{\cite{WilsonOrtiz2024Review,NeupertEtAl2022,JiangEtAl2023KagomeReview,YinLianHasan2022Review,YinEtAl2026,DiSanteEtAl2026KagomeMetals}}.

\section{Electromagnetic vertices of multicomponent superconducting fluctuations}

We consider the magnetic response of a multicomponent superconductor
above the superconducting transition, \(T>T_c\), where the saddle-point
value of the Cooper-pair field vanishes,
\(\langle \Delta \rangle = 0\). We work in the Gaussian approximation and
neglect interactions between short-lived Cooper pairs.
The components may originate from multidimensional irreducible representations of the crystal point group, competing pairing channels, or multiband degrees of
freedom \cite{SigristUeda1991,MaitiHirschfeld2015,Leggett1966}.
Throughout this work, we use units in which $\hbar=c=k_B=1$.
The fluctuation action in the presence of an external electromagnetic field is written as
\begin{equation}\label{eq:fla}
    S_{\mathrm{fl}}[\Delta,A]
    =
    \sum_{P,P'}
    \Delta^\dagger(P)\,
    \hat L^{-1}_{P,P'}[A]\,
    \Delta(P').
\end{equation}
Here \(\Delta=(\Delta_1,\Delta_2,\ldots,\Delta_N)^{\mathsf T}\) is the
\(N\)-component superconducting order-parameter vector.
The variable \(P=(i\Omega_n,\mathbf p)\) denotes the bosonic frequency and momentum
of the pair field. The inverse propagator \(\hat L^{-1}[A]\) depends on the
external four-potential \(A_\mu=(A_0,\mathbf A)\); hats denote matrices in
the internal pair-component space. For static magnetic response, $A_0=0$
and the four-momentum carried by the external field is $Q=(0,\mathbf q)$, with
$\mathbf A(\mathbf q)$ denoting a spatial Fourier component of the vector potential.
Near the transition temperature $T_c$, the classical Gaussian
fluctuations with zero bosonic Matsubara frequency ($i\Omega_0=0$) give the
leading singular magnetic response, while quantum fluctuations with nonzero
Matsubara frequencies ($i\Omega_n\neq0$) contribute only a nonsingular
background, which we neglect as $T\to T_c^+$
\cite{LarkinVarlamov2005,Altland_Simons_2023}.

We now expand the inverse pair propagator in powers of the external field around \(\mathbf A=0\):
\begin{equation}
    \hat L^{-1}[\mathbf A]
    =
    \hat L_0^{-1}
    +
    \hat V_1[\mathbf A]
    +
    \hat V_2[\mathbf A]
    +
    O(\mathbf A^3),
\end{equation}
where \(\hat L_0^{-1}\equiv \hat L^{-1}[\mathbf A=0]\) is the zero-field
static inverse pair propagator, Hermitian in the internal pair-component
space: $[\hat L_0^{-1}(\mathbf p)]^\dagger=\hat L_0^{-1}(\mathbf p)$.
We introduce the spectral representation of the inverse pair propagator,
$\hat L_0^{-1}|u_\nu\rangle=\varepsilon_\nu|u_\nu\rangle$,
with $\nu=1,\ldots,N$ and $\varepsilon_\nu>0$ above $T_c$.
The critical temperature is fixed by
$\min_{\nu,\mathbf p}\varepsilon_\nu(\mathbf p,T_c)=0$,
and the corresponding eigenvector $|u_\nu(\mathbf p)\rangle$ identifies
the favored pairing channel.
As we will show below, this inverse propagator plays a role analogous to
the electronic Bloch Hamiltonian, and its eigensystem can be used to define
the pair quantum geometry~\cite{sun2026quantum,keskiner2026superconductivity,KalkanIskin2026CollectivePairing}.
In the long-wavelength and weak-field regime, we assume that the minimally
coupled inverse pair propagator,
$\hat L^{-1}[\mathbf A]\simeq\hat L_0^{-1}(\mathbf p-e^*\mathbf A)$,
admits a local gauge-covariant expansion~\cite{Gorkov1959GL}.
We take $e<0$ to be the signed electron charge, so $e^*=2e<0$ is
the charge of the fluctuating Cooper pair.
We retain terms through $O(\mathbf A^2)$, which suffices for linear response,
and define the linear (paramagnetic) and quadratic (diamagnetic) vertices as
\begin{equation}\label{eq:Verdef}
\begin{aligned}
    &[\hat V_1]_{\mathbf p+\frac{\mathbf q}{2},\,\mathbf p-\frac{\mathbf q}{2}}
    =
    \frac{1}{\sqrt{\mathcal V}}\,
    \hat\Lambda_c(\mathbf p;\mathbf q)\,
    A_c(\mathbf q),
    \\
    &[\hat V_2]_{\mathbf p,\mathbf p}
    =
    \frac{1}{2\mathcal V}
    \sum_\mathbf q
    \hat\Gamma_{bc}(\mathbf p;\mathbf q,-\mathbf q)\,
    A_b(\mathbf q)A_c(-\mathbf q),
\end{aligned}
\end{equation}
where $\mathcal V$ is the system volume.
The paramagnetic vertex parametrizes arbitrary incoming and outgoing pair momenta
$\mathbf k,\mathbf k'$ by $\mathbf p=(\mathbf k'+\mathbf k)/2$ and
$\mathbf q=\mathbf k'-\mathbf k$.
Since $\hat L_0$ is diagonal in momentum,
the quadratic free-energy term $\mathrm{Tr}(\hat L_0\hat V_2)$ selects
the elements of $\hat V_2$ diagonal in Cooper-pair momentum, for which
the total momentum transferred by the photons is zero.
The one-photon vertex is even in
$\mathbf q$, i.e., $\hat\Lambda_c(\mathbf p;\mathbf q)=\hat\Lambda_c(\mathbf p;-\mathbf q)$,
whereas the two-photon vertex is symmetric under exchange of the photon legs,
$\hat\Gamma_{bc}
(\mathbf p;\mathbf q_1,\mathbf q_2)
=
\hat\Gamma_{cb}
(\mathbf p;\mathbf q_2,\mathbf q_1)$. Here we have adopted the Einstein summation convention.
The minimal coupling, detailed in
\textbf{Supplemental \mbox{Material} (SM)~S1.1}~\cite{SupplementaryMaterial}, yields the full vertices.
With $\partial_c\equiv\partial_{p_c}$, the one-photon vertex expands as
\begin{equation}
\begin{aligned}
&\hat\Lambda_c(\mathbf p;\mathbf q)=-e^*\partial_c\hat L_0^{-1}(\mathbf p)\\
&\mathord{-}\frac{e^*}{24}q_aq_b\partial_a\partial_b\partial_c\hat L_0^{-1}(\mathbf p)
  +O(\mathbf q^4).\mkern8mu
\end{aligned}
\label{eq:derivative_form}
\end{equation}
The quadratic coupling to the vector potential is described by the
two-photon vertex,
\begin{equation}
\begin{aligned}
&\hat\Gamma_{bc}(\mathbf p;\mathbf q,-\mathbf q)=e^{*2}\partial_b\partial_c\hat L_0^{-1}(\mathbf p)\\
&\mathord{+}\frac{e^{*2}}{12}q_aq_d\partial_a\partial_d\partial_b\partial_c\hat L_0^{-1}(\mathbf p)
  +O(\mathbf q^4).\mkern8mu
\end{aligned}
\label{eq:derivative_form_two_photon}
\end{equation}
As shown in \textbf{SM~S1.2}~\cite{SupplementaryMaterial},
the vertices in Eqs.~\eqref{eq:derivative_form} and~\eqref{eq:derivative_form_two_photon} automatically satisfy the
Ward--Takahashi identities to the retained order, ensuring gauge invariance
of the response theory.

\section{Fluctuation free energy and response kernel}

Then we consider the response to the static external magnetic field.
The fluctuation partition function associated with Eq.~\eqref{eq:fla} is
\begin{equation}
    Z[\mathbf A]
    =
    \int \mathcal D\vec{\Delta}^\dagger \mathcal D\vec{\Delta}\,
    \exp\!\left\{
        -\beta S_{\mathrm{fl}}[\Delta,\mathbf A]
    \right\},
\end{equation}
where $\beta=1/T$. Using the Gaussian result
$Z[\mathbf A]=1/\det\{\beta \hat L^{-1}[\mathbf A]\}$, the fluctuation
contribution to the free energy is therefore
\begin{equation}
    F_{\mathrm{fl}}[\mathbf A]
    =
    -T\ln Z[\mathbf A]
    =
    T\,\mathrm{Tr}\ln\!\left[\beta\hat L^{-1}[\mathbf A]\right],
\end{equation}
where \(\mathrm{Tr}\) denotes the full trace over both momenta and
other internal degrees of freedom.

Expanding the fluctuation free energy to quadratic order in the external field gives
\begin{equation}
\begin{aligned}
&\Delta F_{\mathrm{fl}}[\mathbf A]
\equiv F_{\mathrm{fl}}[\mathbf A]-F_{\mathrm{fl}}[\mathbf0]
 =T\mathrm{Tr}(\hat L_0\hat V_1)\\
&+T\mathrm{Tr}\Big(\hat L_0\hat V_2-\frac12\hat L_0\hat V_1\hat L_0\hat V_1\Big)
 +O(\mathbf A^3).
\end{aligned}
\label{eq:free_energy_expansion}
\end{equation}
The term linear in \(\mathbf A\) vanishes because equilibrium pairing fluctuations cannot sustain a net bulk transport current \cite{bohm1949note}. Substituting Eq.~\eqref{eq:Verdef} into Eq.~\eqref{eq:free_energy_expansion}, we write the quadratic free energy in the standard form
\begin{equation}
    \Delta F_{\mathrm{fl}}[\mathbf A]
    =
    \frac{1}{2}
    \sum_\mathbf q
    K_{bc}^{\mathrm{fl}}(\mathbf q)\,
    A_b(\mathbf q)A_c(-\mathbf q),
\label{eq:free_energy_kernel}
\end{equation}
with the Gaussian fluctuation response kernel
\begin{align}
K_{bc}^{\mathrm{fl}}(\mathbf q)
&=\frac{T}{\mathcal V}\sum_\mathbf p\mathrm{tr}_{\mathrm{mat}}\Big[
 \hat L_0(\mathbf p)\hat\Gamma_{bc}(\mathbf p;\mathbf q,-\mathbf q)\label{eq:Kmunu_general}\\
&\mathrel{-}\hat L_0\!\left(\mathbf p+\tfrac{\mathbf q}{2}\right)\hat\Lambda_b(\mathbf p;\mathbf q)
 \hat L_0\!\left(\mathbf p-\tfrac{\mathbf q}{2}\right)\hat\Lambda_c(\mathbf p;-\mathbf q)\Big].
\notag
\end{align}
Here \(\mathrm{tr}_{\mathrm{mat}}\) denotes the trace over internal pair-component indices only.

To second order in $\mathbf q$, the long-wavelength vertices in
Eqs.~\eqref{eq:derivative_form} and~\eqref{eq:derivative_form_two_photon} give
\begin{align}
K_{bc}^{\mathrm{fl}}(\mathbf q)
&=e^{*2}T\int\frac{d^dp}{(2\pi)^d}\mathrm{tr}_{\mathrm{mat}}\Big[
 \hat L_0(\mathbf p)\partial_b\partial_c\hat L_0^{-1}(\mathbf p)\label{eq:Kbc_derivative}\\
&\mathrel{-}\hat L_0\!\left(\mathbf p+\tfrac{\mathbf q}{2}\right)\partial_b\hat L_0^{-1}(\mathbf p)
 \hat L_0\!\left(\mathbf p-\tfrac{\mathbf q}{2}\right)\partial_c\hat L_0^{-1}(\mathbf p)\Big].
\notag
\end{align}
The first term comes from the diamagnetic vertex and the second from the
paramagnetic vertex. We have taken the thermodynamic limit,
$\frac{1}{\mathcal V}\sum_{\mathbf p}\to\int\frac{d^d p}{(2\pi)^d}$.
The $O(\mathbf q^2)$ vertices in Eqs.~\eqref{eq:derivative_form} and~\eqref{eq:derivative_form_two_photon} contribute
only vanishing surface terms after integration by parts
(see \textbf{SM~S1.3}~\cite{SupplementaryMaterial}) and are therefore omitted from
Eq.~\eqref{eq:Kbc_derivative}. To identify the magnetic response, we expand the linear response kernel
in the external wavevector $\mathbf q$:
\begin{equation}
K_{ab}^{\mathrm{fl}}(\mathbf q)
=K_{ab}^{\mathrm{fl}}(0)+C_{ab;c}q_c+C_{ab;cd}q_cq_d+O(\mathbf q^3).
\label{eq:kernel_q_expansion}
\end{equation}
The semicolon separates the external field indices from the wavevector indices, and
$C_{ab;cd}=C_{ab;dc}$.
Reciprocity, $K_{ab}^{\mathrm{fl}}(\mathbf q)=K_{ba}^{\mathrm{fl}}(-\mathbf q)$,
also gives $C_{ab;c}=-C_{ba;c}$ and $C_{ab;cd}=C_{ba;cd}$.
The coefficients $C_{ab;cd}$ are responsible for the orbital magnetic susceptibility because
$\mathbf B(\mathbf q)=i\mathbf q\times\mathbf A(\mathbf q)$, while
both $K_{ab}^{\mathrm{fl}}(0)$ and
$C_{ab;c}$ vanish, as discussed in
\textbf{SM~S1.3}~\cite{SupplementaryMaterial}.

Expanding the two shifted propagators in Eq.~\eqref{eq:Kbc_derivative}
and regrouping their matrix products gives
\begin{align}
&C_{ab;cd}=-\frac{e^{*2}T}{16}\int\frac{d^dp}{(2\pi)^d}
\operatorname{tr}_{\mathrm{mat}}\Big[(\partial_c\hat R_d+\partial_d\hat R_c)\label{eq:georesponse}\\
&\mathord{\times}(\hat R_a\hat R_b+\hat R_b\hat R_a)
 -[\hat R_a,\hat R_c][\hat R_b,\hat R_d]
 -[\hat R_a,\hat R_d][\hat R_b,\hat R_c]\Big].\notag
\end{align}
Here, we describe the momentum dependence of the propagator by its right
logarithmic derivative (RLD) \cite{LiuJingWang2015RLD}, defined by
$\partial_a\hat L_0=\hat L_0\hat R_a$, or equivalently
$\hat R_a=\hat L_0^{-1}\partial_a\hat L_0$.
Since $\hat L_0(\mathbf p)=\beta\langle\Delta_{\mathbf p}\Delta_{\mathbf p}^\dagger\rangle_{\mathrm{eq}}$
in the Gaussian regime \cite{sun2026quantum}, $\hat R_a$ measures the
variation of $\hat L_0(\mathbf p)$ with respect to $p_a$.
Geometrically, $\hat{\mathcal R}\equiv\hat L_0^{-1}d\hat L_0
=\hat R_a\,dp_a$ is the Maurer--Cartan one-form pulled back to Cooper-pair
momentum space. It consequently satisfies the Maurer--Cartan equation~\cite{Nakahara2003}
$\partial_a\hat R_b-\partial_b\hat R_a+[\hat R_a,\hat R_b]=0$.

Gauge invariance requires
$q_aK_{ab}^{\mathrm{fl}}(\mathbf q)=K_{ab}^{\mathrm{fl}}(\mathbf q)q_b=0$.
These Ward identities also imply $q_aC_{ab;cd}q_cq_d=0$, with the analogous identity on index $b$.
Their explicit verification is given in
\textbf{SM~S1.3}~\cite{SupplementaryMaterial}.
Although $C_{ab;cd}$ originates entirely from the term containing two
paramagnetic vertices in Eq.~\eqref{eq:Kbc_derivative}, it encodes both
paramagnetic and diamagnetic responses, as we show below.

\section{Fluctuation orbital magnetic susceptibility}

The susceptibility is a thermodynamic derivative with respect to the
magnetic field,
$\chi_{ij}^{\mathrm{fl}}=-\mathcal V^{-1}\partial_{B_i}\partial_{B_j}F_{\mathrm{fl}}|_{\mathbf B=0}$.
Since $\mathbf B(\mathbf q)=i\mathbf q\times\mathbf A(\mathbf q)$, the
relevant part in Eq.~\eqref{eq:free_energy_kernel} is
$-\tfrac12\sum_{\mathbf q}\chi_{ij}^{\mathrm{fl}}B_i(\mathbf q)B_j(-\mathbf q)$.
Matching the coefficients of $q_cq_dA_a(\mathbf q)A_b(-\mathbf q)$ gives
\begin{align}
\chi_{ij}^{\mathrm{fl}}
&=-\frac13\epsilon_{ica}\epsilon_{jdb}C_{ab;cd}\label{eq:chi_from_C}\\
&=-\frac{e^{*2}T}{24}\int\frac{d^dp}{(2\pi)^d}
 \epsilon_{iab}\epsilon_{jcd}\operatorname{tr}_{\mathrm{mat}}\Big[\hat J_{ac}\hat J_{bd}+\hat\Xi_{ab}\hat\Xi_{cd}\Big].\notag
\end{align}
Here, $\hat J_{ab}\equiv(\partial_a\partial_b\hat L_0^{-1})\hat L_0
=\tfrac12(\hat R_a\hat R_b+\hat R_b\hat R_a-\partial_a\hat R_b-\partial_b\hat R_a)$
is related to the Hessian of the inverse propagator, and
$\hat\Xi_{ab}\equiv[\hat R_a,\hat R_b]$ is the commutator of the RLD
components. The Hessian term generalizes the inverse-mass contribution
of a single-component fluctuation. A nonzero commutator
signals momentum-dependent mixing between pairing channels that cannot
be captured by independent scalar propagators.

For a two-dimensional thin film in a perpendicular magnetic field, only
$\chi_{zz}^{\mathrm{fl}}$ is needed. Using
\begin{align}
\epsilon_{zab}\epsilon_{zcd}\operatorname{tr}_{\mathrm{mat}}(\hat J_{ac}\hat J_{bd})
&=2\operatorname{tr}_{\mathrm{mat}}(\hat J_{xx}\hat J_{yy}-\hat J_{xy}^2),\notag\\
\epsilon_{zab}\epsilon_{zcd}\operatorname{tr}_{\mathrm{mat}}
\bigl(\hat\Xi_{ab}\hat\Xi_{cd}\bigr)
&=4\operatorname{tr}_{\mathrm{mat}}\hat\Xi_{xy}^2,\label{eq:chi_2d_contractions}
\end{align}
Eq.~\eqref{eq:chi_from_C} reduces to
$\chi_{zz}^{\mathrm{fl}}=\chi_{zz}^{\mathrm K}+\chi_{zz}^{\mathrm C}$, in which
\begin{equation}
\begin{aligned}
\chi_{zz}^{\mathrm K}
&=-\frac{e^{*2}T}{12d_s}\int\frac{d^2p}{(2\pi)^2}
 \operatorname{tr}_{\mathrm{mat}}(\hat J_+^2-\hat J_-^2-\hat J_{xy}^2),\\
\chi_{zz}^{\mathrm C}
&=-\frac{e^{*2}T}{6d_s}\int\frac{d^2p}{(2\pi)^2}
 \operatorname{tr}_{\mathrm{mat}}\hat\Xi_{xy}^2.
\end{aligned}
\label{eq:chi_2d}
\end{equation}
Here $\hat J_\pm\equiv(\hat J_{xx}\pm\hat J_{yy})/2$ and $d_s$ is the film thickness.
Eq.~\eqref{eq:chi_2d} is the central result of this work: it separates the orbital
magnetic response of multicomponent superconducting fluctuations into
kinetic (K) and coherent (C) contributions.

In a multicomponent superconductor, the relative amplitudes and phases
of the pairing components can vary with momentum, giving rise to pair
quantum geometry. A magnetic field couples to both the center-of-mass
motion and the internal movement of a fluctuating Cooper pair. The kinetic
cost of the overall motion raises the fluctuation free energy, giving
rise to diamagnetism ($\chi<0$), whereas the internal movement lowers it,
producing paramagnetism ($\chi>0$), as shown in Fig.~\ref{fig:mechanism}.
The kinetic term mainly describes the former process: the isotropic
component $\hat J_+$ contributes diamagnetically, while the anisotropic
components $\hat J_-$ and $\hat J_{xy}$ contribute paramagnetically.
Although the net kinetic contribution can have either sign, the isotropic
$\hat J_+$ term typically dominates, yielding a diamagnetic response.
The coherent term describes the latter process; it is non-negative and
controlled by the RLD commutator. As we show below, it provides the main
mechanism for turning fluctuation diamagnetism into paramagnetism.

We then discuss the general conditions for the emergence of a net
paramagnetic fluctuation response, i.e., $\chi_{zz}^{\mathrm{fl}}>0$,
from the perspective of pair quantum geometry:
the strength of the coherent contribution is constrained by the internal
geometry of the pair propagator. Define a Gram tensor
$\mathcal G_{ab}=\operatorname{tr}_{\mathrm{mat}}(\bar R_a\bar R_b)$,
where $\bar R_a=\hat R_a-\hat{1}\operatorname{tr}_{\mathrm{mat}}(\hat R_a)/N$ is the
traceless part of $\hat R_a$, and $N$ is the number of pairing components.
As shown in \textbf{SM~S2.3}~\cite{SupplementaryMaterial},
the B\"ottcher--Wenzel
inequality~\cite{BottcherWenzel2008Commutator} implies
$0\leq-\operatorname{tr}_{\mathrm{mat}}\hat\Xi_{xy}^2
\leq2\det\mathcal G$, and hence
\begin{equation}
    0\leq\chi_{zz}^{\mathrm C}\leq U_{\mathrm{geom}}
    \equiv\frac{e^{*2}T}{3d_s}
    \int\frac{d^2p}{(2\pi)^2}\det\mathcal G(\mathbf p).
    \label{eq:main_commutator_bound}
\end{equation}
When $\chi_{zz}^{\mathrm K}<0$, a necessary condition for a net paramagnetic
fluctuation response is
$U_{\mathrm{geom}}>|\chi_{zz}^{\mathrm K}|$.

As a benchmark, take a single-component order parameter with anisotropic
stiffness $L_0^{-1}(\mathbf p)=\alpha+\gamma_{ab}p_ap_b$.
Here $\alpha>0$ above $T_c$, and Gaussian stability requires
$\boldsymbol\gamma=(\gamma_{ab})$ to be positive definite:
$\gamma_x,\gamma_y>0$ and
$\det\boldsymbol\gamma=\gamma_x\gamma_y-\gamma_{xy}^2>0$.
All quantities are scalars, so the coherent contribution vanishes and
$J_{ab}=2\gamma_{ab}L_0$.
Using $\int\frac{d^2p}{(2\pi)^2}L_0^2=1/(4\pi\alpha\sqrt{\det\boldsymbol\gamma})$
in Eq.~\eqref{eq:chi_2d} gives
\begin{equation}
\begin{aligned}
\chi_{zz}^{\mathrm{fl}}
&=-\frac{e^{*2}T}{3d_s}\det\boldsymbol\gamma
  \int\frac{d^2p}{(2\pi)^2}L_0^2\\
&=-\frac{e^{*2}T}{12\pi\alpha d_s}\sqrt{\det\boldsymbol\gamma}<0.
\end{aligned}
\label{eq:single_component_three_terms}
\end{equation}
Thus stability enforces diamagnetism in the single-component Gaussian
theory even with anisotropic stiffness.
The mean coherence length obeys
$\xi_{\mathrm{CL}}^2=\sqrt{\det\boldsymbol\gamma}/\alpha$~\cite{LarkinVarlamov2005},
and Eq.~\eqref{eq:single_component_three_terms} becomes the standard diamagnetic result
$\chi_{zz}^{\mathrm{fl}}=-e^{*2}T\xi_{\mathrm{CL}}^2/(12\pi d_s)$
\cite{LarkinVarlamov2005,schmid1969diamagnetic,prange1970diamagnetic,UssishkinSondhiHuse2002},
as shown in Fig.~\ref{fig:mechanism}(a).
The stiffness can include Bloch-state geometric contributions
\cite{Iskin2023GLGeometry,ChenLaw2024GLGeometry,Iskin2024CoherenceGeometry},
while the scalar propagator has no intermode pair geometry.

\section{Pair quantum geometry in two-component superconductors}
\label{sec:two_component_geometry}

The minimal nontrivial examples are two-component superconductors,
with the pair propagator reading
\begin{equation}
 \hat L_0(\mathbf p)
 =\sum_{\nu=\pm}
 \frac{|u_\nu(\mathbf p)\rangle\langle u_\nu(\mathbf p)|}
 {\varepsilon_\nu(\mathbf p)}.
 \label{eq:pair_spectral_representation}
\end{equation}
The momentum dependence of the eigenmode basis is encoded in the
matrix-valued pair Berry connection
$\mathcal A_a^{\nu\nu'}\equiv i\langle u_\nu|\partial_a u_{\nu'}\rangle$,
whose off-diagonal elements describe intermode mixing.
The pair quantum geometry of these fluctuation eigenmodes
\cite{sun2026quantum,keskiner2026superconductivity} is characterized by
the pair quantum geometric tensor
$Q_{ab}=\mathcal A_a^{+-}\mathcal A_b^{-+}$.
Thus $g_{ab}^{+}=g_{ab}^{-}\equiv g_{ab}
=\operatorname{Re}Q_{ab}$ is the quantum
metric of either eigenmode, and
$\omega_{xy}^{+}=-\omega_{xy}^{-}\equiv\omega_{xy}
=-2\operatorname{Im}Q_{xy}$ is the corresponding Berry
curvature.

Here $W=(\varepsilon_+-\varepsilon_-)^2/(\varepsilon_+\varepsilon_-)$
is a dimensionless measure of the separation between the two fluctuation
eigenvalues, and the adjugate of the metric tensor is defined by
$(\operatorname{adj}g)_{xx}=g_{yy}$,
$(\operatorname{adj}g)_{yy}=g_{xx}$, and
$(\operatorname{adj}g)_{xy}=(\operatorname{adj}g)_{yx}=-g_{xy}$.
In this spectral representation, the kinetic contribution is
\begingroup\predisplaypenalty=0
\begin{align}
\chi_{zz}^{\mathrm K}
&=-\frac{e^{*2}T}{12d_s}\int\frac{d^2p}{(2\pi)^2}
\Biggl\{\sum_{\nu=\pm}\det\!\left[
\frac{\partial_a\partial_b\varepsilon_\nu}{\varepsilon_\nu}+2Wg_{ab}
\right]\notag\\
&\mathrel{-}\frac{\partial_a\partial_b\!\left[
(\varepsilon_+-\varepsilon_-)^2(\operatorname{adj}g)_{ab}\right]}
{\varepsilon_+\varepsilon_-}
-W^2\omega_{xy}^2\Biggr\}.
\label{eq:two_mode_H_main}
\end{align}
\endgroup
And the coherent contribution is
\begin{equation}
\chi_{zz}^{\mathrm C}
\!=\!\frac{e^{*2}T}{3d_s}\!\int\!\frac{d^2p}{(2\pi)^2}\!\biggl[
W^2\omega_{xy}^2\!+\!
\frac{(\operatorname{adj}g)_{ab}\partial_aW\partial_bW}{W+4}\biggr].
\label{eq:two_mode_C_main}
\end{equation}
The two non-negative terms in Eq.~\eqref{eq:two_mode_C_main} are
governed by the Berry curvature and quantum metric, respectively.
Adding Eqs.~\eqref{eq:two_mode_H_main} and \eqref{eq:two_mode_C_main} gives
\begin{align}
\chi_{zz}^{\mathrm{fl}}
&=\frac{e^{*2}T}{12d_s}\int\frac{d^2p}{(2\pi)^2}
\Biggl\{\frac{\partial_a\partial_b\left[
(\varepsilon_+-\varepsilon_-)^2(\operatorname{adj}g)_{ab}\right]}
{\varepsilon_+\varepsilon_-}\notag\\
&\mathrel{+}\frac{4(\operatorname{adj}g)_{ab}\partial_aW\partial_bW}{W+4}
+5W^2\omega_{xy}^2\notag\\
&\mathrel{-}\sum_{\nu=\pm}\det\left[
\frac{\partial_a\partial_b\varepsilon_\nu}{\varepsilon_\nu}+2Wg_{ab}
\right]\Biggr\}.
\label{eq:two_mode_total_main}
\end{align}
The spectral representation thus makes explicit that the orbital magnetic
response probes both the fluctuation spectrum and the pair quantum geometry
due to the coherence between its eigenmodes.
It is easy to show that Eq.~\eqref{eq:two_mode_C_main} saturates the upper
bound in Eq.~\eqref{eq:main_commutator_bound}, giving
$\chi_{zz}^{\mathrm C}=U_{\mathrm{geom}}$, as shown in
\textbf{SM~S2.3}~\cite{SupplementaryMaterial}.
Consequently,
$U_{\mathrm{geom}}>-\chi_{zz}^{\mathrm K}$ becomes both necessary and sufficient
for $\chi_{zz}^{\mathrm{fl}}>0$ in this situation.
The general-$N$ eigenmode formulation is given in
\textbf{SM~S2.1}~\cite{SupplementaryMaterial}.

We now use Eqs.~\eqref{eq:two_mode_H_main} and \eqref{eq:two_mode_C_main}
to examine how pair quantum geometry modifies the single-component relation
between susceptibility and coherence length in
Eq.~\eqref{eq:single_component_three_terms}.
We take the lower branch $\nu=-$ to be the isolated soft mode as $T\to T_c^+$,
while the upper branch $\bar\nu=+$ remains massive.
Write $\varepsilon_-(\mathbf p)=\epsilon+\gamma_{ab}p_ap_b+\cdots$,
with $\epsilon\equiv\varepsilon_-(\mathbf0)>0$ and positive-definite $\boldsymbol\gamma$.
Near $T_c$, we treat the massive eigenvalue $\varepsilon_+(\mathbf0)$
as nearly temperature independent.
From \textbf{SM~S2.1}~\cite{SupplementaryMaterial}, we have
\begin{equation}
 \langle u_-|\hat J_{ab}|u_-\rangle
 =\frac{\partial_a\partial_b\varepsilon_-}{\varepsilon_-}
 +2\frac{\varepsilon_+-\varepsilon_-}{\varepsilon_-}g_{ab}.
 \label{eq:two_mode_J_lower}
\end{equation}
Evaluating it at $\mathbf p=\mathbf0$, where
$\partial_a\partial_b\varepsilon_-=2\gamma_{ab}$, gives the
critical-mode gradient tensor for $\epsilon\ll\varepsilon_+(\mathbf0)$:
\begin{equation}
 \gamma_{ab}
 \simeq\frac{\epsilon}{2}\langle u_-|\hat J_{ab}|u_-\rangle_{\mathbf p=\mathbf0}
 -\varepsilon_+(\mathbf0)g_{ab}(\mathbf0).
 \label{eq:soft_mode_coherence}
\end{equation}
The coherence length is given by
$\xi_{\mathrm{CL}}^2=\sqrt{\det\boldsymbol\gamma}/\epsilon$~\cite{keskiner2026superconductivity,KalkanIskin2026CollectivePairing},
so the pair quantum metric enters $\boldsymbol\gamma$ as well as $\xi_{\mathrm{CL}}$
through virtual mixing with the massive mode.

In the limit $\epsilon/\varepsilon_+(\mathbf0)\to0$ ($T\to T_c^+$), the leading singular terms
of Eqs.~\eqref{eq:two_mode_H_main} and \eqref{eq:two_mode_C_main} give
\begin{equation}
 \chi_{zz}^{\mathrm{fl}}
 \simeq-\frac{e^{*2}T}{12\pi d_s}\xi_{\mathrm{CL}}^2
 \left[1-\frac{3\varepsilon_+^{\,2}(\mathbf0)\omega_{xy}^{\,2}(\mathbf0)}
 {4\det\boldsymbol\gamma}\right],
 \label{eq:single_soft_response}
\end{equation}

At order $1/\epsilon$, the remaining explicit geometric terms reduce
to the displayed Berry-curvature correction through
$g_{xx}g_{yy}-g_{xy}^2=\omega_{xy}^2/4$.
Near $T_c$, with intermode coupling matrix elements and soft-mode stiffness fixed,
$\omega_{xy}(\mathbf0)\propto\varepsilon_+^{-2}(\mathbf0)$, so the relative
Berry-curvature correction decreases as $\varepsilon_+^{-2}(\mathbf0)$.
A nonzero Berry curvature produces an additional paramagnetic correction,
so the coherence length no longer determines the leading susceptibility as in
Eq.~\eqref{eq:single_component_three_terms}, even when only one
of the eigenmodes eventually becomes critical.
The derivation is given in \textbf{SM~S2.2}~\cite{SupplementaryMaterial}.

For an isolated nondegenerate soft mode at $\mathbf p=0$, time-reversal
symmetry forces $\omega_{xy}(\mathbf0)=0$, so a nonzero
Berry-curvature correction requires broken time-reversal symmetry.
Thus, with time-reversal symmetry, the fluctuation diamagnetism
of an isolated soft mode in a multicomponent superconductor near $T_c$
coincides with the single-component result.
When several low-energy modes are degenerate, however, the quantum metric
in Eq.~\eqref{eq:two_mode_C_main} can still produce a large paramagnetic
contribution even with time-reversal symmetry, as we will show in the next section.

\section{Paramagnetism from intercomponent coherence}
\label{sec:kagome}

\begin{figure}[t]
 \centering
 \includegraphics[width=\columnwidth]{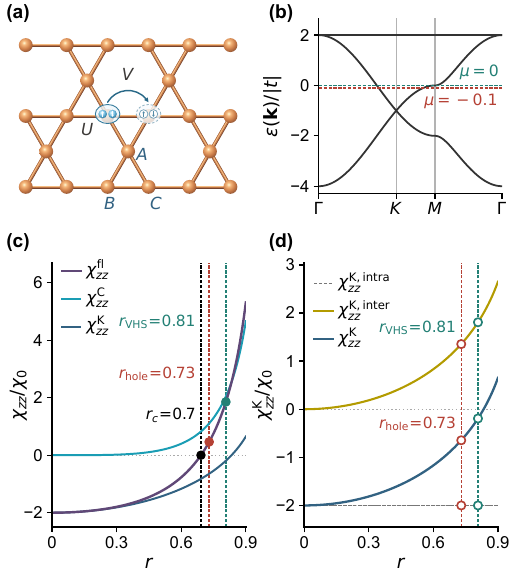}
 \caption[Kagome model and coherence-induced Gaussian paramagnetism]{%
 (a) Schematic of a kagome superconductor with onsite spin-singlet attraction
 $U<0$ and nearest-neighbor pair transfer $V>0$.
 (b) Band structure of a kagome lattice model with hopping constant $t=-1$
 along the high-symmetry path $\Gamma$--$K$--$M$--$\Gamma$.
 The green dashed line marks $\mu=0$, at the van Hove singularity (VHS)
 of the upper dispersive band at $M$. The red dashed line marks
 $\mu=-0.1$, with slight hole doping away from the VHS.
 (c) Total ($\chi_{zz}^{\rm fl}$, purple), coherent ($\chi_{zz}^{\rm C}$, cyan)
 and kinetic ($\chi_{zz}^{\rm K}$, blue) orbital magnetic susceptibilities at fixed $T$ and
 common mass $\alpha$, plotted against $r=|\gamma_1|/\gamma_0$ and normalized
 by $\chi_0=e^{*2}T\gamma_0/(12\pi d_s\alpha)$.
 The black dashed line and dot mark the sign change of the total
 susceptibility at $r_c=0.7$.
 (d) Intramode ($\chi_{zz}^{\mathrm{K,intra}}$, gray dashed),
 intermode ($\chi_{zz}^{\mathrm{K,inter}}$, gold)
 and total ($\chi_{zz}^{\mathrm K}$, blue) kinetic contributions, resolved using the Hessian operator
 $\hat J_{ab}$ in the pair eigenbasis.
 Red and green vertical dashed lines mark $r_{\rm hole}=0.73$ and
 $r_{\rm VHS}=0.81$, respectively, for the chemical potentials in (b).}
 \label{fig:kagome_bands}
\end{figure}

In this section, we show how intercomponent coherence in a multicomponent
superconductor gives rise to fluctuation paramagnetism.
As a concrete realization, we apply the pair quantum geometric formula of orbital magnetic susceptibility to a
nearest-neighbor tight-binding model on a kagome lattice with onsite spin-singlet
pairing [Fig.~\ref{fig:kagome_bands}(a)]. The full Bloch Hamiltonian is
given in \textbf{SM~S3.1}~\cite{SupplementaryMaterial},
and its band structure is shown in Fig.~\ref{fig:kagome_bands}(b).
This application is motivated by evidence for unconventional
superconductivity~\cite{WuEtAl2021,HossainEtAl2025,WangEtAl2026TRSB,YinEtAl2026} and
by experimental indications and theoretical proposals of
time-reversal-breaking loop-current order~\cite{MielkeEtAl2022,FengEtAl2021,DennerThomaleNeupert2021,ChaEtAl2026}
in kagome metals $A$V$_3$Sb$_5$ ($A=\mathrm{K},\mathrm{Rb},\mathrm{Cs}$).
In these materials, especially $\mathrm{CsV}_3\mathrm{Sb}_5$~\cite{KangEtAl2022},
van Hove singularities lie near the Fermi level. In our model, the green dashed
line in Fig.~\ref{fig:kagome_bands}(b) marks the chemical potential at this
van Hove saddle point of the upper dispersive band.

The three sublattice pairing fields $\Delta_\alpha$ ($\alpha=A,B,C$)
are defined as the gap coefficients in the local pairing term
$\sum_\alpha(\Delta_\alpha c_{\alpha\uparrow}^\dagger
c_{\alpha\downarrow}^\dagger+\mathrm{h.c.})$.
They form the reducible representation
$A_{1g}\oplus E_{2g}$ of the $D_{6h}$ point group~\cite{NagashimaEtAl2024,HolbaekFischer2025}.
The symmetry-adapted $A_{1g}$ field
$s=(\Delta_A+\Delta_B+\Delta_C)/\sqrt3$ denotes the $s$-wave pairing channel,
and the $E_{2g}$ doublet,
$\eta_x=(2\Delta_A-\Delta_B-\Delta_C)/\sqrt6$ and
$\eta_y=(\Delta_B-\Delta_C)/\sqrt2$,
denotes the $(d_{x^2-y^2},d_{xy})$-wave channel~\cite{WuEtAl2021}.
Here we use the chiral basis
$\eta_\pm=(\eta_y\pm i\eta_x)/\sqrt2$ for the $E_{2g}$ channel
to represent $d\pm id$ pairing.
To model the pairing interaction, we choose onsite attraction $U<0$~\cite{NagashimaEtAl2024}
and repulsive pair transfer $V>0$ on every nearest-neighbor
bond,
\begin{equation}
H_{\rm int}=U\sum_iP_i^\dagger P_i
 +V\sum_{\langle ij\rangle}(P_i^\dagger P_j+\mathrm{h.c.}),
\label{eq:kagome_interaction}
\end{equation}
where $P_i=c_{i\downarrow}c_{i\uparrow}$ is the pair annihilation operator.
At zero pair momentum, the effective coupling constants
for the $A_{1g}$ and $E_{2g}$ channels are $g_s=-U-4V$ and $g_E=-U+2V$, respectively.
For simplicity, we choose $g_s=0$, so $g_E=6V>0$ and the $E_{2g}$ doublet is soft. Leading two-dimensional $d$-wave pairing
channels~\cite{RomerEtAl2022} and chiral $d\pm id$ superconducting
states~\cite{YuLi2012,WangEtAl2013} have been reported in kagome lattice models
with repulsive interaction near the $5/12$ filling, i.e., the van Hove
singularity. These findings support this choice. In this situation, its inverse pair propagator
is a $2\times2$ matrix.

\begin{samepage}
To quadratic order in $\mathbf p$, the $2\times2$ effective inverse pair
propagator in the chiral basis, consistent with hexagonal and
time-reversal symmetries, is
\begin{equation}
\hat L_E^{-1}(\mathbf p)=
\begin{pmatrix}
\alpha+\gamma_0p^2 & \gamma_1p_+^2\\
\gamma_1p_-^2 & \alpha+\gamma_0p^2
\end{pmatrix},
\label{eq:kagome_kernel}
\end{equation}
where $p_\pm=p_x\pm ip_y$ and $\alpha\simeq a_E(T-T_c)$ with $a_E>0$. The Ginzburg--Landau
coefficients $a_E$, $\gamma_0$ and $\gamma_1$ can be obtained directly
from microscopic calculations
(see \textbf{SM~S3.3}~\cite{SupplementaryMaterial}).
The coefficient $\gamma_0$ is the intracomponent stiffness,
while $\gamma_1$ couples the chiral components through
$p_\pm^2=p^2e^{\pm2i\theta}$, where $\theta$ is the momentum polar angle.
\par\end{samepage}

For this degenerate doublet, evaluating
Eq.~\eqref{eq:chi_2d} with Eq.~\eqref{eq:kagome_kernel} yields (see \textbf{SM~S3.4}~\cite{SupplementaryMaterial})
$\chi^{\mathrm K}_{zz}=\chi_0\mathcal S_E^{\mathrm K}(r)$,
$\chi^{\mathrm C}_{zz}=\chi_0\mathcal S_E^{\mathrm C}(r)$ and
$\chi^{\mathrm{fl}}_{zz}=\chi_0\mathcal S_E(r)$, where
$r=|\gamma_1|/\gamma_0$, $\chi_0=e^{*2}T\gamma_0/(12\pi d_s\alpha)$ and
\begin{equation}
\begin{aligned}
\mathcal S_E^{\mathrm K}(r)&=2\big[r\,\operatorname{atanh}r-1\big],\\
\mathcal S_E^{\mathrm C}(r)&=8\bigg[\Big(\frac3r-r\Big)\operatorname{atanh}r-3\bigg],\\
\mathcal S_E(r)&=\mathcal S_E^{\mathrm K}+\mathcal S_E^{\mathrm C}=6(4-r^2)\frac{\operatorname{atanh}r}{r}-26.
\end{aligned}
\label{eq:kagome_sign_parts}
\end{equation}

\noindent Both the kinetic and coherent contributions to the susceptibility shift
toward more positive values as the mixing ratio $r$ increases, until the model becomes unstable at $r=1$.
The total susceptibility $\chi_{zz}^{\mathrm{fl}}$ also increases and
changes sign at $r_c=0.7$, i.e., fluctuation
paramagnetism rather than diamagnetism occurs when $r>r_c$, as shown in
Fig.~\ref{fig:kagome_bands}(c).

To clarify the origin of net fluctuation paramagnetism in this model,
we resolve Eq.~\eqref{eq:chi_2d} in the pair eigenbasis, as we did in
Sec.~\ref{sec:two_component_geometry}, but further identify the intramode
and intermode contributions as the diagonal and off-diagonal matrix-element
products of the Hessian matrix, respectively, i.e.,
\begin{equation}
\operatorname{tr}_{\mathrm{mat}}(\hat J_{ac}\hat J_{bd})
=\sum_\nu J_{ac}^{\nu\nu}J_{bd}^{\nu\nu}
+\sum_{\nu\ne\nu'}J_{ac}^{\nu\nu'}J_{bd}^{\nu'\nu},
\label{eq:kagome_intra_inter_trace}
\end{equation}
where $J_{ab}^{\nu\nu'}=\langle u_\nu|\hat J_{ab}|u_{\nu'}\rangle$,
with an analogous decomposition for
$\operatorname{tr}_{\mathrm{mat}}\hat\Xi_{xy}^2$
(see \textbf{SM~S2.1}~\cite{SupplementaryMaterial}).
The intramode kinetic terms sum to the constant diamagnetic susceptibility
$\chi_{zz}^{\mathrm{K,intra}}=-2\chi_0$ for every $r$
[Fig.~\ref{fig:kagome_bands}(d)]. This sum
coincides with the combined response of two independent diamagnetic modes
with stiffnesses $\gamma_0\pm|\gamma_1|$.
The sign reversal is therefore entirely an intermode effect
[Fig.~\ref{fig:kagome_bands}(d)]: the intermode
kinetic contribution is
$\chi_{zz}^{\mathrm{K,inter}}=2r\operatorname{atanh}r\,\chi_0$.
The coherent contribution in Eq.~\eqref{eq:kagome_sign_parts} is also
entirely intermode in this model because of the vanishing Berry curvature,
$\omega_{xy}=0$. Both intermode contributions are positive and
increase monotonically with $r$, while the intramode diamagnetic
contribution is negative and independent of $r$.
For sufficiently large $r$, the intermode contributions outweigh
$2\chi_0$, reversing the sign of the total susceptibility.

\begin{figure}[t]
 \centering
 \includegraphics[width=\columnwidth]{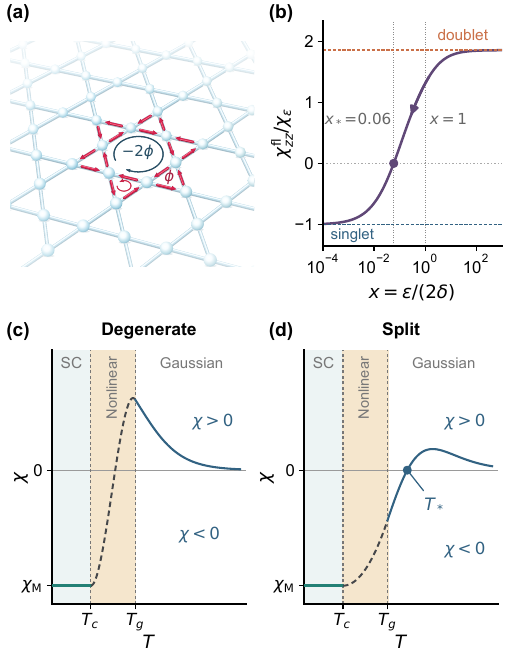}
 \caption[Loop-current splitting, Gaussian response and schematic Meissner connection]{%
 (a) Schematic of one domain of the chiral flux
 phase~\cite{OhgushiMurakamiNagaosa2000} in the kagome lattice.
 Bond arrows specify the hopping-phase orientation. Red and blue
 circular arrows show circulation around triangles and hexagons, respectively.
 Counterclockwise hopping-phase sums are $\phi$ on triangles and $-2\phi$
 on hexagons, giving zero net flux in each unit cell.
 (b) Normalized Gaussian fluctuation susceptibility
 $\chi_{zz}^{\rm fl}/\chi_\epsilon$ with fixed chiral splitting $\delta>0$,
 where $\epsilon=\alpha-\delta$, $x=\epsilon/(2\delta)$ and
 $\chi_\epsilon=e^{*2}T\gamma_0/(12\pi d_s\epsilon)$.
 The arrow indicates cooling toward smaller $x$.
 Orange and blue dashed lines mark the degenerate-doublet (paramagnetic)
 and single-soft-mode (diamagnetic) limits, respectively.
 Vertical dotted lines indicate the sign change at $x_*=0.06$
 and the splitting scale $x=1$.
 (c,d) Possible connections between Gaussian fluctuations and Meissner
 screening for degenerate and split doublets, respectively.
 Blue solid curves show the Gaussian response for $T>T_g$, and
 the point at $T_*$ in (d) marks its sign change.
 Green plateaus represent ideal Meissner screening, $\chi_{\rm M}<0$, below $T_c$.
 Gold shading denotes the nonlinear interval $T_c<T<T_g$, where black
 dashed curves schematically connect the Gaussian response to the
 Meissner plateau.}
 \label{fig:kagome_main}
\end{figure}

\begingroup\clubpenalty=10000
An illustrative set of microscopic parameters gives a stable $E_{2g}$-doublet
inverse pair propagator with paramagnetic susceptibility
(\textbf{SM~S3.3--S3.4}~\cite{SupplementaryMaterial}).
Measuring energies in units of the hopping magnitude $|t|$ and lengths
in units of the nearest-neighbor distance $a$, we choose $t=-1$,
$\mu=0$, $U=-4.4$ and $V=1.1$, giving $g_E=6.6$ and $g_s=0$.
These parameters give the critical temperature
$T_c=0.77$ and, at this temperature, $\gamma_0=0.0104$,
$\gamma_1=0.0084$ and $a_E=0.0255$, so that $r_{\mathrm{VHS}}=0.81$. Microscopically, pair hopping $V$ contributes equally to $\gamma_0$ and $\gamma_1$, accounting for a large mixing
ratio $r$ (see \textbf{SM~S3.3}~\cite{SupplementaryMaterial}).
At these microscopic coefficients, $r_{\mathrm{VHS}}>r_c$ and
$(\mathcal S_E^{\mathrm K},\mathcal S_E^{\mathrm C},\mathcal S_E)
=(-0.19,2.04,1.85)$, as indicated by the green dashed line in Fig.~\ref{fig:kagome_bands}(c).
The coherent contribution outweighs the
diamagnetic kinetic contribution, so the total susceptibility becomes positive.
For comparison, slight hole doping shifts the chemical potential away from the
van Hove singularity to $\mu=-0.1$
[red dashed line in Fig.~\ref{fig:kagome_bands}(b)], reducing the mixing ratio
to $r_{\mathrm{hole}}=0.73$ at the corresponding $T_c$.
The paramagnetic response is weaker, but we still have $r_{\mathrm{hole}}>r_c$
[Fig.~\ref{fig:kagome_bands}(c)], showing that the predicted paramagnetism
persists under this small change in filling near the van Hove point.
\par\endgroup

\begingroup\clubpenalty=10000\brokenpenalty=10000\displaywidowpenalty=10000
We now assume that the normal state already hosts a weak loop-current
order, as proposed for kagome superconductors $A$V$_3$Sb$_5$
($A=\mathrm{K},\mathrm{Rb},\mathrm{Cs}$)~\cite{FengEtAl2021,DennerThomaleNeupert2021}.
Here we consider the simplest translation-preserving chiral flux
pattern~\cite{OhgushiMurakamiNagaosa2000} illustrated in
Fig.~\ref{fig:kagome_main}(a), and investigate its interplay with chiral
superconducting fluctuations.
The chiral flux order induces a time-reversal-odd, quadratic $A_{2g}$
chiral splitting term in the Ginzburg--Landau free energy~\cite{HolbaekFischer2025,TazaiYamakawaKontani2026},
parametrized by $\delta$ as
\begin{equation}
 f_{\rm LC}=\delta(|\eta_+|^2-|\eta_-|^2).
 \label{eq:kagome_loop_current}
\end{equation}
\endgroup
The loop-current order thus adds a $\delta\sigma_z$ term to
Eq.~\eqref{eq:kagome_kernel}, lifting the degeneracy between the two chiral
branches and selecting the lower-mass branch as the favored pairing state.
The sign of the splitting $\delta$ depends on the relative chirality between
the pairing and loop-current orders.
We take $\delta>0$, which favors $\eta_-$ for the positive triangle hopping
phase $\phi=0.03$ in the microscopic model, as presented in
\textbf{SM~S4.1}~\cite{SupplementaryMaterial}.
Reversing the loop-current chirality sends $\delta\to-\delta$ and favors
$\eta_+$ instead.
Near the pairing transition of $\eta_-$, the two masses are parametrized as
$\epsilon=\alpha-\delta>0$ and $\epsilon+2\delta$, with
$\epsilon\simeq a_E(T-T_c)$.
Within this approximation, $T_c$ is shifted upward by
$\delta/a_E$ from its original value.

\begin{samepage}
The temperature dependence of the orbital magnetic susceptibility in this case is
shown in Fig.~\ref{fig:kagome_main}(b). With
$\chi_\epsilon=e^{*2}T\gamma_0/(12\pi d_s\epsilon)$ and
$\mathcal X(x)=\chi^{\mathrm{fl}}_{zz}/\chi_\epsilon$, we obtain
\begingroup
\thinmuskip=2mu
\medmuskip=1mu
\thickmuskip=2mu
\begin{equation}
 \mathcal X(x)=\frac{12r^2x(4u^2-D^2)}{D^3}\operatorname{atanh}\frac{D}{u}
 -\frac{u(13D^2-12)}{(1+x)D^2},
 \label{eq:kagome_split_limits}
\end{equation}
\endgroup
where $u=1+2x$ and $D=\sqrt{1+4r^2x(x+1)}$.
Here, the temperature-dependent variable $x=\epsilon/(2\delta)$ controls the
low-energy distinction between the two chiral branches.
\end{samepage}
The branch separation sets the temperature scale $T_{\mathrm{split}}=2\delta/a_E$
(corresponding to $x\simeq1$), which controls whether
the superconducting fluctuations resolve the chiral splitting.
In the limit $x\gg1$, $\mathcal X(x)\simeq\mathcal S_E(r)$:
both branches are effectively degenerate, so the fluctuations are insensitive
to the splitting. Thus the diamagnetic and paramagnetic contributions to the
susceptibility both scale as $1/x$, so their relative amplitudes determine the sign,
giving fluctuation paramagnetism for the microscopic parameters used above.
In the limit $x\ll1$, where $\mathcal X(x)\simeq-1$,
the fluctuations resolve the two chiral branches and only one branch becomes soft.
For the kagome model of Eq.~\eqref{eq:kagome_kernel} with the chiral splitting in
Eq.~\eqref{eq:kagome_loop_current}, we notice that
$g_{ab}(\mathbf0)=\omega_{xy}(\mathbf0)=0$. Thus, the intramode diamagnetic
contribution retains its $1/x$ divergence, whereas the intermode
paramagnetic contribution crosses over to the weaker logarithmic divergence
$\ln(1/x)$.
The diamagnetic term therefore dominates sufficiently close to $T_c$,
restoring the traditional single-component diamagnetic singularity according to
Eq.~\eqref{eq:single_soft_response}.
A crossover from fluctuation paramagnetism to fluctuation diamagnetism
occurs between these two regimes [Fig.~\ref{fig:kagome_main}(b)],
with an inevitable zero crossing $\chi_{zz}^{\mathrm{fl}}(T_*)=0$ at the crossover
temperature $T_*$.
For $r=r_{\mathrm{VHS}}$, the curve in
Fig.~\ref{fig:kagome_main}(b) changes sign at $x_*=0.06$.
A detailed analysis of these two scaling regimes and the crossover is
given in \textbf{SM~S4.3}~\cite{SupplementaryMaterial}.
The orbital magnetic susceptibility $\chi^{\mathrm{fl}}$ is even in $\delta$,
so domains of the loop-current order related by time reversal
make identical contributions that do not cancel each other.
Such a distinctive critical crossover may provide indirect evidence for the
emergence of chiral superconductivity.
\par

\section{Conclusions and Discussion}
\label{sec:discussion}

In this work, we show that the orbital magnetic response of Gaussian fluctuations
in a multicomponent superconductor can be paramagnetic: the quantum geometry of the pair
propagator produces a coherent contribution that can overcome the conventional
diamagnetic response, establishing a mechanism for fluctuation paramagnetism
beyond a description based on coherence lengths alone. The $d\pm id$ doublet
in the microscopic model of a kagome superconductor provides a concrete realization.
If a loop-current background pre-exists in the normal state, it will split
the chiral pairing modes and restore the leading diamagnetic response near
the transition.
For the spin-singlet pairing considered here, the uniform spin response
to Zeeman coupling is generically less singular than the orbital susceptibility near $T_c$ \cite{LarkinVarlamov2005}.
We therefore neglect the spin contribution in our analysis of the
leading singular magnetic response.

Nevertheless, this raises the question of how complete diamagnetism in the
superconductor is recovered from fluctuation paramagnetism.
Panels (c) and (d) of Fig.~\ref{fig:kagome_main} schematically illustrate two routes from
the Gaussian fluctuation susceptibility toward the Meissner effect.
We take $T_g$ to mark the temperature scale below which
nonlinear and quantum corrections can no longer be neglected,
with the Gaussian description valid for
$T>T_g$. For a doublet that remains effectively degenerate down to $T_g$,
the sign reversal is expected to occur within the nonlinear region $T_c<T<T_g$
[Fig.~\ref{fig:kagome_main}(c)]. For a split doublet with $T_g<T_*$,
the Gaussian response already becomes diamagnetic before the nonlinear
critical region is entered [Fig.~\ref{fig:kagome_main}(d)].
Nonlinear effects then connect this diamagnetic response smoothly to the
Meissner effect.
The dashed curves are schematic continuations through the nonlinear
region.
In either case, the change from paramagnetism to diamagnetism would unavoidably appear
on cooling to the superconducting critical temperature.
A quantitative description requires an interacting treatment of the
multicomponent fluctuations, for example an extension of self-consistent
fluctuation theory (SCFT) \cite{AdachiIkeda2016,JiangLiRosenstein2014}.

\begingroup\brokenpenalty=10000
The modern theory of orbital magnetization identifies contributions from
the orbital magnetic moment and the Berry curvature of Bloch states
\cite{XiaoShiNiu2005,ShiVignaleXiaoNiu2007,XiaoChangNiu2010}.
The orbital magnetic susceptibility of Bloch electrons has a
Green-function formulation~\cite{Fukuyama1971} and contains dispersion
and interband geometric contributions in the spectral
representation~\cite{OgataFukuyama2015}.
Specifically, for two-band models~\cite{GaoYangNiu2015,PiechonEtAl2016},
it can be decomposed into an intraband Landau--Peierls term and interband
Berry-curvature ($\chi_\Omega$) and quantum-metric
($\chi_g$, $\widetilde\chi_g$) terms.
In the static Gaussian regime, fluctuating Cooper pairs obey the
Rayleigh--Jeans distribution $n_{\mathrm{RJ}}(\varepsilon)=T/\varepsilon$.
As shown in \textbf{SM~S2.4}~\cite{SupplementaryMaterial},
we establish a one-to-one correspondence between these two different
scenarios: the two-band Bloch-electron orbital magnetic susceptibility
is recovered from two-component superconducting fluctuations by
replacing $n_{\mathrm{RJ}}$ with the Fermi--Dirac distribution
$n_F(E)=[e^{(E-\mu)/T}+1]^{-1}$, the pair eigenvalue spectrum and pair quantum
geometry with the electronic band dispersions and Bloch quantum geometry,
and the Cooper-pair charge and center-of-mass momentum $(e^*,\mathbf p)$
with the single-electron charge and momentum $(e,\mathbf k)$.
We expect this correspondence to extend to arbitrary multicomponent
superconductors and multiband electronic systems under the same replacement
rules. This mathematical similarity reveals the universality of the
quantum-geometric description of orbital magnetism, even for two very
different physical systems.
\par\endgroup

\begingroup
The present theory is based on the weak-field limit and minimal orbital
coupling of the pair propagator. Two extensions for future work are to
determine how fluctuation paramagnetism evolves at finite magnetic fields and to include
orbital magnetic couplings beyond minimal coupling.
Magnetic fields can reduce the effective dimensionality of fluctuations
near the field-dependent superconducting transition, thereby enhancing
the singular behavior of critical fluctuations~\cite{LeeShenoy1972}.
In the context of Eq.~\eqref{eq:kagome_kernel}, exact Landau quantization
is possible and provides a direct route to finite-field magnetic response
and pairing instabilities within the GL regime. For the second scenario,
the Ward--Takahashi identities satisfied by the vertices in
Eqs.~\eqref{eq:derivative_form} and~\eqref{eq:derivative_form_two_photon}
constrain only their longitudinal parts.
Additional transverse vertices, however, cannot be uniquely determined
from minimal coupling alone, but require additional microscopic input.
For the model in Eq.~\eqref{eq:kagome_kernel}, such magnetic couplings can
generate an extra symmetry-allowed term $\kappa B_z(|\eta_+|^2-|\eta_-|^2)$
beyond the minimal-coupling formula.
This term directly splits the chiral fluctuation channels.
Ref.~\cite{SumiyoshiFujimoto2014} showed that such field-induced splitting
can also produce a paramagnetic fluctuation contribution, which, however,
is reported to be parametrically small compared with conventional
fluctuation diamagnetism~\cite{SumiyoshiFujimoto2014}.
\widowpenalty=10000\brokenpenalty=10000\par\endgroup
\FloatBarrier

\section*{Acknowledgments}

We thank F. C. Zhang, K. T. Law, Harold Y. Hwang, Achim Rosch,
and Qian Niu for inspiring discussions.
N.N. was supported by JSPS KAKENHI Grant Numbers 24H00197, 24H02231 and 24K00583, and by the RIKEN TRIP initiative.
Y.M.X. acknowledges support from the Yangyang Development Fund and startup funds from SJTU.
C.H. was supported by the National Science Foundation CAREER Award No.~DMR-2541471. C.H. acknowledges the hospitality of RIKEN, where part of this work was carried out during a research visit.

Claude Opus 5.5, \mbox{OpenAI} GPT-6.1 Sol, and \mbox{OpenAI} GPT-6 Astra assisted with formula simplification,
literature research, numerical calculation of the Ginzburg--Landau
coefficients from the microscopic model, and language polishing.
We directed these tasks and
manually checked the analytical derivations, literature sources,
and numerical code. We take full responsibility for the content and results.

\section*{Data availability}

The data supporting the findings of this study are available from the
corresponding author upon request.

% Continue the references in the current column, as in the supplied PRX example.
\makeatletter
\renewcommand{\bibsection}{%
  \par\begingroup\baselineskip26pt
  \bib@device{\columnwidth}{.28\columnwidth}%
  \endgroup\nobreak\@nobreaktrue\addvspace{19pt}\par
}
\makeatother
% Pre-generated, independent bibliographies for the two parts.
\ifArxivSupplementary

\else

\fi

\clearpage
\onecolumngrid
\setcounter{page}{1}
\renewcommand{\thepage}{S\arabic{page}}
\setcounter{section}{0}
\setcounter{subsection}{0}
\setcounter{equation}{0}
\setcounter{figure}{0}
\setcounter{table}{0}
\renewcommand{\thesection}{S\arabic{section}}
\renewcommand{\thesubsection}{S\arabic{section}.\arabic{subsection}}
\renewcommand{\theequation}{S\arabic{equation}}
\renewcommand{\thefigure}{S\arabic{figure}}
\renewcommand{\thetable}{S\arabic{table}}
\renewcommand{\theHsection}{supp.\arabic{section}}
\renewcommand{\theHsubsection}{supp.\arabic{section}.\arabic{subsection}}
\renewcommand{\theHequation}{supp.\arabic{equation}}
\renewcommand{\theHfigure}{supp.\arabic{figure}}
\renewcommand{\theHtable}{supp.\arabic{table}}
\makeatletter
\renewcommand{\p@subsection}{}
\renewcommand{\bibsection}{\section*{References}}
\makeatother
\ArxivSupplementarytrue
\addtocontents{toc}{\protect\ArxivTocSupplement}
\pdfbookmark[0]{Supplemental Material}{supplement-start}
\begin{center}
{\large\bfseries Supplemental Material for ``Quantum geometry drives orbital paramagnetism in superconducting fluctuations''}\par
\vspace{6pt}
Zi-Ting Sun$^{1}$, Ying-Ming Xie$^{3}$, Chunli Huang$^{4}$, and Naoto Nagaosa$^{1,2,*}$\par
\vspace{4pt}
{\small\itshape $^{1}$RIKEN Center for Emergent Matter Science (CEMS), Wako, Saitama 351-0198, Japan\\
$^{2}$Fundamental Quantum Science Program (FQSP), TRIP Headquarters, RIKEN, Wako 351-0198, Japan\\
$^{3}$Institute of Condensed Matter Physics, School of Physics and Astronomy, Shanghai Jiao Tong University, Shanghai 200240, China\\
$^{4}$Department of Physics and Astronomy, University of Kentucky, Lexington, Kentucky 40506-0055, USA}\par
{\footnotesize *nagaosa@riken.jp}
\end{center}
{\small\let\ArxivRealContentsline\contentsline\tableofcontents}
\section{Ward--Takahashi identities and gauge invariance}
\label{app:ward}

Gauge invariance relates electromagnetic vertices and pair propagators through
Ward--Takahashi identities~\cite{SM:Nambu1960quasi}. Here we show that the
electromagnetic vertices in Eqs.~(4) and (5) of the \textbf{Main Text} satisfy these
identities upon contraction through cubic order in $\mathbf q$,
where $\mathbf q$ is the momentum transferred by the external field.
These identities fix the longitudinal part of the vertices.
We use units $\hbar=c=1$, and take $e<0$ to be the electron charge, so $e^*=2e<0$ is the
Cooper-pair charge and the kinetic momentum is $-i\boldsymbol\nabla-e^*\mathbf A$.
Repeated spatial indices are summed according to the Einstein convention,
as in the \textbf{Main Text}, and $\partial_a\equiv\partial/\partial p_a$.
We work in the stable Gaussian fluctuation regime, where $\hat L_0^{-1}(\mathbf p)$ is
smooth and positive definite. We assume that momentum integration is performed
either over a full periodic Brillouin zone or, for a continuum description,
with a regularization that preserves momentum shifts and eliminates the
surface terms displayed below.

\subsection{Electromagnetic vertices}
\label{app:ward_vertices}

We first derive the full vertices in Eqs.~(4) and (5) of the \textbf{Main Text}.
For a spatially varying vector potential, we construct the minimally coupled
inverse pair propagator by attaching the electromagnetic phase accumulated
along the straight segment from $\mathbf y$ to $\mathbf x$
to its zero-field real-space form~\cite{SM:IsmailBeigiChangLouie2001,SM:MantoiuPurice2004}:
\begin{equation}
 \hat L^{-1}[\mathbf A](\mathbf x,\mathbf y)
 =\exp\!\left[ie^*\int_{\mathbf y}^{\mathbf x}
       \mathbf A(\boldsymbol\ell)\cdot d\boldsymbol\ell\right]
   \hat L_0^{-1}(\mathbf x-\mathbf y).
 \label{eq:A_gauge_covariant_kernel}
\end{equation}
Here $\hat L_0^{-1}(\mathbf x-\mathbf y)$ is the Fourier transform of
$\hat L_0^{-1}(\mathbf p)$, and the line integral is taken along the straight segment
from $\mathbf y$ to $\mathbf x$. The electromagnetic phase is a scalar
in the pair-component space. Under a gauge transformation it acquires the
endpoint phases needed for covariance of the inverse pair propagator. For uniform
$\mathbf A$, its Fourier transform is
$\hat L_0^{-1}(\mathbf p-e^*\mathbf A)$, consistent with the minimal
coupling used in the \textbf{Main Text}.
The straight-segment prescription also specifies the transverse parts of
the vertices.

Parametrize the segment by
$(\mathbf x+\mathbf y)/2+s(\mathbf x-\mathbf y)$ with
$-1/2\leq s\leq1/2$. Expanding the exponential to first and second order
in $\mathbf A$ gives factors $ie^*(\mathbf x-\mathbf y)_c$ and
$-e^{*2}(\mathbf x-\mathbf y)_b(\mathbf x-\mathbf y)_c/2$, respectively,
integrated over the insertion positions on this segment. With the Fourier
convention
\begin{equation}
 \hat L_0^{-1}(\mathbf p)
 =\int d^d(\mathbf x-\mathbf y)\,
       e^{-i\mathbf p\cdot(\mathbf x-\mathbf y)}
       \hat L_0^{-1}(\mathbf x-\mathbf y),
 \label{eq:A_Fourier_convention}
\end{equation}
each factor $(\mathbf x-\mathbf y)_c$ becomes $i\partial_c$.
For momenta $\mathbf q$ and $-\mathbf q$ of the external field, the two
insertion positions supply the relative phase
$\exp[i(s-t)\mathbf q\cdot(\mathbf x-\mathbf y)]$.
Identifying the coefficients with the vertices defined in Eq.~(3) of
the \textbf{Main Text} and reversing $s,t$ within the symmetric integration
intervals gives
\begin{equation}
\begin{aligned}
 \hat\Lambda_c(\mathbf p;\mathbf q)
 &=-e^*\int_{-1/2}^{1/2}ds\,
       \partial_c\hat L_0^{-1}(\mathbf p+s\mathbf q),
 \\
 \hat\Gamma_{bc}(\mathbf p;\mathbf q,-\mathbf q)
 &=e^{*2}\int_{-1/2}^{1/2}ds\int_{-1/2}^{1/2}dt\,
       \partial_b\partial_c\hat L_0^{-1}(\mathbf p+(s-t)\mathbf q).
\end{aligned}
 \label{eq:A_line_Lambda}
\end{equation}
The changes of variables $s\mapsto-s$ and $(s,t)\mapsto(-s,-t)$ leave the
integration domains invariant and give
\begin{equation}
 \hat\Lambda_c(\mathbf p;-\mathbf q)=\hat\Lambda_c(\mathbf p;\mathbf q),
 \qquad
 \hat\Gamma_{bc}(\mathbf p;-\mathbf q,\mathbf q)
 =\hat\Gamma_{bc}(\mathbf p;\mathbf q,-\mathbf q).
    \label{eq:A_vertices_even_q}
\end{equation}
Both vertices are therefore even in $\mathbf q$, so all terms odd in
$\mathbf q$ vanish when the vertices are expanded in $\mathbf q$. To obtain the small-$\mathbf q$ vertex expansions
used in the \textbf{Main Text}, we use the Taylor expansion of the
$\hat\Lambda_c$ integrand in Eq.~\eqref{eq:A_line_Lambda}:
\begin{equation}
 \partial_c\hat L_0^{-1}(\mathbf p+s\mathbf q)
 =\partial_c\hat L_0^{-1}(\mathbf p)
       +s q_a\partial_a\partial_c\hat L_0^{-1}(\mathbf p)
       +\frac{s^2}{2}q_aq_b
       \partial_a\partial_b\partial_c\hat L_0^{-1}(\mathbf p)
       +O(\mathbf q^3).
 \label{eq:A_line_Taylor}
\end{equation}
The corresponding expansion for $\hat\Gamma_{bc}$ in Eq.~\eqref{eq:A_line_Lambda} just replaces $s$
by $s-t$ and $\partial_c$ by $\partial_b\partial_c$ in
Eq.~\eqref{eq:A_line_Taylor}. The first moments
vanish on the symmetric intervals, while the second moments are
\begin{equation}
\begin{aligned}
 \int_{-1/2}^{1/2}ds\,s^2
   &=\left[\frac{s^3}{3}\right]_{-1/2}^{1/2}=\frac1{12},\\
 \int_{-1/2}^{1/2}ds\int_{-1/2}^{1/2}dt\,(s-t)^2
   &=\int_{-1/2}^{1/2}ds\int_{-1/2}^{1/2}dt\,(s^2+t^2-2st)
   =\frac16.
\end{aligned}
 \label{eq:A_path_second_moments}
\end{equation}
Together with the Taylor factor $1/2$, these give the expanded vertices
in Eqs.~(4) and (5) of the \textbf{Main Text}:
\begin{equation}
\begin{aligned}
 \hat\Lambda_c(\mathbf p;\mathbf q)
 &=-e^*\left[\partial_c\hat L_0^{-1}(\mathbf p)
      +\frac1{24}q_aq_b
       \partial_a\partial_b\partial_c\hat L_0^{-1}(\mathbf p)\right]
       +O(\mathbf q^4),\\
 \hat\Gamma_{bc}(\mathbf p;\mathbf q,-\mathbf q)
 &=e^{*2}\left[\partial_b\partial_c\hat L_0^{-1}(\mathbf p)
      +\frac1{12}q_aq_d
       \partial_a\partial_d\partial_b\partial_c\hat L_0^{-1}(\mathbf p)\right]
       +O(\mathbf q^4).
\end{aligned}
 \label{eq:A_expanded_vertices}
\end{equation}

\subsection{Ward--Takahashi identities}
\label{app:ward_identities}

We now check the Ward--Takahashi identities. Gauge covariance of the
inverse pair propagator implies the one-photon identity
\begin{equation}
    q_c \hat{\Lambda}_c(\mathbf p;\mathbf q)
    =
    -e^*
    \left[
    \hat L_0^{-1}\!\left(\mathbf p+\frac{\mathbf q}{2}\right)
    -
    \hat L_0^{-1}\!\left(\mathbf p-\frac{\mathbf q}{2}\right)
    \right] .
    \label{eq:one_photon_WI}
\end{equation}
Our convention gives
$\hat\Lambda_c(\mathbf p;\mathbf0)=-e^*\partial_c\hat L_0^{-1}$.
Expanding the right-hand side through cubic order in $\mathbf q$ gives
\begin{equation}
    \hat L_0^{-1}\!\left(\mathbf p+\frac{\mathbf q}{2}\right)
    -
    \hat L_0^{-1}\!\left(\mathbf p-\frac{\mathbf q}{2}\right)
    =
    q_c \partial_c \hat L_0^{-1}(\mathbf p)
    +\frac{1}{24}q_aq_bq_c\partial_a\partial_b\partial_c\hat L_0^{-1}(\mathbf p)
    +O(\mathbf q^5),
    \label{eq:one_photon_central_difference}
\end{equation}
Contracting the one-photon vertex in Eq.~\eqref{eq:A_expanded_vertices} gives
\begin{equation}
    q_c\hat{\Lambda}_c(\mathbf p;\mathbf q)
    =
    -e^* q_c\partial_c \hat L_0^{-1}(\mathbf p)
    -\frac{e^*}{24}q_aq_bq_c\partial_a\partial_b\partial_c\hat L_0^{-1}(\mathbf p)
    +O(\mathbf q^5).
    \label{eq:Lambda_long_wavelength}
\end{equation}
The two expansions agree through cubic order in $\mathbf q$, with an
$O(\mathbf q^5)$ residual, verifying Eq.~\eqref{eq:one_photon_WI} to the order
used in the \textbf{Main Text}.

The two-photon Ward--Takahashi identity reads
\begin{equation}
    q_b\hat\Gamma_{bc}(\mathbf p;\mathbf q,-\mathbf q)
    =
    -e^\ast
    \left[
        \hat\Lambda_c\left(\mathbf p+\frac{\mathbf q}{2};-\mathbf q\right)
        -
        \hat\Lambda_c\left(\mathbf p-\frac{\mathbf q}{2};-\mathbf q\right)
    \right].
    \label{eq:two_photon_Ward_full}
\end{equation}
For the two-photon vertex in Eq.~\eqref{eq:A_expanded_vertices}, contraction of the retained terms gives
\begin{equation}
 q_b\hat\Gamma_{bc}(\mathbf p;\mathbf q,-\mathbf q)
 =e^{*2}q_b\partial_b\partial_c\hat L_0^{-1}
 +\frac{e^{*2}}{12}q_aq_dq_b\partial_a\partial_d\partial_b\partial_c\hat L_0^{-1}
 +O(\mathbf q^5),
 \label{eq:two_photon_q3_check}
\end{equation}
where the derivatives are evaluated at $\mathbf p$.
Expanding the right-hand side of Eq.~\eqref{eq:two_photon_Ward_full}
produces the same coefficients as Eq.~\eqref{eq:two_photon_q3_check}.
In detail, the two momentum differences entering the one-photon vertex are
\begin{equation}
\begin{aligned}
\partial_c\hat L_0^{-1}(\mathbf p+\mathbf q/2)
 -\partial_c\hat L_0^{-1}(\mathbf p-\mathbf q/2)
 &=q_b\partial_b\partial_c\hat L_0^{-1}
 +\frac{q_aq_dq_b}{24}\partial_a\partial_d\partial_b\partial_c\hat L_0^{-1}
 +O(\mathbf q^5),\\
\partial_a\partial_d\partial_c\hat L_0^{-1}(\mathbf p+\mathbf q/2)
 -\partial_a\partial_d\partial_c\hat L_0^{-1}(\mathbf p-\mathbf q/2)
 &=q_b\partial_a\partial_d\partial_b\partial_c\hat L_0^{-1}
 +O(\mathbf q^3).
\end{aligned}
\label{eq:A_two_photon_differences}
\end{equation}
Substituting the one-photon vertex in Eq.~\eqref{eq:A_expanded_vertices}
into the right-hand side of Eq.~\eqref{eq:two_photon_Ward_full} therefore gives
\begin{equation}
-e^*\big[\hat\Lambda_c(\mathbf p+\mathbf q/2;-\mathbf q)
          -\hat\Lambda_c(\mathbf p-\mathbf q/2;-\mathbf q)\big]
 =e^{*2}q_b\partial_b\partial_c\hat L_0^{-1}
  +\frac{e^{*2}}{12}q_aq_dq_b
       \partial_a\partial_d\partial_b\partial_c\hat L_0^{-1}
  +O(\mathbf q^5).
\label{eq:A_two_photon_rhs}
\end{equation}
Thus both identities hold through $O(\mathbf q^3)$, with residuals of
$O(\mathbf q^5)$ for the truncated vertices of the \textbf{Main Text}.

We next contract the response kernel in Eq.~(10) of the
\textbf{Main Text} to examine its Ward identity.
Using the one-photon Ward identity in the second term of Eq.~(10),
the adjacent propagators cancel the two inverse propagators separately:
\begin{equation}
\hat L_0(\mathbf p+\mathbf q/2)
 \big[\hat L_0^{-1}(\mathbf p+\mathbf q/2)
      -\hat L_0^{-1}(\mathbf p-\mathbf q/2)\big]
 \hat L_0(\mathbf p-\mathbf q/2)
 =\hat L_0(\mathbf p-\mathbf q/2)
       -\hat L_0(\mathbf p+\mathbf q/2).
\label{eq:A_resolvent_Ward}
\end{equation}
Using this identity and also the two-photon Ward identity yields
\begin{equation}
\begin{aligned}
q_bK_{bc}^{\mathrm{fl}}(\mathbf q)
={}&-e^*T\int\frac{d^dp}{(2\pi)^d}\operatorname{tr}_{\mathrm{mat}}\Big\{
 \hat L_0(\mathbf p)
 \big[\hat\Lambda_c(\mathbf p+\mathbf q/2;-\mathbf q)
      -\hat\Lambda_c(\mathbf p-\mathbf q/2;-\mathbf q)\big]\\
&-\big[\hat L_0(\mathbf p-\mathbf q/2)
             -\hat L_0(\mathbf p+\mathbf q/2)\big]
             \hat\Lambda_c(\mathbf p;-\mathbf q)\Big\}.
\end{aligned}
\label{eq:A_contracted_kernel}
\end{equation}
Here $T$ is the temperature in units $k_B=1$, and
$\operatorname{tr}_{\mathrm{mat}}$ is the trace over component indices.
In the contact contribution, which is the first line of
Eq.~\eqref{eq:A_contracted_kernel}, set $\mathbf k=\mathbf p+\mathbf q/2$
in the term containing $\hat\Lambda_c(\mathbf p+\mathbf q/2;-\mathbf q)$,
and $\mathbf k=\mathbf p-\mathbf q/2$ in the other term.
Relabeling $\mathbf k$ as $\mathbf p$ gives
\begin{equation}
\begin{alignedat}{2}
&\int\frac{d^dp}{(2\pi)^d}\operatorname{tr}_{\mathrm{mat}}\Big\{
 \hat L_0(\mathbf p)
 \big[\hat\Lambda_c(\mathbf p+\mathbf q/2;-\mathbf q)
      -\hat\Lambda_c(\mathbf p-\mathbf q/2;-\mathbf q)\big]&&\Big\}\\
&\quad=\int\frac{d^dp}{(2\pi)^d}\operatorname{tr}_{\mathrm{mat}}\Big\{
 \big[\hat L_0(\mathbf p-\mathbf q/2)
      -\hat L_0(\mathbf p+\mathbf q/2)\big]
       \hat\Lambda_c(\mathbf p;-\mathbf q)&&\Big\}.
\end{alignedat}
\label{eq:A_contact_shift}
\end{equation}
These shifts preserve the integral under the boundary assumptions stated above.
Thus the right-hand side exactly cancels the bubble contribution in
Eq.~\eqref{eq:A_contracted_kernel}.
The other contraction follows from
$K_{bc}^{\mathrm{fl}}(\mathbf q)=K_{cb}^{\mathrm{fl}}(-\mathbf q)$,
the symmetry of the second derivatives of the free energy.
Since these contractions vanish for all sufficiently small $\mathbf q$,
their Taylor coefficients vanish at each power of $\mathbf q$.
In particular, the quadratic coefficient $C_{ab;cd}$ in Eq.~(12) of the
\textbf{Main Text} satisfies the cubic constraint. Hence
\begin{equation}
 q_bK_{bc}^{\mathrm{fl}}(\mathbf q)=0,
 \qquad K_{bc}^{\mathrm{fl}}(\mathbf q)q_c=0,
 \qquad q_aC_{ab;cd}q_cq_d=0.
 \label{eq:responsetensor}
\end{equation}
These are the Ward identities for the electromagnetic response kernel,
which establish the gauge invariance of the theory.

\subsection{Long-wavelength response kernel}
\label{app:ward_response}

We next show explicitly how Eqs.~(4), (5), and (10) of the \textbf{Main Text}
lead to Eq.~(11). Substituting the vertices in Eqs.~(4) and (5) into Eq.~(10) gives
\begin{equation}
\begin{aligned}
K_{bc}^{\mathrm{fl}}(\mathbf q)
&=e^{*2}T\int\frac{d^dp}{(2\pi)^d}\operatorname{tr}_{\mathrm{mat}}
 \Big\{\hat L_0\partial_b\partial_c\hat L_0^{-1}
 -\hat L_0(\mathbf p+\mathbf q/2)\partial_b\hat L_0^{-1}
  \hat L_0(\mathbf p-\mathbf q/2)\partial_c\hat L_0^{-1}\\
&+\frac{q_iq_j}{24}
 \Big[2\hat L_0\partial_b\partial_c\partial_i\partial_j\hat L_0^{-1}
 -\hat L_0(\partial_b\partial_i\partial_j\hat L_0^{-1})\hat L_0(\partial_c\hat L_0^{-1})
 -\hat L_0(\partial_b\hat L_0^{-1})\hat L_0(\partial_c\partial_i\partial_j\hat L_0^{-1})\Big]\Big\}
 +O(\mathbf q^3).
\end{aligned}
\label{eq:A_kernel_expanded}
\end{equation}
The propagators in the last bracket are evaluated at $\mathbf p$ because
their momentum shifts affect only terms of $O(\mathbf q^3)$ and higher.
Differentiating $\hat L_0\hat L_0^{-1}=\hat1$ gives
$\partial_b\hat L_0=-\hat L_0(\partial_b\hat L_0^{-1})\hat L_0$.
The product rule then yields
\begin{equation}
\begin{aligned}
\partial_b\operatorname{tr}_{\mathrm{mat}}
 [\hat L_0\partial_c\partial_i\partial_j\hat L_0^{-1}]
&=\operatorname{tr}_{\mathrm{mat}}\Big[
 \hat L_0\partial_b\partial_c\partial_i\partial_j\hat L_0^{-1}
 -\hat L_0(\partial_b\hat L_0^{-1})\hat L_0
 (\partial_c\partial_i\partial_j\hat L_0^{-1})\Big],\\
\partial_c\operatorname{tr}_{\mathrm{mat}}
 [\hat L_0\partial_b\partial_i\partial_j\hat L_0^{-1}]
&=\operatorname{tr}_{\mathrm{mat}}\Big[
 \hat L_0\partial_b\partial_c\partial_i\partial_j\hat L_0^{-1}
 -\hat L_0(\partial_b\partial_i\partial_j\hat L_0^{-1})
 \hat L_0(\partial_c\hat L_0^{-1})\Big],
\end{aligned}
\label{eq:A_surface_product_b}
\end{equation}
where the second equality uses trace cyclicity. Adding these identities converts the $q_iq_j$ contribution
in the expanded response kernel into the two total derivatives
\begin{equation}
 \frac{e^{*2}T}{24}q_iq_j \int\frac{d^dp}{(2\pi)^d} \Big\{\partial_b\operatorname{tr}_{\mathrm{mat}} [\hat L_0\partial_c\partial_i\partial_j\hat L_0^{-1}] +\partial_c\operatorname{tr}_{\mathrm{mat}} [\hat L_0\partial_b\partial_i\partial_j\hat L_0^{-1}] \Big\}=0 .
 \label{eq:A_vertex_surface}
\end{equation}
Thus the explicit $O(\mathbf q^2)$ vertex corrections make no net
contribution, and the remaining terms give Eq.~(11) of the \textbf{Main Text}.

We write the right logarithmic derivative (RLD) of the pair propagator as
$\hat R_a=\hat L_0^{-1}\partial_a\hat L_0$.
Differentiating this definition gives the Maurer--Cartan identity
$\partial_a\hat R_b-\partial_b\hat R_a=-[\hat R_a,\hat R_b]$.
The zero-momentum response kernel is
\begin{equation}
K_{bc}^{\mathrm{fl}}(0) =-e^{*2}T
\int\frac{d^dp}{(2\pi)^d}
\partial_b
\operatorname{tr}_{\mathrm{mat}}\hat R_c=0.
\end{equation}
The diamagnetic and paramagnetic contributions therefore cancel exactly
at zero external momentum.

For nonzero $\mathbf q$, we expand $K_{bc}^{\mathrm{fl}}(\mathbf q)$ in
$\mathbf q$. To do so, we first expand the pair propagator as
\begin{equation}
\hat{L}_0\left(\mathbf p \pm \frac{\mathbf q}{2}\right) = \hat{L}_0(\mathbf p) \pm \frac{ q_a}{2}\partial_a \hat{L}_0(\mathbf p) + \frac{ q_a  q_b}{8} \partial_a \partial_b \hat{L}_0(\mathbf p)+O(\mathbf q^3).
\label{eq:Taylor_exp}
\end{equation}
After substituting Eq.~\eqref{eq:Taylor_exp} into Eq.~(11) of the \textbf{Main Text}, the $O(\mathbf q)$ term $K_{bc}^{\mathrm{fl}(1)}(\mathbf q)=C_{bc;a}q_a$ has coefficient
\begin{equation}\label{eq:linear_q_RLD}
C_{bc;a}
=
-\frac{e^{*2}T}{2}
\int\frac{d^dp}{(2\pi)^d}
\operatorname{tr}_{\mathrm{mat}}
\left(
    [\hat R_a,\hat R_b]\hat R_c
\right),
\end{equation}
which is the coefficient of a static current response odd in wavevector.
Trace cyclicity makes $\operatorname{tr}_{\mathrm{mat}}
([\hat R_a,\hat R_b]\hat R_c)$ totally antisymmetric in the three spatial indices.
It therefore vanishes identically in two dimensions, where three distinct
spatial indices are unavailable.
For a smooth periodic positive-definite propagator, its momentum integral
also vanishes in three dimensions. To see this, a smooth variation of the
propagator gives
$\delta\hat R_a=\partial_a(\hat L_0^{-1}\delta\hat L_0)
+[\hat R_a,\hat L_0^{-1}\delta\hat L_0]$.
The Maurer--Cartan identity and trace cyclicity then give
\begin{equation}
\delta\operatorname{tr}_{\mathrm{mat}}([\hat R_a,\hat R_b]\hat R_c)
=\partial_a\operatorname{tr}_{\mathrm{mat}}(\hat L_0^{-1}\delta\hat L_0[\hat R_b,\hat R_c])
 +\partial_b\operatorname{tr}_{\mathrm{mat}}(\hat L_0^{-1}\delta\hat L_0[\hat R_c,\hat R_a])
 +\partial_c\operatorname{tr}_{\mathrm{mat}}(\hat L_0^{-1}\delta\hat L_0[\hat R_a,\hat R_b]).
\label{eq:A_triple_trace_variation}
\end{equation}
Consider the path
$\hat L_t=(1-t)\hat L_0+t\hat L_{\mathrm{ref}}$, $0\leq t\leq1$,
where $\hat L_{\mathrm{ref}}$ is a constant positive-definite matrix,
and define $\hat R_a(t)=\hat L_t^{-1}\partial_a\hat L_t$.
Both endpoints are positive definite, so their convex combination stays
positive definite and the inverse exists throughout the path.
Along this path, Eq.~\eqref{eq:A_triple_trace_variation}, evaluated at
$\hat L_t$ with $\delta$ replaced by $\partial_t$, expresses the $t$
derivative of the integrand in Eq.~\eqref{eq:linear_q_RLD} as a sum of
total momentum derivatives.
Their integrals vanish on the periodic Brillouin zone, so $C_{bc;a}(t)$, the coefficient
in Eq.~\eqref{eq:linear_q_RLD} evaluated along this path, is independent
of $t$ and satisfies $C_{bc;a}(0)=C_{bc;a}(1)$.
At $t=1$, $\hat L_t=\hat L_{\mathrm{ref}}$ is momentum independent,
hence $\partial_a\hat L_t|_{t=1}=0$ and $\hat R_a(1)=0$.
Therefore $C_{bc;a}(1)=0$, and path independence gives
$C_{bc;a}=C_{bc;a}(0)=0$.
This result can be explained by the fact that a uniform static magnetic
field cannot induce an equilibrium bulk transport current.

To obtain the quadratic coefficient $C_{ab;cd}$, the same Taylor expansion
of Eq.~(11) of the \textbf{Main Text} gives
\begin{equation}
\begin{aligned}
C_{ab;cd}={}&\mathrel{-}\frac{e^{*2}T}{8}\int\frac{d^dp}{(2\pi)^d}
\operatorname{tr}_{\mathrm{mat}}\Big[
(\partial_c\partial_d\hat L_0)(\partial_a\hat L_0^{-1})
 \hat L_0(\partial_b\hat L_0^{-1})
 +\hat L_0(\partial_a\hat L_0^{-1})
 (\partial_c\partial_d\hat L_0)(\partial_b\hat L_0^{-1})\\
&\mathrel{-}(\partial_c\hat L_0)(\partial_a\hat L_0^{-1})
 (\partial_d\hat L_0)(\partial_b\hat L_0^{-1})
 -(\partial_d\hat L_0)(\partial_a\hat L_0^{-1})
 (\partial_c\hat L_0)(\partial_b\hat L_0^{-1})\Big].
\end{aligned}
\label{eq:A_quadratic_Taylor_coefficient}
\end{equation}
The first two terms in Eq.~\eqref{eq:A_quadratic_Taylor_coefficient}
come from the second derivatives of the individual
propagators, and the last two from the product of their first derivatives,
symmetrized in $c,d$. We now use
$\partial_a\hat L_0=\hat L_0\hat R_a$,
$\partial_a\hat L_0^{-1}=-\hat R_a\hat L_0^{-1}$, and
\begin{equation}
\partial_c\partial_d\hat L_0
=\frac{\hat L_0}{2}
\left(\partial_c\hat R_d+\partial_d\hat R_c
      +\hat R_c\hat R_d+\hat R_d\hat R_c\right).
\label{eq:A_second_derivative_RLD}
\end{equation}
After symmetrizing $c,d$, trace cyclicity combines the products of four
$\hat R$ matrices as
\begin{equation}
\begin{aligned}
\operatorname{tr}_{\mathrm{mat}}\Big[
&(\hat R_c\hat R_d+\hat R_d\hat R_c)(\hat R_a\hat R_b+\hat R_b\hat R_a)
 -2\hat R_c\hat R_a\hat R_d\hat R_b-2\hat R_d\hat R_a\hat R_c\hat R_b\Big]\\
&=-\operatorname{tr}_{\mathrm{mat}}\Big[
[\hat R_a,\hat R_c][\hat R_b,\hat R_d]
+[\hat R_a,\hat R_d][\hat R_b,\hat R_c]\Big].
\end{aligned}
\label{eq:A_quartic_trace_identity}
\end{equation}
The remaining derivative terms give
$(\partial_c\hat R_d+\partial_d\hat R_c)(\hat R_a\hat R_b+\hat R_b\hat R_a)$,
yielding Eq.~(13) of the \textbf{Main Text}.
Using $\hat J_{ab}=\tfrac12(\hat R_a\hat R_b+\hat R_b\hat R_a-\partial_a\hat R_b-\partial_b\hat R_a)$
and $\hat\Xi_{ab}=[\hat R_a,\hat R_b]$, integration by parts gives
\begin{equation}
\begin{aligned}
&\int\frac{d^dp}{(2\pi)^d}\operatorname{tr}_{\mathrm{mat}}
 \left(2\hat J_{ab}\hat J_{cd}-\hat J_{ac}\hat J_{bd}-\hat J_{ad}\hat J_{bc}\right)\\
&\quad=-\frac32\int\frac{d^dp}{(2\pi)^d}\operatorname{tr}_{\mathrm{mat}}
 \left[(\partial_c\hat R_d+\partial_d\hat R_c)
       (\hat R_a\hat R_b+\hat R_b\hat R_a)\right]\\
&\qquad-\frac12\int\frac{d^dp}{(2\pi)^d}\operatorname{tr}_{\mathrm{mat}}
 \left(\hat\Xi_{ac}\hat\Xi_{bd}+\hat\Xi_{ad}\hat\Xi_{bc}\right).
\end{aligned}
\label{eq:A_Hessian_trace_rearrangement}
\end{equation}
The boundary terms vanish under the assumptions stated at the beginning
of this section. Combining this identity with Eq.~\eqref{eq:A_quartic_trace_identity}
gives the Hessian--commutator form used to obtain Eq.~(14) of the
\textbf{Main Text}.

We can verify the Ward identity of $C_{ab;cd}$ in Eq.~\eqref{eq:responsetensor}
directly from Eq.~(13) of the \textbf{Main Text}.
Both commutator products in Eq.~(13) vanish on contraction because
$q_aq_c[\hat R_a,\hat R_c]=q_aq_d[\hat R_a,\hat R_d]=0$.
For the remaining term, the product rule of differentiation and trace cyclicity give
\begin{equation}
\begin{aligned}
q_aC_{ab;cd}q_cq_d
&=-\frac{e^{*2}T}{8}\int\frac{d^dp}{(2\pi)^d}
 \operatorname{tr}_{\mathrm{mat}}\Big\{
 q_c\partial_c(q_i\hat R_i)
 \big[(q_j\hat R_j)\hat R_b+\hat R_b(q_j\hat R_j)\big]\Big\}\\
&=-\frac{e^{*2}T}{8}\int\frac{d^dp}{(2\pi)^d}\Big\{
 q_c\partial_c\operatorname{tr}_{\mathrm{mat}}[(q_i\hat R_i)^2\hat R_b]
 -\operatorname{tr}_{\mathrm{mat}}[(q_i\hat R_i)^2q_c\partial_c\hat R_b]\Big\}\\
&=-\frac{e^{*2}T}{8}\int\frac{d^dp}{(2\pi)^d}\Big\{
 q_c\partial_c\operatorname{tr}_{\mathrm{mat}}[(q_i\hat R_i)^2\hat R_b]
 -\frac13\partial_b\operatorname{tr}_{\mathrm{mat}}[(q_i\hat R_i)^3]\Big\}=0.
\end{aligned}
\label{eq:A_C_ward_verification}
\end{equation}
In the last line, the Maurer--Cartan identity gives
$q_c\partial_c\hat R_b=\partial_b(q_i\hat R_i)-[q_i\hat R_i,\hat R_b]$.
The commutator has zero trace when multiplied by $(q_i\hat R_i)^2$,
and trace cyclicity gives the factor $1/3$ in the remaining derivative.

\section{Pair quantum geometry interpretation}
\label{app:geometry}

\subsection{Eigenmode formulation and two-component case}
\label{app:two_components}

We recast Eqs.~(14) and (16) of the \textbf{Main Text}
in the eigenmode basis of the Hermitian inverse pair propagator,
\begin{equation}
    \hat L_0^{-1}(\mathbf p)|u_n(\mathbf p)\rangle
    =
    \varepsilon_n(\mathbf p)|u_n(\mathbf p)\rangle,
    \qquad
    \varepsilon_n>0.
    \label{eq:E_eigen}
\end{equation}
The transition temperature $T_c$ is determined by
$\min_{n,\mathbf p}\varepsilon_n(\mathbf p;T_c)=0$.
This representation exposes the quantum-geometric content of the response.
All matrix elements below are taken between the eigenstates $|u_n\rangle$.

The Berry connection is $\mathcal A_a^{nm}\equiv i\langle u_n|\partial_au_m\rangle$,
with the index convention of the \textbf{Main Text}.
For $n\neq m$, the intermode quantum geometric tensor
$Q_{ab}^{nm}\equiv\mathcal A_a^{nm}\mathcal A_b^{mn}$ has symmetric and
antisymmetric parts~\cite{SM:ProvostVallee1980,SM:Berry1984}:
\begin{equation}
    g^{nm}_{ab}
    \equiv
    \tfrac12\left(Q^{nm}_{ab}+Q^{nm}_{ba}\right)
    =
    \operatorname{Re}Q^{nm}_{ab},
    \qquad
    \omega^{nm}_{ab}
    \equiv
    i\left(Q^{nm}_{ab}-Q^{nm}_{ba}\right)
    =
    -2\operatorname{Im}Q^{nm}_{ab}.
    \label{eq:E_gomega}
\end{equation}

To simplify subsequent expressions, we define the dimensionless spectral factors
\begin{equation}
 W_{nm}\equiv\frac{(\varepsilon_n-\varepsilon_m)^2}
                       {\varepsilon_n\varepsilon_m},\qquad
 \ell_{nm}\equiv\ln\frac{\varepsilon_n}{\varepsilon_m},\qquad
 r_{nm}\equiv\frac{\varepsilon_n}{\varepsilon_m}-1.
 \label{eq:E_spectral_factors}
\end{equation}
Here $W_{nm}=W_{mn}\geq0$ and $\ell_{nm}=-\ell_{mn}$,
with the ordering of the indices in $r_{nm}$ kept explicitly below.

In this language, the RLD $\hat R_a$ has the matrix elements
\begin{equation}
    R_a^{nm}
    =
    -(\partial_a\ln\varepsilon_n)\delta_{nm}
    -
    i r_{nm}\mathcal A_a^{nm},
    \label{eq:E_R}
\end{equation}
which separates $\hat R_a$ into a spectral contribution, governed by the
momentum dependence of $\varepsilon_n$, and an intermode geometric contribution,
governed by the rotation of the eigenvectors. The commutator
$\hat\Xi_{ab}\equiv[\hat R_a,\hat R_b]$ entering the coherent contribution
of the orbital magnetic susceptibility follows immediately. Its diagonal
element is determined by a $W_{nm}$-weighted Berry curvature,
\begin{equation}
    \Xi_{ab}^{nn}
    =
    -i\sum_{m\neq n}
    W_{nm}
    \,\omega^{nm}_{ab},
    \label{eq:E_Omega_diag}
\end{equation}
while for $n\neq m$ Eq.~\eqref{eq:E_R} gives the exact off-diagonal element
\begin{equation}
 \begin{aligned}
 \Xi_{ab}^{nm}
 &=i r_{nm}
 \left[(\partial_a\ell_{nm})\mathcal A_b^{nm}
       -(\partial_b\ell_{nm})\mathcal A_a^{nm}\right]\\
 &\mathrel{-}\sum_{l\neq n,m}
 r_{nl}r_{lm}
 \left(\mathcal A_a^{nl}\mathcal A_b^{lm}-\mathcal A_b^{nl}\mathcal A_a^{lm}\right).
 \end{aligned}
 \label{eq:E_Omega_offdiag}
\end{equation}
The tensor $\hat J_{ab}=(\partial_a\partial_b\hat L_0^{-1})\hat L_0$
entering the kinetic contribution can also be decomposed into spectral and
geometric contributions, where the diagonal element is
\begin{equation}
 J^{nn}_{ab}
 =\frac{\partial_a\partial_b\varepsilon_n}{\varepsilon_n}
    +2\sum_{m\neq n}r_{mn}g^{nm}_{ab}.
 \label{eq:E_J_diag}
\end{equation}
The first term arises purely from the Hessian of the inverse-propagator eigenvalue.
The second is the quantum-metric correction generated
by intermode mixing.
The off-diagonal Hessian elements follow from
$\hat J_{ab}=\hat R_{(a}\hat R_{b)}-\partial_{(a}\hat R_{b)}$ as
\begin{equation}
 J^{nm}_{ab}
 =\frac12\sum_l \left(R^{nl}_aR^{lm}_b+R^{nl}_bR^{lm}_a\right)
   -\nabla_{(a}R^{nm}_{b)}.
 \label{eq:E_J_full}
\end{equation}
For an operator $\hat O$, differentiation in the moving eigenbasis gives
\begin{equation}
 \nabla_aO^{nm}\equiv\langle u_n|\partial_a\hat O|u_m\rangle
 =\partial_aO^{nm}
 -i\sum_l\left(\mathcal A_a^{nl}O^{lm}-O^{nl}\mathcal A_a^{lm}\right).
 \label{eq:E_eigenbasis_covariant_derivative}
\end{equation}
The connection terms remove the derivatives of the basis vectors from
$\partial_aO^{nm}$. Parentheses denote symmetrization with weight $1/2$, in particular,
$\nabla_{(a}R^{nm}_{b)}=(\nabla_aR^{nm}_b+\nabla_bR^{nm}_a)/2$.
Using completeness,
$\operatorname{tr}_{\mathrm{mat}}(\hat A\hat B)=\sum_{n,m}A^{nm}B^{mn}$,
Eq.~(14) of the \textbf{Main Text} becomes
\begin{equation}
    \chi^{\mathrm{fl}}_{ij}
    =
    -\frac{e^{*2}T}{24}
    \int\frac{d^dp}{(2\pi)^d}
    \epsilon_{iab}\epsilon_{jcd}
    \sum_{n,m}
    \left[
        J^{nm}_{ac}J^{mn}_{bd}
        +
        \Xi^{nm}_{ab}\Xi^{mn}_{cd}
    \right].
    \label{eq:E_chi_exact}
\end{equation}

\noindent\textit{Two components.}

For a two-component propagator the remaining sums can be performed explicitly.
There is a single independent quantum geometric tensor
$Q_{ab}=Q_{ab}^{+-}=\mathcal A_a^{+-}\mathcal A_b^{-+}$.
We abbreviate $g_{ab}\equiv g_{ab}^{+-}$ and
$\omega_{ab}\equiv\omega_{ab}^{+-}$, with the convention of
Eq.~\eqref{eq:E_gomega}. Since
$Q_{ab}=\mathcal A_a^{+-}(\mathcal A_b^{+-})^*$, it has rank at most one and obeys
\begin{equation}
 g_{ac}g_{bd}-g_{ad}g_{bc}=\tfrac14\omega_{ab}\omega_{cd}.
 \label{eq:E_rank_one}
\end{equation}
In the same notation Eq.~\eqref{eq:E_J_diag} reduces to
$J^{\nu\nu}_{ab}=(\partial_a\partial_b\varepsilon_\nu)/\varepsilon_\nu+2r_{\bar\nu\,\nu}g_{ab}$, where $\nu=\pm$ and
$\bar\nu$ is opposite to $\nu$. The off-diagonal elements obey
$J^{-+}_{ab}=(\varepsilon_-/\varepsilon_+)(J^{+-}_{ab})^*$,
which follows from
$\hat J^\dagger_{ab}=\hat L_0\hat J_{ab}\hat L_0^{-1}$.
We use $\ell\equiv\ell_{+-}$ and $W\equiv W_{+-}\geq0$.
There are no intermediate-mode terms in Eq.~\eqref{eq:E_Omega_offdiag}, so that
\begin{equation}
 \Xi^{++}_{ab}=-\Xi^{--}_{ab}=-iW\omega_{ab},\qquad
 \Xi^{+-}_{ab}=2i r_{+-}(\partial_{[a}\ell)\mathcal A_{b]}^{+-}.
 \label{eq:E_two_Omega}
\end{equation}
Hermiticity of $\hat L_0$ gives
$\hat R_a^\dagger=\hat L_0\hat R_a\hat L_0^{-1}$ and hence
$\hat\Xi_{ab}^\dagger=-\hat L_0\hat\Xi_{ab}\hat L_0^{-1}$.
In the eigenmode basis this means
$\Xi_{ab}^{\nu'\nu}=-(\varepsilon_{\nu'}/\varepsilon_\nu)(\Xi_{ab}^{\nu\nu'})^*$.
Using this weighted conjugation for $\Xi_{ab}^{-+}$ in
$\operatorname{tr}_{\mathrm{mat}}(\hat\Xi_{ab}\hat\Xi_{cd})
=\sum_{\nu,\nu'=\pm}\Xi_{ab}^{\nu\nu'}\Xi_{cd}^{\nu'\nu}$ gives
\begin{equation}
 -\tfrac12\operatorname{tr}_{\mathrm{mat}}(\hat\Xi_{ab}\hat\Xi_{cd})
 =W^2\omega_{ab}\omega_{cd}
 +4Wg_{[a[c}(\partial_{d]}\ell)(\partial_{b]}\ell).
 \label{eq:E_two_Omega_trace}
\end{equation}
Square brackets antisymmetrize each enclosed index pair with weight $1/2$,
$X_{[a}Y_{b]}=(X_aY_b-X_bY_a)/2$.

For a two-dimensional film, the kernel of the kinetic contribution in Eq.~(20) of the \textbf{Main Text} is
$K=\operatorname{tr}_{\mathrm{mat}}(\hat J_{xx}\hat J_{yy}-\hat J_{xy}^2)$.
Define the covariant derivative on intermode quantities by
$\mathcal D_a\equiv\partial_a-i(\mathcal A_a^{++}-\mathcal A_a^{--})$.
Away from degeneracies, we choose a smooth local eigenframe; in this region, $W>0$.
Completeness of the two eigenvectors gives
\begin{equation}
 \partial_x\mathcal A_y^{\nu\nu'}-\partial_y\mathcal A_x^{\nu\nu'}
 =i\sum_{\mu=\pm}\left(
 \mathcal A_x^{\nu\mu}\mathcal A_y^{\mu\nu'}
 -\mathcal A_y^{\nu\mu}\mathcal A_x^{\mu\nu'}\right).
 \label{eq:E_two_frame_completeness}
\end{equation}
Its $+-$ entry gives
$\mathcal D_x\mathcal A_y^{+-}=\mathcal D_y\mathcal A_x^{+-}$.
The difference of its diagonal entries instead gives
\begin{equation}
 \partial_x(\mathcal A_y^{++}-\mathcal A_y^{--})
 -\partial_y(\mathcal A_x^{++}-\mathcal A_x^{--})=2\omega_{xy},
 \qquad [\mathcal D_x,\mathcal D_y]=-2i\omega_{xy}.
 \label{eq:E_two_connection_curl}
\end{equation}
The Berry-connection terms cancel when differentiating $g_{ab}=\operatorname{Re}[\mathcal A_a^{+-}(\mathcal A_b^{+-})^*]$:
\begin{equation}
 \partial_c g_{ab}
 =\operatorname{Re}\!\left[
 (\mathcal D_c\mathcal A_a^{+-})(\mathcal A_b^{+-})^*
 +\mathcal A_a^{+-}(\mathcal D_c\mathcal A_b^{+-})^*\right].
 \label{eq:E_two_metric_derivative}
\end{equation}
Differentiating once more and using the symmetry of
$\mathcal D_x\mathcal A_y^{+-}$ gives
\begin{equation}
\begin{aligned}
 \partial_a\partial_b(\operatorname{adj}g)_{ab}
 &=\partial_x^2g_{yy}+\partial_y^2g_{xx}-2\partial_x\partial_yg_{xy}\\
 &=2|\mathcal D_x\mathcal A_y^{+-}|^2
 -2\operatorname{Re}\!\left[
 (\mathcal D_x\mathcal A_x^{+-})(\mathcal D_y\mathcal A_y^{+-})^*\right]
 +2\operatorname{Re}\!\left[
 ([\mathcal D_y,\mathcal D_x]\mathcal A_y^{+-})(\mathcal A_x^{+-})^*\right].
\end{aligned}
 \label{eq:E_two_metric_double_derivative}
\end{equation}
Here $\operatorname{adj}g$ is the adjugate matrix, with
$(\operatorname{adj}g)_{xx}=g_{yy}$,
$(\operatorname{adj}g)_{yy}=g_{xx}$, and
$(\operatorname{adj}g)_{xy}=(\operatorname{adj}g)_{yx}=-g_{xy}$.
Since $\omega_{xy}=-2\operatorname{Im}[\mathcal A_x^{+-}(\mathcal A_y^{+-})^*]$,
the last term is $-2\omega_{xy}^2$. Rearranging therefore gives
\begin{equation}
 \operatorname{Re}\!\left[(\mathcal D_x\mathcal A_x^{+-})
 (\mathcal D_y\mathcal A_y^{+-})^*\right]
 -|\mathcal D_x\mathcal A_y^{+-}|^2
 =-\frac12\partial_a\partial_b(\operatorname{adj}g)_{ab}-\omega_{xy}^2.
 \label{eq:E_two_metric_covariant_identity}
\end{equation}

To connect this identity to $K$, differentiate
$\langle u_+|\partial_b\hat L_0^{-1}|u_-\rangle
=i(\varepsilon_+-\varepsilon_-)\mathcal A_b^{+-}$,
including the derivatives of both eigenvectors. The off-diagonal Hessian is
\begin{equation}
\begin{aligned}
 \varepsilon_-J_{ab}^{+-}
 &=\mathcal D_a\!\left[i(\varepsilon_+-\varepsilon_-)\mathcal A_b^{+-}\right]
 +i\partial_b(\varepsilon_+-\varepsilon_-)\mathcal A_a^{+-}\\
 &=i\left[
 \partial_a(\varepsilon_+-\varepsilon_-)\mathcal A_b^{+-}
 +\partial_b(\varepsilon_+-\varepsilon_-)\mathcal A_a^{+-}
 +(\varepsilon_+-\varepsilon_-)\mathcal D_a\mathcal A_b^{+-}\right].
\end{aligned}
 \label{eq:E_two_J_offdiag}
\end{equation}
Combining
Eq.~\eqref{eq:E_two_metric_derivative} with
$\mathcal D_a\mathcal A_b^{+-}=\mathcal D_b\mathcal A_a^{+-}$ gives
\begin{equation}
 2\operatorname{Re}\!\left[
 (\mathcal D_a\mathcal A_b^{+-})(\mathcal A_c^{+-})^*\right]
 =\partial_ag_{bc}+\partial_bg_{ac}-\partial_cg_{ab}.
 \label{eq:E_two_metric_cross_terms}
\end{equation}
In the pair eigenbasis, we decompose the kernel as $K=K_{\mathrm{intra}}+K_{\mathrm{inter}}$.
Equal and unequal eigenmode indices define the intramode and intermode kernels, respectively.
The diagonal Hessian elements in Eq.~\eqref{eq:E_J_diag} give
\begin{equation}
 K_{\mathrm{intra}}
 =\sum_{\nu=\pm}\left[
 J_{xx}^{\nu\nu}J_{yy}^{\nu\nu}-(J_{xy}^{\nu\nu})^2\right]
 =\sum_{\nu=\pm}\det\!\left[
 \frac{\partial_a\partial_b\varepsilon_\nu
       +2(\varepsilon_{\bar\nu}-\varepsilon_\nu)g_{ab}}{\varepsilon_\nu}
 \right].
 \label{eq:E_K_intra_inter}
\end{equation}
Here the determinant is taken over the spatial indices $a,b=x,y$.

For the intermode kernel, use $J_{ab}^{-+}=(\varepsilon_-/\varepsilon_+)(J_{ab}^{+-})^*$
in the off-diagonal trace. Products quadratic in
$\partial_a(\varepsilon_+-\varepsilon_-)$ give metric terms,
while mixed products reduce by Eq.~\eqref{eq:E_two_metric_cross_terms} to the
first derivatives of $\operatorname{adj}g$. Products quadratic in
$\mathcal D_a\mathcal A_b^{+-}$ reduce by
Eq.~\eqref{eq:E_two_metric_covariant_identity} to the second derivative
of $\operatorname{adj}g$ and the curvature term $-2W\omega_{xy}^2$.
The product rule applied to
$\partial_a\partial_b[(\varepsilon_+-\varepsilon_-)^2(\operatorname{adj}g)_{ab}]$
combines these derivative terms, giving
\begin{equation}
 K_{\mathrm{inter}}
 =\sum_{\nu\ne \nu'}\left(J_{xx}^{\nu\nu'}J_{yy}^{\nu'\nu}-J_{xy}^{\nu\nu'}J_{xy}^{\nu'\nu}\right)
 =-\frac{\partial_a\partial_b\!\left[
       (\varepsilon_+-\varepsilon_-)^2(\operatorname{adj}g)_{ab}\right]
       -2(\varepsilon_+-\varepsilon_-)(\operatorname{adj}g)_{ab}
       \partial_a\partial_b(\varepsilon_+-\varepsilon_-)}
      {\varepsilon_+\varepsilon_-}
       -2W\omega_{xy}^2.
 \label{eq:E_two_K_offdiag}
\end{equation}
Adding Eqs.~\eqref{eq:E_K_intra_inter} and \eqref{eq:E_two_K_offdiag} and using
$\det g=\omega_{xy}^2/4$ gives the full kinetic kernel:
\begin{samepage}
\begin{equation}
K=\sum_{\nu=\pm}
 \det\!\left[
 \frac{\partial_a\partial_b\varepsilon_\nu}{\varepsilon_\nu}+2Wg_{ab}
 \right]
 -\frac{\partial_a\partial_b\!\left[
       (\varepsilon_+-\varepsilon_-)^2(\operatorname{adj}g)_{ab}\right]}
      {\varepsilon_+\varepsilon_-}
 -W^2\omega_{xy}^2.
 \label{eq:E_KH_2d}
\end{equation}
\end{samepage}

The kernel of the coherent contribution satisfies
$\operatorname{tr}_{\mathrm{mat}}\hat\Xi_{xy}^2=-2\mathcal P_{xy}$.
Its intramode and intermode parts follow from the diagonal and off-diagonal commutator matrix elements:
\begin{equation}
\begin{aligned}
 \mathcal P_{xy}
 &=\mathcal P_{xy}^{\mathrm{intra}}+\mathcal P_{xy}^{\mathrm{inter}},\\
 \mathcal P_{xy}^{\mathrm{intra}}
 &=-\frac12\sum_{\nu=\pm}(\Xi_{xy}^{\nu\nu})^2
 =W^2\omega_{xy}^2,\\
 \mathcal P_{xy}^{\mathrm{inter}}
 &=-\frac12\sum_{\nu\ne\nu'}\Xi_{xy}^{\nu\nu'}\Xi_{xy}^{\nu'\nu}
 =\frac{(\operatorname{adj}g)_{ab}(\partial_aW)(\partial_bW)}{W+4}.
\end{aligned}
 \label{eq:E_Pxy}
\end{equation}
Both terms are non-negative, and the intramode part vanishes when $\omega_{xy}=0$.
Eqs.~\eqref{eq:E_KH_2d} and \eqref{eq:E_Pxy} give the separate
kinetic and coherent susceptibilities in Eq.~(16) of the \textbf{Main Text}:
\begin{equation}
\begin{aligned}
 \chi^{\mathrm K}_{zz}
 &=-\frac{e^{*2}T}{12d_s}\int\frac{d^2p}{(2\pi)^2}
 \Biggl\{\sum_{\nu=\pm}\det\!\left[
 \frac{\partial_a\partial_b\varepsilon_\nu}{\varepsilon_\nu}+2Wg_{ab}
 \right]
 -\frac{\partial_a\partial_b\!\left[
 (\varepsilon_+-\varepsilon_-)^2(\operatorname{adj}g)_{ab}\right]}
 {\varepsilon_+\varepsilon_-}
 -W^2\omega_{xy}^2\Biggr\},\\
 \chi^{\mathrm C}_{zz}
 &=\frac{e^{*2}T}{3d_s}\int\frac{d^2p}{(2\pi)^2}
 \left[W^2\omega_{xy}^2
 +\frac{(\operatorname{adj}g)_{ab}(\partial_aW)(\partial_bW)}{W+4}\right].
\end{aligned}
 \label{eq:E_two_chi_parts_2d}
\end{equation}
Adding the two contributions gives Eq.~(22) of the \textbf{Main Text}:
\begin{equation}
\begin{aligned}
 \chi_{zz}^{\mathrm{fl}}
 &=\chi_{zz}^{\mathrm K}+\chi_{zz}^{\mathrm C}
 =\frac{e^{*2}T}{12d_s}\int\frac{d^2p}{(2\pi)^2}
 \Biggl\{\frac{\partial_a\partial_b\!\left[
 (\varepsilon_+-\varepsilon_-)^2(\operatorname{adj}g)_{ab}\right]}
 {\varepsilon_+\varepsilon_-}
 +5W^2\omega_{xy}^2\\
 &\mathrel{+}\frac{4(\operatorname{adj}g)_{ab}\partial_aW\partial_bW}{W+4}
 -\sum_{\nu=\pm}\det\!\left[
 \frac{\partial_a\partial_b\varepsilon_\nu}{\varepsilon_\nu}+2Wg_{ab}
 \right]\Biggr\}.
\end{aligned}
 \label{eq:E_two_chi_2d}
\end{equation}

We can also use a Pauli representation in this two-component scenario.
For $\hat L_0^{-1}=d_0\sigma_0+\mathbf d\cdot\boldsymbol\sigma$, define
$\hat{\mathbf n}=\mathbf d/|\mathbf d|$ and $\mathbf m=\mathbf d/d_0$.
The propagator is $\hat L_0=D(d_0\sigma_0-\mathbf d\cdot\boldsymbol\sigma)$,
where $D=(d_0^2-\mathbf d^2)^{-1}$.
Positive definiteness requires $d_0>|\mathbf d|$, equivalently
$d_0>0$ and $|\mathbf m|<1$.
Away from $\mathbf d=0$, label the branches so that
$\hat{\mathbf n}\cdot\boldsymbol\sigma|u_\nu\rangle=\nu|u_\nu\rangle$.
The spectral quantities then obey
\begin{equation}
 \varepsilon_\nu=d_0(1+\nu|\mathbf m|),\qquad
 W=\frac{4|\mathbf m|^2}{1-|\mathbf m|^2},\qquad
 g_{ab}=\tfrac14\partial_a\hat{\mathbf n}\cdot\partial_b\hat{\mathbf n},\qquad
 \omega_{xy}=-\tfrac12\hat{\mathbf n}\cdot(\partial_x\hat{\mathbf n}\times\partial_y\hat{\mathbf n}).
 \label{eq:two_mode_Pauli_dictionary_main}
\end{equation}
To evaluate the coherent response, substitute
$\hat L_0^{-1}=d_0(\sigma_0+\mathbf m\cdot\boldsymbol\sigma)$ into
$\hat R_a=-(\partial_a\hat L_0^{-1})\hat L_0$.
The scalar part drops out of the commutator, and the Pauli multiplication
rule gives
\begin{equation}
 -\operatorname{tr}_{\mathrm{mat}}[\hat R_x,\hat R_y]^2
 =\frac{8\left\{|\partial_x\mathbf m\times\partial_y\mathbf m|^2
 -[\mathbf m\cdot(\partial_x\mathbf m\times\partial_y\mathbf m)]^2\right\}}
 {(1-\mathbf m^2)^3}.
 \label{eq:E_Pauli_commutator_norm}
\end{equation}
Inserting this into the coherent response and using
$(1-\mathbf m^2)^{-3}=D^3d_0^6$ gives
\begin{equation}
 \chi_{zz}^{\mathrm C}
 =\frac{4e^{*2}T}{3d_s}\int\frac{d^2p}{(2\pi)^2}
 D^3d_0^6\left\{
 |\partial_x\mathbf m\times\partial_y\mathbf m|^2
 -\left[\mathbf m\cdot
 (\partial_x\mathbf m\times\partial_y\mathbf m)\right]^2\right\}.
 \label{eq:main_Pauli_C}
\end{equation}
Since $|\mathbf m|<1$, the integrand vanishes pointwise if and only if
$\partial_x\mathbf m\times\partial_y\mathbf m=0$.
The scalar chirality
$\mathbf m\cdot(\partial_x\mathbf m\times\partial_y\mathbf m)$ then also vanishes.
These chiralities characterize the momentum-space pseudospin texture of
the inverse propagator and determine the paramagnetic coherent contribution.

For the kinetic contribution, define
$X_{ab}=d_0\partial_a\partial_b d_0-
\mathbf d\cdot\partial_a\partial_b\mathbf d$ and
$M_{ab,cd}=(\partial_a\partial_b d_0)(\partial_c\partial_d d_0)
-(\partial_a\partial_b\mathbf d)\cdot(\partial_c\partial_d\mathbf d)$.
The Pauli trace gives
\begin{equation}
 \operatorname{tr}_{\mathrm{mat}}(\hat J_{ab}\hat J_{cd})
 =2D(2DX_{ab}X_{cd}-M_{ab,cd}).
 \label{eq:E_Pauli_H_trace}
\end{equation}
Inserting this identity into Eq.~(16) of the \textbf{Main Text} gives
\begin{equation}
 \chi_{zz}^{\mathrm K}
 =-\frac{e^{*2}T}{6d_s}\int\frac{d^2p}{(2\pi)^2}
 D\left[2D(X_{xx}X_{yy}-X_{xy}^2)
              -M_{xx,yy}+M_{xy,xy}\right].
 \label{eq:main_Pauli_H}
\end{equation}

\subsection{Single soft mode and its coherence length}
\label{app:soft_mode}

We consider Gaussian fluctuations of a generic $N$-component superconductor in a two-dimensional film.
We assume that only one nondegenerate mode $s$ becomes soft near $T_c$
and corresponds to a stable zero-momentum pairing state.
Sufficiently close to $T_c$, the soft-mode eigenvalue has the quadratic expansion
\begin{equation}
 \varepsilon_s(\mathbf p)=\epsilon+\gamma_{ab}p_ap_b+\cdots,
 \qquad \epsilon=\varepsilon_s(\mathbf0)>0.
 \label{eq:E_soft_pole}
\end{equation}
We assume that $\boldsymbol\gamma$ is positive definite and has a finite,
nonsingular limit at $T_c$. The $N-1$ other modes $(m\neq s)$ remain
separated by finite gaps $\varepsilon_m(\mathbf0)>0$ as $\epsilon\to0^+$,
and their dispersions and the local eigenvectors are smooth near
$\mathbf p=\mathbf0$ and $T_c$.
The length $\xi_s$, defined by $\xi_s^2=\sqrt{\det\boldsymbol\gamma}/\epsilon$,
is the geometric mean of the two principal coherence lengths of the critical
mode~\cite{SM:keskiner2026superconductivity,SM:KalkanIskin2026CollectivePairing}.
We also define the eigenvalue-weighted quantum metric and Berry curvature of the soft mode by
\begin{equation}
 \widetilde g_{ab}=\sum_{m\neq s}\varepsilon_m g^{sm}_{ab},
 \qquad
 \widetilde\omega_{xy}=\sum_{m\neq s}\varepsilon_m\omega^{sm}_{xy}.
 \label{eq:E_soft_weighted_geometry}
\end{equation}
At zero momentum, Eq.~\eqref{eq:E_J_diag} gives
$\epsilon J_{ab}^{ss}(\mathbf0)
=2[\gamma_{ab}+\widetilde g_{ab}(\mathbf0)]+O(\epsilon)$.
In the two-dimensional film, the kinetic part of Eq.~\eqref{eq:E_chi_exact} reads
\begin{equation}
 \chi_{zz}^{\mathrm K}
 =-\frac{e^{*2}T}{12d_s}\int\frac{d^2p}{(2\pi)^2}
 \sum_{n,m}\left(J_{xx}^{nm}J_{yy}^{mn}-J_{xy}^{nm}J_{xy}^{mn}\right).
 \label{eq:E_soft_kinetic_2d}
\end{equation}
Since $J_{ab}^{nm}=\langle u_n|\partial_a\partial_b\hat L_0^{-1}|u_m\rangle/\varepsilon_m$,
each product has denominator $\varepsilon_n\varepsilon_m$ and a finite numerator.
Thus the $n=m=s$ term contains two divergent factors $1/\varepsilon_s$, and all other terms integrate at most logarithmically in two dimensions.
Substitution into the kinetic contribution therefore yields
\begin{equation}
 \chi_{zz}^{\mathrm K}
 \simeq-\frac{e^{*2}T}{3d_s}\det[\boldsymbol\gamma+\widetilde{\boldsymbol g}(\mathbf0)]
 \int\frac{d^2p}{(2\pi)^2}\frac1{\varepsilon_s^2}.
 \label{eq:E_soft_sing_H}
\end{equation}
For the coherent term, commutator elements within the massive subspace
also contain the soft mode as an intermediate state. From
Eqs.~\eqref{eq:E_Omega_diag} and~\eqref{eq:E_Omega_offdiag}, the terms
that contribute to the leading singularity are, for $m,n\ne s$,
\begin{equation}
\resizebox{0.92\linewidth}{!}{$\displaystyle
 \Xi_{xy}^{ss}
 \simeq-\frac{i\widetilde\omega_{xy}(\mathbf0)}{\varepsilon_s},\quad
 \Xi_{xy}^{mn}
 \simeq\frac{\varepsilon_m(\mathbf0)}{\varepsilon_s}
 \left[\mathcal A_x^{ms}(\mathbf0)\mathcal A_y^{sn}(\mathbf0)
       -\mathcal A_y^{ms}(\mathbf0)\mathcal A_x^{sn}(\mathbf0)\right],\quad
 \Xi_{xy}^{sm}
 \simeq-\frac{i}{\varepsilon_s}
 \left[(\partial_x\varepsilon_s)\mathcal A_y^{sm}(\mathbf0)
       -(\partial_y\varepsilon_s)\mathcal A_x^{sm}(\mathbf0)\right].
$}
 \label{eq:E_soft_commutator_blocks}
\end{equation}
Here $\varepsilon_s$ and its gradients retain their momentum dependence.
The trace of the squared commutator in the coherent contribution separates into three terms,
\begin{equation}
 -\operatorname{tr}_{\mathrm{mat}}\hat\Xi_{xy}^2
 =|\Xi_{xy}^{ss}|^2
 +\sum_{m,n\ne s}\frac{\varepsilon_n}{\varepsilon_m}|\Xi_{xy}^{mn}|^2
 +2\sum_{m\ne s}\frac{\varepsilon_m}{\varepsilon_s}|\Xi_{xy}^{sm}|^2.
 \label{eq:E_soft_weighted_trace}
\end{equation}
The second term includes $m=n$ as well as $m\ne n$.
Its intermediate soft mode supplies the singular denominator displayed in
Eq.~\eqref{eq:E_soft_commutator_blocks}.

For the leading singular coefficients, we evaluate the massive eigenvalues
and Berry connections at zero momentum, as in Eq.~\eqref{eq:E_soft_commutator_blocks}.
Using $\mathcal A_a^{ms}=(\mathcal A_a^{sm})^*$ and
Eq.~\eqref{eq:E_gomega}, the required weighted overlaps are
\begin{equation}
\begin{aligned}
 \sum_{m\ne s}\varepsilon_m(\mathbf0)
 \mathcal A_x^{sm}(\mathbf0)\mathcal A_y^{ms}(\mathbf0)
 &=\sum_{m\ne s}\varepsilon_m(\mathbf0)
 \left[g_{xy}^{sm}(\mathbf0)-\tfrac i2\omega_{xy}^{sm}(\mathbf0)\right]
 =\widetilde g_{xy}(\mathbf0)-\tfrac i2\widetilde\omega_{xy}(\mathbf0),\\
 \sum_{m\ne s}\varepsilon_m(\mathbf0)|\mathcal A_x^{sm}(\mathbf0)|^2
 &=\widetilde g_{xx}(\mathbf0),\qquad
 \sum_{m\ne s}\varepsilon_m(\mathbf0)|\mathcal A_y^{sm}(\mathbf0)|^2
 =\widetilde g_{yy}(\mathbf0).
\end{aligned}
 \label{eq:E_soft_weighted_overlap}
\end{equation}
The soft-soft contribution (the first term on the right-hand side of
Eq.~\eqref{eq:E_soft_weighted_trace}) follows directly from the first entry of
Eq.~\eqref{eq:E_soft_commutator_blocks}, giving
$\varepsilon_s^2|\Xi_{xy}^{ss}|^2\simeq\widetilde\omega_{xy}^2(\mathbf0)$.
For the massive block (the second term in Eq.~\eqref{eq:E_soft_weighted_trace}),
substitution of the second entry of Eq.~\eqref{eq:E_soft_commutator_blocks} gives
\begin{equation}
\begin{aligned}
 &\varepsilon_s^2\sum_{m,n\ne s}
 \frac{\varepsilon_n(\mathbf0)}{\varepsilon_m(\mathbf0)}|\Xi_{xy}^{mn}|^2
 \simeq\sum_{m,n\ne s}\varepsilon_m(\mathbf0)\varepsilon_n(\mathbf0)
 \left|\mathcal A_x^{ms}(\mathbf0)\mathcal A_y^{sn}(\mathbf0)
       -\mathcal A_y^{ms}(\mathbf0)\mathcal A_x^{sn}(\mathbf0)\right|^2\\
 &\quad=2\widetilde g_{xx}(\mathbf0)\widetilde g_{yy}(\mathbf0)
 -2\operatorname{Re}\!\left[
 \left(\widetilde g_{xy}(\mathbf0)-\tfrac i2\widetilde\omega_{xy}(\mathbf0)\right)^2\right]
 =2\det\widetilde{\boldsymbol g}(\mathbf0)
 +\tfrac12\widetilde\omega_{xy}^2(\mathbf0).
\end{aligned}
 \label{eq:E_soft_massive_block_expansion}
\end{equation}
For the mixed block (the last term in Eq.~\eqref{eq:E_soft_weighted_trace}),
the derivatives of $\varepsilon_s$ are real,
so the imaginary overlap cancels from the expanded modulus,
\begin{equation}
\begin{aligned}
 &\sum_{m\ne s}\varepsilon_m(\mathbf0)
 \left|(\partial_x\varepsilon_s)\mathcal A_y^{sm}(\mathbf0)
       -(\partial_y\varepsilon_s)\mathcal A_x^{sm}(\mathbf0)\right|^2\\
 &\quad=\widetilde g_{yy}(\mathbf0)(\partial_x\varepsilon_s)^2
 +\widetilde g_{xx}(\mathbf0)(\partial_y\varepsilon_s)^2
 -2\widetilde g_{xy}(\mathbf0)(\partial_x\varepsilon_s)(\partial_y\varepsilon_s)
 =(\operatorname{adj}\widetilde g(\mathbf0))_{ab}
 (\partial_a\varepsilon_s)(\partial_b\varepsilon_s).
\end{aligned}
 \label{eq:E_soft_mixed_block_expansion}
\end{equation}
Combining these results, the leading singular part of the coherent contribution is
\begin{equation}
 \chi_{zz}^{\mathrm C}
 \simeq\frac{e^{*2}T}{3d_s}\int\frac{d^2p}{(2\pi)^2}
 \left[
 \frac{\det\widetilde{\boldsymbol g}(\mathbf0)+\tfrac34\widetilde\omega_{xy}^2(\mathbf0)}{\varepsilon_s^2}
  +\frac{(\operatorname{adj}\widetilde g(\mathbf0))_{ab}
  (\partial_a\varepsilon_s)(\partial_b\varepsilon_s)}{\varepsilon_s^3}
 \right].
 \label{eq:E_soft_sectors}
\end{equation}
For the leading singular terms we use
$\varepsilon_s=\epsilon+\gamma_{ab}p_ap_b$ over $\mathbb R^2$.
We thus obtain the following integrals:
\begin{equation}
 \begin{aligned}
 &\int_{\mathbb R^2}\frac{d^2p}{(2\pi)^2}\frac1{\varepsilon_s^2}
 =\frac1{4\pi\epsilon\sqrt{\det\boldsymbol\gamma}},\\
 &\int_{\mathbb R^2}\frac{d^2p}{(2\pi)^2}
 \frac{(\operatorname{adj}\widetilde g(\mathbf0))_{ab}
       (\partial_a\varepsilon_s)(\partial_b\varepsilon_s)}{\varepsilon_s^3}
 =\frac{(\operatorname{adj}\widetilde g(\mathbf0))_{ab}\gamma_{ab}}
       {4\pi\epsilon\sqrt{\det\boldsymbol\gamma}}.
 \end{aligned}
 \label{eq:E_soft_integrals}
\end{equation}
The two singular contributions therefore read
\begin{equation}
 \begin{aligned}
 \chi_{zz}^{\mathrm K}
 &\simeq-\frac{e^{*2}T\det[\boldsymbol\gamma+\widetilde{\boldsymbol g}(\mathbf0)]}
 {12\pi d_s\epsilon\sqrt{\det\boldsymbol\gamma}},\\
 \chi_{zz}^{\mathrm C}
 &\simeq\frac{e^{*2}T\left[\det\widetilde{\boldsymbol g}(\mathbf0)
 +\tfrac34\widetilde\omega_{xy}^2(\mathbf0)
  +(\operatorname{adj}\widetilde g(\mathbf0))_{ab}\gamma_{ab}\right]}
 {12\pi d_s\epsilon\sqrt{\det\boldsymbol\gamma}}.
 \end{aligned}
 \label{eq:E_soft_singular_parts}
\end{equation}
Expanding the determinant in the first line cancels both the terms linear
in $\widetilde{\boldsymbol g}(\mathbf0)$ and $\det\widetilde{\boldsymbol g}(\mathbf0)$ in the second line.
For arbitrary $N$, the resulting leading susceptibility is
\begin{equation}
 \chi_{zz}^{\mathrm{fl}}
 =-\frac{e^{*2}T\sqrt{\det\boldsymbol\gamma}}{12\pi d_s\epsilon}
 +\frac{e^{*2}T\widetilde\omega_{xy}^2(\mathbf0)}
       {16\pi d_s\epsilon\sqrt{\det\boldsymbol\gamma}}
 +\chi_{zz}^{\mathrm{sub}},
 \qquad \epsilon\chi_{zz}^{\mathrm{sub}}\to0.
 \label{eq:E_single_soft_general}
\end{equation}
For $N=1$ the weighted Berry curvature vanishes, giving the usual diamagnetic susceptibility.

For $N=2$, we follow the convention of the \textbf{Main Text}:
the soft mode is $\nu=-$ and the massive mode is $\bar\nu=+$.
The preceding general-$N$ formulas then use $s=-$, and the geometric mean
of the principal coherence lengths is denoted by $\xi_{\mathrm{CL}}=\xi_s$.
Eq.~\eqref{eq:E_soft_weighted_geometry} gives
$\widetilde\omega_{xy}^{\,2}(\mathbf0)=\varepsilon_+^{\,2}(\mathbf0)\omega_{xy}^2(\mathbf0)$,
so Eq.~\eqref{eq:E_single_soft_general} reduces to
\begin{equation}
 \chi_{zz}^{\mathrm{fl}}
 =-\frac{e^{*2}T\sqrt{\det\boldsymbol\gamma}}
 {12\pi d_s\epsilon}
 +\frac{e^{*2}T\varepsilon_+^{\,2}(\mathbf0)\omega_{xy}^2(\mathbf0)}
 {16\pi d_s\epsilon\sqrt{\det\boldsymbol\gamma}}
 +\chi_{zz}^{\mathrm{sub}}.
 \label{eq:E_single_soft_response}
\end{equation}
This reproduces Eq.~(25) of the \textbf{Main Text}.
The first term gives the conventional fluctuation diamagnetism proportional to
$\xi_{\mathrm{CL}}^2$. The second term is a paramagnetic correction from pair quantum
geometry that breaks this conventional relation between the orbital
susceptibility and the coherence length when $\omega_{xy}(\mathbf0)\ne0$.

\subsection{Geometric bound on the coherent contribution}
\label{app:bound}
As $\chi_{zz}^{\mathrm C}$ has a pair quantum geometric origin,
we derive a geometric bound on this contribution for an arbitrary
number $N$ of fluctuation modes. To do so, we define the traceless part
of $\hat R_a$ and its Gram matrix by
\begin{equation}
 \bar R_a\equiv\hat R_a-\frac{\operatorname{tr}_{\mathrm{mat}}\hat R_a}{N}\hat1,
 \qquad
 \mathcal G_{ab}\equiv\operatorname{tr}_{\mathrm{mat}}(\bar R_a\bar R_b).
 \label{eq:metric_general}
\end{equation}
Using the spectral representation of $\hat R_a$ in Eq.~\eqref{eq:E_R},
together with $\operatorname{tr}_{\mathrm{mat}}\hat R_a=-\sum_n\partial_a\ln\varepsilon_n$,
we obtain
\begin{equation}
    \mathcal G_{ab}
    =
    \frac1N\sum_{n<m}
    (\partial_a\ell_{nm})(\partial_b\ell_{nm})
    +
    2\sum_{n<m}
    W_{nm}
    g_{ab}^{nm}.
    \label{eq:metric_spectral}
\end{equation}
We next bound the coherent contribution.
Hermiticity and positivity of $\hat L_0$ give
$\hat R_a^\dagger=\hat L_0\hat R_a\hat L_0^{-1}$.
It follows that
$[\hat R_x,\hat R_y]^\dagger=-\hat L_0[\hat R_x,\hat R_y]\hat L_0^{-1}$,
which gives
\begin{equation}
    -\operatorname{tr}_{\mathrm{mat}}
    [\hat R_x,\hat R_y]^2
    =
    \sum_{n,m}
    \frac{\varepsilon_m}{\varepsilon_n}
    \left|\Xi_{xy}^{nm}\right|^2
    \geq 0,
    \label{eq:comm_positive_general}
\end{equation}
so that the coherent contribution is non-negative for arbitrary $N$.
For the upper bound, the B\"ottcher--Wenzel
inequality~\cite{SM:BottcherWenzel2008Commutator} applies in trace form since
$\bar R_a^\dagger=\hat L_0\bar R_a\hat L_0^{-1}$ with $\hat L_0>0$.
For $\mathcal G_{xx}>0$, subtracting
$(\mathcal G_{xy}/\mathcal G_{xx})\bar R_x$ from $\bar R_y$
leaves the commutator unchanged. Hence
\begin{equation}
 \begin{aligned}
 0&\leq-\operatorname{tr}_{\mathrm{mat}}[\hat R_x,\hat R_y]^2
 =-\operatorname{tr}_{\mathrm{mat}}
 \left[\bar R_x,\bar R_y-\frac{\mathcal G_{xy}}{\mathcal G_{xx}}\bar R_x\right]^2\\
 &\leq2\mathcal G_{xx}\operatorname{tr}_{\mathrm{mat}}
 \left(\bar R_y-\frac{\mathcal G_{xy}}{\mathcal G_{xx}}\bar R_x\right)^2
 =2(\mathcal G_{xx}\mathcal G_{yy}-\mathcal G_{xy}^2)
 =2\det\mathcal G.
 \end{aligned}
 \label{eq:comm_metric_bound}
\end{equation}
The last line follows by expanding the square using
Eq.~\eqref{eq:metric_general}.
Consequently,
\begin{equation}
    0
    \leq
    \chi_{zz}^{\mathrm C}
    \leq U_{\mathrm{geom}}
    \equiv\frac{e^{*2}T}{3d_s}
    \int\frac{d^2p}{(2\pi)^2}
    \det\mathcal G(\mathbf p).
    \label{eq:chiC_metric_bound}
\end{equation}

\begin{samepage}
For $N=2$, the Pauli representation of $\hat R_a$ gives
\begin{equation}
 \bar R_a\equiv\hat R_a-\tfrac12\operatorname{tr}_{\mathrm{mat}}(\hat R_a)\hat1
 =-\frac{[\partial_a\mathbf m
 +i\mathbf m\times\partial_a\mathbf m]\cdot\boldsymbol\sigma}
 {1-\mathbf m^2}.
 \label{eq:E_Pauli_traceless_R}
\end{equation}
Inserting this expression into Eq.~\eqref{eq:metric_general} and taking
the Pauli trace yields
\begin{equation}
 \mathcal G_{ab}
 =\frac{2\left[(1-\mathbf m^2)\partial_a\mathbf m\cdot\partial_b\mathbf m
 +(\mathbf m\cdot\partial_a\mathbf m)
  (\mathbf m\cdot\partial_b\mathbf m)\right]}
 {(1-\mathbf m^2)^2}.
 \label{eq:E_Pauli_Gram}
\end{equation}
The determinant of its $x,y$ block is
\begin{equation}
 \det\mathcal G
 =\frac{4\left\{|\partial_x\mathbf m\times\partial_y\mathbf m|^2
 -[\mathbf m\cdot(\partial_x\mathbf m\times\partial_y\mathbf m)]^2\right\}}
 {(1-\mathbf m^2)^3}.
 \label{eq:two_band_saturation}
\end{equation}
Since $(1-\mathbf m^2)^{-3}=D^3d_0^6$, comparison with the coherent kernel
in Eq.~\eqref{eq:main_Pauli_C} gives
$\det\mathcal G=\mathcal P_{xy}$ pointwise. Hence
\begin{equation}
 \chi_{zz}^{\mathrm C}=\frac{e^{*2}T}{3d_s}
 \int\frac{d^2p}{(2\pi)^2}\det\mathcal G
 =U_{\mathrm{geom}},\qquad N=2.
 \label{eq:chiC_two_band_metric}
\end{equation}
For a diamagnetic kinetic term, the condition
$U_{\mathrm{geom}}>|\chi_{zz}^{\mathrm K}|$ is therefore the exact criterion
for a net paramagnetic response, as stated in the \textbf{Main Text}.

\end{samepage}
\subsection{Comparison with electronic orbital susceptibility}
\label{app:electronic_comparison}

Here we compare the results of a two-component model of superconducting fluctuations
with the orbital magnetic susceptibility of a two-band model of Bloch electrons.
Generic Green-function formulas~\cite{SM:Fukuyama1971,SM:RaouxEtAl2015} and their
geometric reorganization~\cite{SM:GaoYangNiu2015,SM:PiechonEtAl2016} make this comparison
explicit. We use the same Pauli matrix notation for the electronic Bloch Hamiltonian
and inverse pair propagator:
\begin{equation}
 \begin{aligned}
 \hat h(\mathbf k)&=d_0(\mathbf k)\sigma_0+\mathbf d(\mathbf k)\cdot\boldsymbol\sigma,
 &E_\nu(\mathbf k)&=d_0(\mathbf k)+\nu d(\mathbf k),\\
 \hat L_0^{-1}(\mathbf p)&=d_0(\mathbf p)\sigma_0+\mathbf d(\mathbf p)\cdot\boldsymbol\sigma,
 &\varepsilon_\nu(\mathbf p)&=d_0(\mathbf p)+\nu d(\mathbf p)>0.
 \end{aligned}
 \label{eq:comparison_Pauli_kernels}
\end{equation}
Here $\nu=\pm$, while $d=|\mathbf d|$ and $\hat{\mathbf n}=\mathbf d/d$ are understood in
the corresponding system and momentum variable. Away from degeneracies,
the projectors have the common form
$\hat P_\nu=(\sigma_0+\nu\hat{\mathbf n}\cdot\boldsymbol\sigma)/2$, giving
\begin{equation}
 g_{ab}=\tfrac14\partial_a\hat{\mathbf n}\cdot\partial_b\hat{\mathbf n},
 \qquad
 \omega_{xy}=-\tfrac12\hat{\mathbf n}\cdot
 (\partial_x\hat{\mathbf n}\times\partial_y\hat{\mathbf n}).
 \label{eq:comparison_two_band_geometry}
\end{equation}
The tensors and connection use the same symbols in the corresponding system,
and $\omega_{xy}^2=4\det g$ for both.
With $\mathcal A_a^{+-}(\mathbf k)=i\langle v_+|\partial_av_-\rangle$,
the interband electronic velocity and pair RLD are
\begin{equation}
 \langle v_+|\partial_a\hat h|v_-\rangle
 =2id(\mathbf k)\mathcal A_a^{+-}(\mathbf k),
 \qquad
 R_a^{+-}(\mathbf p)
 =-\frac{2id(\mathbf p)}{d_0(\mathbf p)-d(\mathbf p)}\mathcal A_a^{+-}(\mathbf p).
 \label{eq:comparison_electronic_derivative}
\end{equation}

The thermodynamic weights distinguish the two responses.
For the electronic bands and the static Gaussian pair modes,
respectively, the distribution functions are
\begin{equation}
 n_F(E_\nu)=[e^{(E_\nu-\mu)/T}+1]^{-1},
 \qquad
 n_{\mathrm{RJ}}(\varepsilon_\nu)=T/\varepsilon_\nu.
 \label{eq:comparison_spectral_weights}
\end{equation}
Here $n_F$ is the Fermi--Dirac distribution, and $n_{\mathrm{RJ}}$ denotes the classical Rayleigh--Jeans weight~\cite{SM:MishonovEtAl2003,SM:BoevKovalev2024}.
These weights are the eigenvalues of the corresponding equilibrium
correlation matrices: the electronic single-particle reduced density
matrix and the pairing-fluctuation density matrix~\cite{SM:sun2026quantum}, which read
\begin{equation}
 \begin{aligned}
 {}[\hat\rho_{\mathrm{el}}(\mathbf k)]_{ij}
 &=\langle c^\dagger_{j\mathbf k}c_{i\mathbf k}\rangle
   =[n_F(\hat h(\mathbf k))]_{ij},\\
 \hat C_{\mathrm{pair}}(\mathbf p)
 &=\langle\Delta_{\mathbf p}\Delta_{\mathbf p}^\dagger\rangle
   =T\hat L_0(\mathbf p).
 \end{aligned}
 \label{eq:comparison_equilibrium_correlators}
\end{equation}
The distributions $n_F$ and $n_{\mathrm{RJ}}$ therefore
describe fermionic occupations and static Gaussian pair-field variances,
respectively. The latter are fixed by classical equipartition and grow
as a mode softens.

For the minimally coupled two-band electronic model, the complete orbital magnetic susceptibility is
\begin{equation}
 \begin{aligned}
 \chi_{zz}^{\mathrm{el}}
 &=\frac{e^2}{d_s}\sum_{\nu=\pm}\int_{\mathrm{BZ}}\frac{d^2k}{(2\pi)^2}
 \Biggl\{\frac{n_F'(E_\nu)}{12}\big[E_{\nu,xx}E_{\nu,yy}-E_{\nu,xy}^2\big]\\
 &\mathrel{+}\left[-n_F'(E_\nu)+\frac{\nu\,n_F(E_\nu)}{d}\right]
 \left[d^2\omega_{xy}^2-(\operatorname{adj}g)_{ab}(\partial_aE_\nu)(\partial_bE_{-\nu})\right]
 -\frac{\nu\,n_F(E_\nu)}{2d}\partial_a\partial_b\!\left[d^2(\operatorname{adj}g)_{ab}\right]
 \Biggr\}.
 \end{aligned}
 \label{eq:comparison_electronic_two_band}
\end{equation}
Eq.~\eqref{eq:comparison_electronic_two_band} is the result of
Ref.~\cite{SM:PiechonEtAl2016} expressed in our units and notation.
Its matrix Green-function starting point is the lattice response formula
of Ref.~\cite{SM:RaouxEtAl2015}.
The wave-packet treatment of Ref.~\cite{SM:GaoYangNiu2015} gives a complementary
decomposition into orbital-moment, Van Vleck, Langevin, and geometrical responses.
In Eq.~\eqref{eq:comparison_electronic_two_band}, the non-LP terms are grouped
into the Berry-curvature and quantum-metric contributions of
Ref.~\cite{SM:PiechonEtAl2016}.
The metric contributions include the particle-hole-asymmetry term and can
remain finite when $\omega_{xy}=0$~\cite{SM:PiechonEtAl2016,SM:GaoYangNiu2015}.
By algebraically rearranging Eq.~\eqref{eq:E_two_chi_2d} and integrating by parts,
we obtain the result for Gaussian pairing fluctuations
\begin{equation}
 \begin{aligned}
 \chi_{zz}^{\mathrm{fl}}
 &=\frac{e^{*2}}{d_s}\sum_{\nu=\pm}\int\frac{d^2p}{(2\pi)^2}
 \Biggl\{\frac{n_{\mathrm{RJ}}'(\varepsilon_\nu)}{12}\big[\varepsilon_{\nu,xx}\varepsilon_{\nu,yy}-\varepsilon_{\nu,xy}^2\big]\\
 &\mathrel{+}\left[-n_{\mathrm{RJ}}'(\varepsilon_\nu)+\frac{\nu\,n_{\mathrm{RJ}}(\varepsilon_\nu)}{d}\right]
 \left[d^2\omega_{xy}^2-(\operatorname{adj}g)_{ab}(\partial_a\varepsilon_\nu)(\partial_b\varepsilon_{-\nu})\right]
 -\frac{\nu\,n_{\mathrm{RJ}}(\varepsilon_\nu)}{2d}\partial_a\partial_b\!\left[d^2(\operatorname{adj}g)_{ab}\right]
 \Biggr\}.
 \end{aligned}
 \label{eq:comparison_fluctuation_two_mode}
\end{equation}
The first terms in the two complete expressions above depend only on
the eigenvalue dispersions. They are the electronic Landau--Peierls (LP)
term and the Gaussian fluctuation \emph{dispersion contribution},
denoted by $\chi_{zz}^{\mathrm{el,LP}}$ and
$\chi_{zz}^{\mathrm{fl,disp}}$, respectively.
We use $E_{\nu,ab}=\partial_a\partial_bE_\nu$ and
$\varepsilon_{\nu,ab}=\partial_a\partial_b\varepsilon_\nu$.
Primes denote derivatives with respect to the scalar argument,
in particular, $n_{\mathrm{RJ}}'(\varepsilon_\nu)=-T/\varepsilon_\nu^2$.
The fluctuation term has the LP Hessian structure, with a Gaussian thermodynamic weight
in place of the electronic Fermi-surface weight.
For momentum-independent pair eigenprojectors, $g_{ab}=\omega_{xy}=0$
and the full fluctuation susceptibility reduces to
$\chi_{zz}^{\mathrm{fl,disp}}$.

The two-band and two-component results suggest the following correspondence
between electronic and static Gaussian pairing fluctuation formulas:
\begin{equation}
 \hat h(\mathbf k)\longrightarrow\hat L_0^{-1}(\mathbf p),
 \qquad e\longrightarrow e^*,
 \qquad n_F(E)\longrightarrow n_{\mathrm{RJ}}(\varepsilon).
 \label{eq:comparison_general_rule}
\end{equation}
We expect that this replacement rule can be extended to arbitrary $N$-band
electronic systems and $N$-component superconductors.

\section{Microscopic model on a kagome lattice}
\label{app:kagome}

In this section, we derive the effective doublet pair propagator used in the \textbf{Main Text}
from a microscopic kagome model. After specifying the normal-state bands
and the pairing interaction, we project the Cooper-pair propagator onto
the critical $E_{2g}$ doublet and extract its gradient coefficients.
We then use this kernel to calculate the orbital magnetic susceptibility and
coherence lengths, keeping track of the contribution from intermode coherence.

\subsection{Normal-state bands and real-space interaction}

We choose primitive vectors $\mathbf a_1=(2,0)$ and
$\mathbf a_2=(1,\sqrt3)$, with nearest-neighbor distance $a=1$
and unit-cell area $A_{\rm cell}=2\sqrt3$. The sublattices are at
$\mathbf r_A=\mathbf a_2/2$, $\mathbf r_B=\mathbf a_1/2$, and
$\mathbf r_C=0$. The nearest-neighbor tight-binding Hamiltonian is
\begin{equation}
 H_0=t\sum_{\langle ij\rangle,\sigma}
 (c_{i\sigma}^\dagger c_{j\sigma}+\mathrm{h.c.})
 -\mu\sum_{i,\sigma}c_{i\sigma}^\dagger c_{i\sigma}.
 \label{eq:A_kagome_real_h}
\end{equation}
In the physical-position gauge we use
$c_{a\sigma}(\mathbf k)=N_k^{-1/2}\sum_{\mathbf R}
e^{-i\mathbf k\cdot(\mathbf R+\mathbf r_a)}c_{\mathbf R a\sigma}$,
where $N_k$ is the number of unit cells. With
$\psi_{\mathbf k\sigma}=(c_{A\sigma}(\mathbf k),c_{B\sigma}(\mathbf k),c_{C\sigma}(\mathbf k))^T$,
the Hamiltonian becomes
$H_0=\sum_{\mathbf k,\sigma}\psi_{\mathbf k\sigma}^\dagger
h(\mathbf k)\psi_{\mathbf k\sigma}$, with
\begin{equation}
 h(\mathbf k)=
 \begin{pmatrix}
 -\mu&2t c_{AB}(\mathbf k)&2t c_{CA}(\mathbf k)\\
 2t c_{AB}(\mathbf k)&-\mu&2t c_{BC}(\mathbf k)\\
 2t c_{CA}(\mathbf k)&2t c_{BC}(\mathbf k)&-\mu
 \end{pmatrix}.
 \label{eq:A_kagome_h}
\end{equation}
Here $c_{AB}(\mathbf q)=\cos[(q_x-\sqrt3q_y)/2]$,
$c_{BC}(\mathbf q)=\cos q_x$, and
$c_{CA}(\mathbf q)=\cos[(q_x+\sqrt3q_y)/2]$.
We use uniform nearest-neighbor hopping $t=-1$ and place the chemical
potential at $\mu=0$. The reciprocal vectors are
$\mathbf b_1=(\pi,-\pi/\sqrt3)$ and
$\mathbf b_2=(0,2\pi/\sqrt3)$. At
$\Gamma$, $K$, and $M$, the three band energies are $(-4,2,2)$, $(-1,-1,2)$, and $(-2,0,2)$,
respectively. The chemical potential $\mu=0$ therefore lies at the van Hove
energy of the upper dispersive band, whose saddle points are located at the
$M$ points. The Dirac crossing is at energy $-1$, and the flat band is at $2$.
These features are shown in Fig.~2(b) of the \textbf{Main Text}.
At zero temperature, the van Hove filling is $\nu=5/12$,
or $5/2$ electrons per unit cell.

The interaction between onsite pair operators $P_i=c_{i\downarrow}c_{i\uparrow}$
is given by Eq.~(26) of the \textbf{Main Text}. A negative $U$ gives an onsite
singlet attraction, and a positive $V$ describes repulsive pair transfer with equal
amplitude on each nearest-neighbor bond. The Fourier transform of $P_a(\mathbf p)$ uses the same physical site
positions as the electron fields. Expressing the interaction as
$H_{\rm int}=-\sum_{\mathbf p}P^\dagger(\mathbf p)g(\mathbf p)P(\mathbf p)$
gives
\begin{equation}
 g(\mathbf p)=
 \begin{pmatrix}
 -U&-2V c_{AB}(\mathbf p)&-2V c_{CA}(\mathbf p)\\
 -2V c_{AB}(\mathbf p)&-U&-2V c_{BC}(\mathbf p)\\
 -2V c_{CA}(\mathbf p)&-2V c_{BC}(\mathbf p)&-U
 \end{pmatrix}.
 \label{eq:A_kagome_g}
\end{equation}
At zero pair momentum, the symmetric and doublet channels have coupling constants
$g_s(0)=-U-4V$ and $g_E(0)=-U+2V$, respectively.
For the parameters $U=-4.4$
and $V=1.1$ used in the \textbf{Main Text}, the channel coupling constants are
$g_E(0)=6.6$ and $g_s(0)=0$. Although the symmetric channel has zero
coupling at the origin, $g(\mathbf p)$ is generally full rank at finite
pair momentum. We therefore retain all three sublattice components when
constructing the propagator.

\subsection{Band-basis pairing gap functions and \texorpdfstring{$d$}{d}-wave symmetry}
\label{app:kagome_band_gap}

To identify the symmetry of the pairing gap function in the band basis, let
$\boldsymbol u_n(\mathbf k)=(u_{An},u_{Bn},u_{Cn})^T$ be the normalized
eigenvectors of the Bloch Hamiltonian $h(\mathbf k)$ in Eq.~\eqref{eq:A_kagome_h}.
The onsite pairing gap amplitudes associated with the real $E_{2g}$ doublet are
\begin{equation}
 \boldsymbol\Delta\equiv(\Delta_A,\Delta_B,\Delta_C)^T
 =V_E^T\begin{pmatrix}\eta_x\\\eta_y\end{pmatrix},\qquad
 V_E=\begin{pmatrix}
 2/\sqrt6&-1/\sqrt6&-1/\sqrt6\\
 0&1/\sqrt2&-1/\sqrt2
 \end{pmatrix}.
 \label{eq:A_kagome_doublet_embedding}
\end{equation}
The rows of $V_E$ are orthonormal basis vectors for the $E_{2g}$ pairing
channel in sublattice space.
At zero pair momentum, the band-basis pairing gap matrix is
\begin{equation}
 \Delta_{nm}^{\rm band}(\mathbf k)
 =\sum_{a=A,B,C}u_{an}^*(\mathbf k)\Delta_a u_{am}^*(-\mathbf k).
 \label{eq:A_kagome_band_gap}
\end{equation}
In a time-reversal-compatible gauge,
$u_{an}(-\mathbf k)=u_{an}^*(\mathbf k)$. The intraband pairing gap function of the
$E_{2g}$ field is therefore
\begin{equation}
 \begin{aligned}
 \Delta_n(\mathbf k)&\equiv\Delta_{nn}^{\rm band}(\mathbf k)
 =\eta_xf_{x,n}(\mathbf k)+\eta_yf_{y,n}(\mathbf k),
 \qquad f_{\ell,n}(\mathbf k)=\sum_a(V_E)_{\ell a}|u_{an}(\mathbf k)|^2,
 \quad \ell=x,y,\\
 f_{x,n}&=\frac{2|u_{An}|^2-|u_{Bn}|^2-|u_{Cn}|^2}{\sqrt6},\qquad
 f_{y,n}=\frac{|u_{Bn}|^2-|u_{Cn}|^2}{\sqrt2}.
 \end{aligned}
 \label{eq:A_kagome_dwave_form_factors}
\end{equation}
The sublattice weights $|u_{an}(\mathbf k)|^2$ vary along a band, so an
onsite pairing pattern acquires momentum dependence upon projection into
the band basis. This sublattice interference produces the form factors
in Eq.~\eqref{eq:A_kagome_dwave_form_factors}, which transform as $E_{2g}$
and are the lattice analogs of $(d_{x^2-y^2},d_{xy})$.
At $\mu=0$, the three inequivalent $M$ saddle points of the upper dispersive
band have weights localized on different sublattices~\cite{SM:KieselThomale2012,SM:KieselPlattThomale2013,SM:WuEtAl2021}.
For the chiral states $\eta_y=\pm i\eta_x$, analogous to the chiral $d$-wave state
found at van Hove filling in the kagome Hubbard model~\cite{SM:YuLi2012},
$(\Delta_A,\Delta_B,\Delta_C)\propto
(1,e^{\pm2\pi i/3},e^{\mp2\pi i/3})$. The pairing gap magnitudes are then equal
at all three saddles. In the band basis, the pairing gap function is nodeless
and has winding number $\pm2$ along the $\mu=0$ Fermi contour.

\subsection{Cooper bubble and projection onto the critical channel}

We next calculate the pair propagator from the normal-state bands.
Using the band dispersions $\xi_n(\mathbf k)$ from Eq.~\eqref{eq:A_kagome_h}
and the eigenvectors presented in section~\ref{app:kagome_band_gap}, we write
the static Cooper bubble (pairing susceptibility) as~\cite{SM:LarkinVarlamov2005}
\begin{align}
 \Pi_{ab}(\mathbf p,T)&=\frac1{N_k}\sum_{\mathbf k,n,m}
 u_{an}(\mathbf k+\mathbf p/2)u_{bn}^*(\mathbf k+\mathbf p/2)
 u_{am}(-\mathbf k+\mathbf p/2)u_{bm}^*(-\mathbf k+\mathbf p/2)
 \nonumber\\[-2pt]
 &\mathrel{\times}\frac{1-f[\xi_n(\mathbf k+\mathbf p/2)]
 -f[\xi_m(-\mathbf k+\mathbf p/2)]}
 {\xi_n(\mathbf k+\mathbf p/2)+\xi_m(-\mathbf k+\mathbf p/2)}.
 \label{eq:A_kagome_bubble}
\end{align}
Here $f(\xi)=[e^{\xi/T}+1]^{-1}$.
For energies $\xi,\xi'$, the weight in this expression is
$[\tanh(\xi/2T)+\tanh(\xi'/2T)]/[2(\xi+\xi')]\geq0$.
Its continuous limit at $\xi'=-\xi$ is
$-f'(\xi)=[4T\cosh^2(\xi/2T)]^{-1}$.
Each band-pair term is a non-negative weight times an outer product of its
sublattice vertex. Consequently, $\Pi$ is positive semidefinite. The sum
includes all three bands and the full Brillouin zone: both the band energies
and the momentum-dependent sublattice wave functions enter the gradient coefficients.

The effective doublet pair propagator must retain virtual mixing with the noncritical
symmetric mode, whose sublattice vector is $(1,1,1)^T/\sqrt3$.
We therefore construct the full sublattice propagator before projecting
it onto the real $(\eta_x,\eta_y)$ doublet. For our parameters, $g_s(0)=0$ makes $g(\mathbf0)$ singular.
The usual expression for the inverse of the full sublattice pair propagator,
$\hat L_{\rm sub}^{-1}=g^{-1}-\Pi$, therefore cannot be used at the origin.
The RPA form avoids inverting the singular interaction matrix $g(\mathbf 0)$:
\begin{equation}
\begin{aligned}
 \hat L_{\rm sub}(\mathbf p,T)
 &=[\hat1-g(\mathbf p)\Pi(\mathbf p,T)]^{-1}g(\mathbf p),\\
 \hat L_{E,\mathrm{real}}^{-1}(\mathbf p,T)
 &=[V_E\hat L_{\rm sub}(\mathbf p,T)V_E^T]^{-1}.
\end{aligned}
 \label{eq:A_kagome_projection}
\end{equation}
The projected propagator is regular in the normal state $\alpha>0$,
and its inverse gives the desired doublet kernel. Where $g(\mathbf p)$
is invertible, $\hat L_{\rm sub}^{-1}=g^{-1}-\Pi$. Writing this inverse
in blocks for the doublet $E$ and the symmetric combination
$(1,1,1)^T/\sqrt3$ and eliminating the latter gives
\begin{equation}
 \hat L_{E,\mathrm{real}}^{-1}
 =(\hat L_{\rm sub}^{-1})_{EE}
 -(\hat L_{\rm sub}^{-1})_{Es}
 [(\hat L_{\rm sub}^{-1})_{ss}]^{-1}
 (\hat L_{\rm sub}^{-1})_{sE}.
 \label{eq:A_kagome_Schur}
\end{equation}
The second term is the correction from virtual fluctuations of the
symmetric mode. Eq.~\eqref{eq:A_kagome_projection} retains this
mixing while remaining usable at $\mathbf p=0$.
The susceptibilities in Eqs.~(14) and (16) of the \textbf{Main Text}
follow by minimally coupling the resulting $p^2$ doublet kernel.
The projected Cooper bubble matrix in the doublet channel is defined by
$\hat\Pi_E(\mathbf p,T)=V_E\Pi(\mathbf p,T)V_E^T$.
At zero pair momentum in this model, symmetry gives
$\hat\Pi_E(\mathbf0,T)=\pi_E(T)\hat1$, where
$\pi_E(T)=\tfrac12\operatorname{tr}_{\mathrm{mat}}\hat\Pi_E(\mathbf0,T)$.
The inverse pair propagator at zero momentum then satisfies
$\hat L_{E,\mathrm{real}}^{-1}(\mathbf0,T)=\alpha(T)\hat1$, with
$\alpha(T)=g_E^{-1}-\pi_E(T)$.
Throughout this section, $T_c$ is the unsplit mean-field critical temperature,
determined by $g_E\pi_E(T_c)=1$.
The thermal slope is $a_E=\left.\partial_T\alpha(T)\right|_{T_c}=-\pi_E'(T_c)$.

The $O(p^2)$ mixing contributions in Eq.~\eqref{eq:A_kagome_Schur}
cancel between the $EE$ block and the Schur-complement term.
Expanding the regular RPA expression in Eq.~\eqref{eq:A_kagome_projection}
then gives the quadratic kernel below, with feedback through the
symmetric-sector Cooper bubble entering at $O(p^4)$. We denote Pauli matrices in the
real $(\eta_x,\eta_y)$ basis by $\tau_i$ and reserve $\sigma_i$ for the
chiral $(\eta_+,\eta_-)$ basis.
\begin{equation}
\begin{aligned}
 \hat L_{E,\mathrm{real}}^{-1}
 &=g_E^{-1}\hat1-\hat\Pi_E(\mathbf p,T)
 +\frac{V}{2g_E^2}\left[p^2\hat1
 -(p_x^2-p_y^2)\tau_z+2p_xp_y\tau_x\right]+O(p^4)\\
 &=\alpha\hat1+\gamma_0p^2\hat1
 -\gamma_1(p_x^2-p_y^2)\tau_z+2\gamma_1p_xp_y\tau_x+O(p^4).
\end{aligned}
 \label{eq:A_kagome_real_GL}
\end{equation}
\begin{figure}[!b]
 \centering
 \includegraphics[width=0.65\linewidth]{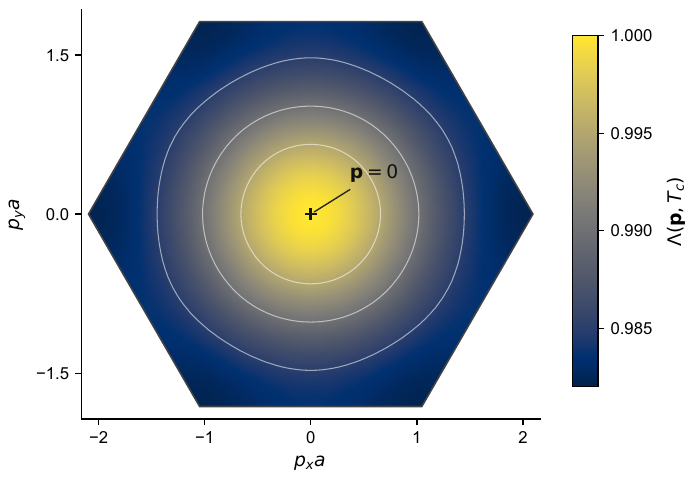}
 \caption[Pairing eigenvalue over two-dimensional momentum]{%
 Largest pairing eigenvalue $\Lambda(\mathbf p,T_c)$ at $\mu=0$ for the microscopic
 parameters used in Fig.~2 of the \textbf{Main Text}.
 The map covers the first Brillouin zone. The marker identifies the maximum
 at $\mathbf p=0$, where $\Lambda=1$. Momentum components are in units of $a^{-1}$.}
 \label{fig:kagome_pairing_lambda}
\end{figure}

Thus the momentum dependence of the interaction contributes
$V/(2g_E^2)$ to both $\gamma_0$ and $\gamma_1$, and
the remaining parts come from the expansion of $-\hat\Pi_E$.
The transformation to the fixed chiral components is
\begin{equation}
 \begin{pmatrix}\eta_+\\\eta_-\end{pmatrix}
 =\mathcal U_{\rm ch}\begin{pmatrix}\eta_x\\\eta_y\end{pmatrix},\qquad
 \mathcal U_{\rm ch}=\frac1{\sqrt2}\begin{pmatrix}i&1\\-i&1\end{pmatrix},\qquad
 \hat L_E^{-1}=\mathcal U_{\rm ch}\hat L_{E,\mathrm{real}}^{-1}
                         \mathcal U_{\rm ch}^\dagger.
 \label{eq:A_kagome_chiral_basis}
\end{equation}
This gives Eq.~(27) of the \textbf{Main Text}, with off-diagonal entries
$\gamma_1p_+^2$ and $\gamma_1p_-^2$. The gradient coefficients can be extracted
from the microscopic kernel in Eq.~\eqref{eq:A_kagome_projection} as
\begin{equation}
 \gamma_0=\frac14\operatorname{tr}_{\mathrm{mat}}
       (\partial_{p_x}^2\hat L_{E,\mathrm{real}}^{-1}),
 \qquad
 \gamma_1=-\frac14\operatorname{tr}_{\mathrm{mat}}
       (\tau_z\partial_{p_x}^2\hat L_{E,\mathrm{real}}^{-1}),
 \label{eq:A_kagome_gradient_derivatives}
\end{equation}
where both derivatives are evaluated at $\mathbf p=0$ and $T=T_c$.
The minus sign in $\gamma_1$ follows from the transformation to the
chiral basis. The derivatives retain the momentum dependence of both
$g(\mathbf p)$ and $\Pi(\mathbf p,T)$.

We also verify that the leading pairing instability at $T_c$ occurs at zero pair momentum
by examining the full finite-momentum pairing condition.
Define
\begin{equation}
 \Lambda(\mathbf p,T)=\lambda_{\max}
 \big[\Pi(\mathbf p,T)^{1/2}g(\mathbf p)\Pi(\mathbf p,T)^{1/2}\big].
 \label{eq:A_kagome_Thouless}
\end{equation}
The identity
$\det(\hat1-g\Pi)=\det(\hat1-\Pi^{1/2}g\Pi^{1/2})$
relates a pole of the RPA propagator to an eigenvalue of the Hermitian
matrix in Eq.~\eqref{eq:A_kagome_Thouless}.
The Thouless criterion identifies the first continuous pairing instability when $\Lambda$
reaches one upon cooling~\cite{SM:Thouless1960perturbation}.
Figure~\ref{fig:kagome_pairing_lambda} shows
$\Lambda(\mathbf p,T_c)$ across the first Brillouin zone.
For our present parameter set, $\Lambda$ is largest at $\mathbf p=0$,
where $\Lambda=1$, so the leading pairing instability is spatially uniform.

\FloatBarrier

\subsection{Radial response and the intermode contribution}
\label{app:kagome_response}

We now evaluate the susceptibility using the doublet pair propagator. To obtain radial
kernels that also apply to the split case in section~\ref{app:chiral_scaling},
we temporarily add a fixed chiral splitting $\delta\sigma_z$ to Eq.~(27) of
the \textbf{Main Text}, where $\sigma_z$ acts on the chiral components
$(\eta_+,\eta_-)$. The unsplit model considered here is recovered by
setting $\delta=0$. Let $y=p^2$ and denote the determinant of the
inverse propagator by $P=(\alpha+\gamma_0y)^2-\gamma_1^2y^2-\delta^2$.
The matrix traces in Eq.~(16) of the \textbf{Main Text} give the kernels
\begin{equation}
 \mathcal I_{\rm K}
 =-\frac1{12}\operatorname{tr}_{\rm mat}
    (\hat J_{xx}\hat J_{yy}-\hat J_{xy}^2),\qquad
 \mathcal I_{\rm C}
 =-\frac16\operatorname{tr}_{\rm mat}[\hat R_x,\hat R_y]^2.
 \label{eq:A_kagome_radial_traces}
\end{equation}
For this kernel the traces depend only on $y$. The angular measure is
$d^2p/(2\pi)^2=dy/(4\pi)$ after angular integration, yielding
\begin{equation}
 \chi_{zz}^{\rm fl}=\frac{e^{*2}T}{4\pi d_s}
 \int_0^\infty dy\,[\mathcal I_{\rm K}(y)+\mathcal I_{\rm C}(y)],
 \label{eq:A_kagome_radial}
\end{equation}
where
\begin{align}
 \mathcal I_{\rm K}(y)
 &=\frac23\left[\frac{\gamma_0^2+2\gamma_1^2}{P}
 -\frac{2(\gamma_0^2(\alpha+\gamma_0y)^2-\gamma_1^4y^2)}{P^2}\right],
 \label{eq:F_IH}\\
 \mathcal I_{\rm C}(y)
 &=\frac{64\gamma_1^2y^2
 [\gamma_1^2\alpha^2+(\gamma_0^2-\gamma_1^2)\delta^2]}{3P^3}.
 \label{eq:F_IC}
\end{align}
These expressions apply for $\alpha>|\delta|$ and $r<1$.
In the continuum approximation, we extend the quadratic GL kernel over
the full momentum plane. A fixed ultraviolet cutoff changes the regular
background and finite constants, but not the coefficients of the critical
pole and logarithm.
For the unsplit case, $\delta=0$, rescaling $y$ by $\alpha/\gamma_0$
brings out the common factor
$\chi_0=e^{*2}T\gamma_0/(12\pi d_s\alpha)$. The remaining integrals give
the dimensionless response weights in Eq.~(28) of the \textbf{Main Text}:
\begin{equation}
 \begin{aligned}
 \frac{\chi_{zz}^{\mathrm K}}{\chi_0}
 &=\mathcal S_E^{\mathrm K}(r)=2\big[r\operatorname{atanh}r-1\big],\\
 \frac{\chi_{zz}^{\mathrm C}}{\chi_0}
 &=\mathcal S_E^{\mathrm C}(r)
 =8\left[\left(\frac3r-r\right)\operatorname{atanh}r-3\right].
 \end{aligned}
 \label{eq:A_kagome_response_weights}
\end{equation}
\begin{figure}[!t]
 \centering
 \includegraphics[width=\linewidth]{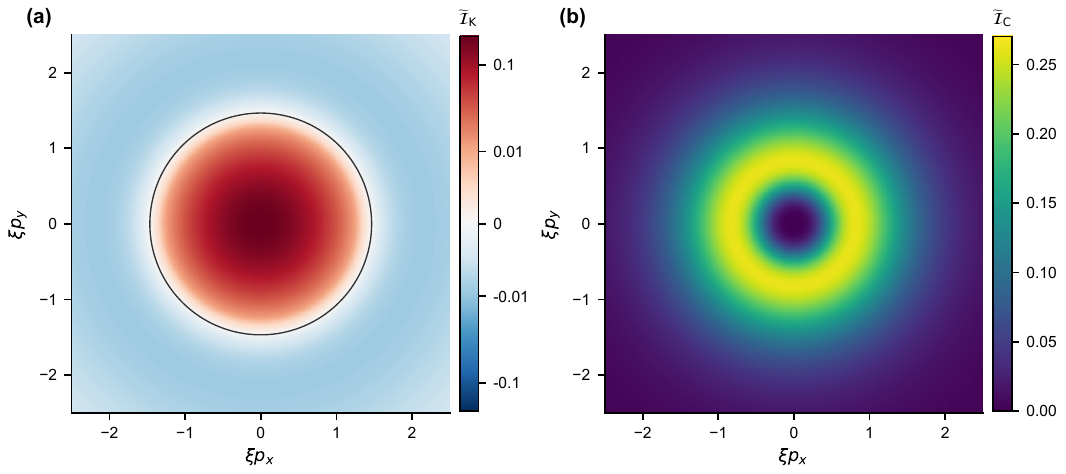}
 \caption[Momentum-space kernels of the two susceptibility sectors]{%
 Momentum kernels of the kinetic (a) and coherent (b) contributions for
 Eq.~(27) of the \textbf{Main Text}, with $\delta=0$ and $r=0.8073$.
 Signed kernels satisfy
 $\chi^{\rm K,C}_{zz}=(e^{*2}T/d_s)\int d^2p/(2\pi)^2\,
 \mathcal I_{\rm K,C}(\mathbf p)$, including the prefactors and signs of
 Eqs.~\eqref{eq:main_Pauli_H} and \eqref{eq:main_Pauli_C}.
 Colors show $\widetilde{\mathcal I}_{\rm K,C}=\xi^{-4}\mathcal I_{\rm K,C}$
 as functions of $\xi p_x$ and $\xi p_y$, with $\xi^2=\gamma_0/\alpha$.
 Both axes and color scales are dimensionless.
 The black contour in (a) marks zero. Panel (a) uses a symmetric logarithmic
 color scale, linear for $|\widetilde{\mathcal I}_{\rm K}|\leq0.01$.
 Panel (b) is non-negative and forms an annulus.
 Full-plane integration gives
 $\chi_{zz}^{\rm K}/\chi_0=-0.19$ and $\chi_{zz}^{\rm C}/\chi_0=2.04$.
 The maps show the kernels within the local GL momentum region.}
 \label{fig:kagome_integrands}
\end{figure}

Their sum is $\chi_{zz}^{\mathrm{fl}}/\chi_0=\mathcal S_E(r)
=\mathcal S_E^{\mathrm K}(r)+\mathcal S_E^{\mathrm C}(r)$. In physical units, the kernels $\mathcal I_{\rm K,C}$ have units of
momentum$^{-4}$. Figure~\ref{fig:kagome_integrands} plots their dimensionless
forms at $r=0.8073$, using the normalization in the caption.
For this unsplit kernel, $\mathbf d$ lies in a plane and winds
twice around the momentum origin, with $|\mathbf d|=|\gamma_1|p^2$ and
$d_0=\alpha+\gamma_0p^2$. The coherent kernel is rotationally symmetric
and forms the non-negative annulus in Fig.~\ref{fig:kagome_integrands}(b).
For $\mathbf p\neq0$, the planar texture has zero local Berry curvature,
so the annulus comes from the quantum-metric term of Eq.~(21) of the
\textbf{Main Text}.

We evaluate the weighted response kernels directly in the fixed component basis,
including at the degenerate point $\mathbf p=0$. The weight $W$
vanishes there and saturates at large $p$ within the extended quadratic kernel,
where stiffness dominates both branches. The identity $W=2(\cosh\ell-1)$ gives
$\partial_aW=2\sinh\ell\,\partial_a\ell$, which connects this variation
to the metric term in Eq.~\eqref{eq:E_Pxy}.
The coherent kernel peaks on an annulus and has algebraically decaying tails.
The kinetic kernel changes sign in momentum space; the full-plane integrals
of both kernels determine the sign of the susceptibility.

\FloatBarrier

\begin{samepage}
To separate the intramode and intermode responses, consider the eigenvalues
of the unsplit inverse propagator,
$\varepsilon_s=\alpha+\gamma_0(1+sr)p^2$, with $s=\pm$ and stiffnesses
$\gamma_0\pm|\gamma_1|$. The label $s=\pm$ orders the momentum-dependent
eigenmodes by stiffness, while $\eta_\pm$ denotes the fixed chiral components.
The eigenvectors wind twice around the origin. Rotational invariance lets
us evaluate the kernel at $p_y=0$ with $y=p_x^2>0$, where
$g_{xx}=g_{xy}=0$ and $g_{yy}=1/y$.
From Eq.~\eqref{eq:E_J_diag},
$J_{xx}^{ss}=2\gamma_0(1+sr)/\varepsilon_s$,
$J_{yy}^{ss}=2\gamma_0(1-sr)/\varepsilon_s$, and $J_{xy}^{ss}=0$.
The diagonal radial Hessian kernel is then
\begin{equation}
 \mathcal I_{{\rm K},s}^{\rm diag}(y)
 =-\frac{\gamma_0^2(1-r^2)}
 {3[\alpha+\gamma_0(1+sr)y]^2}.
 \label{eq:A_kagome_branch_kernel}
\end{equation}
The radial integral follows from
$\int_0^\infty dy/[\alpha+\gamma_0(1+sr)y]^2
 =[\alpha\gamma_0(1+sr)]^{-1}$.
The quantum-metric correction to the diagonal Hessian changes the kinetic
response of each branch relative to the corresponding independent scalar mode:
\begin{equation}
 \frac{\chi_{zz,s}^{\mathrm K,\mathrm{diag}}}{\chi_0}=-(1-sr),
 \qquad
 \frac{\chi_{zz,s}^{\mathrm{scalar}}}{\chi_0}=-(1+sr).
    \label{eq:A_kagome_branch_kinetic}
\end{equation}
\par
\end{samepage}
Summing the intramode kinetic terms gives $-2\chi_0$ for every $r$,
equal to the sum of two independent scalar responses. The intramode
coherent contribution vanishes because the local Berry curvature is zero
away from the degenerate origin, so the coherent response is entirely
intermode. We isolate the intermode kinetic contribution by subtracting
the diagonal sum from $\mathcal S_E^{\mathrm K}(r)\chi_0$:
\begin{equation}
 \frac{\chi_{zz}^{\mathrm K,\mathrm{diag}}}{\chi_0}=-2,\qquad \frac{\chi_{zz}^{\mathrm K,\mathrm{off}}}{\chi_0}=2r\operatorname{atanh}r,\qquad \frac{\chi_{zz}^{\mathrm K}}{\chi_0}=\mathcal S_E^{\mathrm K}(r).
 \label{eq:A_kagome_H_parts}
\end{equation}
Figure~2(d) of the \textbf{Main Text} shows these three curves with the same
$r$ range and susceptibility normalization as panel (c).
At the microscopic ratio $r=0.8073$, the intermode kinetic and coherent
contributions are $1.81\chi_0$ and $2.04\chi_0$, respectively. Together
with the intramode term $-2\chi_0$, they give the paramagnetic response
$\chi_{zz}^{\mathrm{fl}}=1.85\chi_0$. Holding the eigenvectors independent
of momentum while retaining the same two dispersions would instead give
$-2\chi_0$. The sign reversal thus comes from the winding of the doublet
eigenvectors and the resulting intermode coherence.
The hole-doped parameter set in Fig.~2 of the \textbf{Main Text} gives
$r_{\mathrm{hole}}=0.7297$. Substitution into the same response weights
yields $(\mathcal S_E^{\mathrm K},\mathcal S_E^{\mathrm C},\mathcal S_E)
=(-0.65,1.11,0.46)$, so this nearby filling also has a paramagnetic response.

The two modes of this doublet have isotropic fluctuation coherence lengths
given by~\cite{SM:SilaevBabaev2012,SM:OrdVarguninRago2012}
\begin{equation}
 \xi_\pm^{\,2}=\frac{\gamma_0\pm|\gamma_1|}{\alpha}>0.
 \label{eq:kagome_coherence_lengths}
\end{equation}
They describe the spatial decay of the Gaussian pair-field correlations.
The comparison above shows that intermode coherence can reverse the susceptibility
without changing either coherence length.

\FloatBarrier
\section{Critical scaling with a chiral splitting}
\label{app:chiral_scaling}

A chiral splitting changes how the kinetic and coherent contributions
scale near the superconducting transition. We use the minimally coupled model of
Eq.~(27) of the \textbf{Main Text} with the chiral splitting $\delta\sigma_z$,
where $\delta>0$ is induced by pre-existing loop-current order in the normal state,
and is held fixed near the transition [Eq.~(30) of the \textbf{Main Text}].
The zero-momentum masses for the two chiral branches $\eta_-$ and $\eta_+$ are
$\epsilon=\alpha-\delta$ and $\epsilon+2\delta$, respectively.
Thus the inverse pair propagator is given by
\begin{equation}
 \hat L_E^{-1}(\mathbf p)=
 \begin{pmatrix}
  \epsilon+2\delta+\gamma_0p^2&\gamma_1p_+^2\\
  \gamma_1p_-^2&\epsilon+\gamma_0p^2
 \end{pmatrix}.
 \label{eq:G_split_kernel}
\end{equation}
Write the unsplit critical temperature of section~\ref{app:kagome} as $T_c^{(0)}$
and the critical temperature in the fixed-splitting model as $T_c^{(\delta)}$.
The local linearization $\alpha(T)\simeq a_E(T-T_c^{(0)})$ gives
$T_c^{(\delta)}\simeq T_c^{(0)}+\delta/a_E$ and
$\epsilon\simeq a_E(T-T_c^{(\delta)})>0$.
In the temperature formulas below, $T_c\equiv T_c^{(\delta)}$.
The gradient coefficients and the explicit thermal prefactor are smooth
near the transition. We use the radial kernels in Eqs.~\eqref{eq:F_IH}
and \eqref{eq:F_IC} to evaluate the kinetic and coherent contributions in
Eq.~(16) of the \textbf{Main Text}.

\subsection{Microscopic chiral splitting}
\label{app:loop_current_bias}

To determine the chiral splitting microscopically, we add loop-current
order as a background to Eq.~\eqref{eq:A_kagome_h} and recalculate the Cooper bubble
in Eq.~\eqref{eq:A_kagome_bubble}.
Let $\hat L_E^{-1}(\mathbf0,T;\phi)$ be the projected inverse pair propagator
in the $(\eta_+,\eta_-)$ basis for the chosen hopping flux.
The signed chiral-splitting coefficient is
\begin{equation}
 \delta(T,\phi)=\frac12\operatorname{tr}_{\mathrm{mat}}
                    [\sigma_z\hat L_E^{-1}(\mathbf0,T;\phi)].
 \label{eq:G_microscopic_bias_projection}
\end{equation}
Using the bond-arrow convention in Fig.~3(a) of the \textbf{Main Text},
we assign a phase $e^{i\phi/3}$ to electron hopping along
$C\to B\to A\to C$ and the complex-conjugate phase for hopping in the reverse direction.
In the same physical-position gauge and $(A,B,C)$ sublattice basis as
Eq.~\eqref{eq:A_kagome_h}, the Bloch Hamiltonian therefore becomes
\begin{equation}
 h_\phi(\mathbf k)=
 \begin{pmatrix}
 -\mu&2t e^{i\phi/3}c_{AB}&2t e^{-i\phi/3}c_{CA}\\
 2t e^{-i\phi/3}c_{AB}&-\mu&2t e^{i\phi/3}c_{BC}\\
 2t e^{i\phi/3}c_{CA}&2t e^{-i\phi/3}c_{BC}&-\mu
 \end{pmatrix}.
 \label{eq:G_flux_hamiltonian}
\end{equation}
Here $c_{ab}\equiv c_{ab}(\mathbf k)$ are the cosine factors defined
below Eq.~\eqref{eq:A_kagome_h}.
In our convention, $h_{ab}$ multiplies $c_a^\dagger c_b$ in the full Hamiltonian,
so $h_{AB}$, $h_{BC}$, and $h_{CA}$ carry the positive phase. Both nearest-neighbor bonds
between a given pair of sublattices have the same phase.
The additional counterclockwise hopping-phase sums are $\phi$ on each triangle and
$-2\phi$ on each hexagon, as shown in Fig.~3(a) of the \textbf{Main Text}.
They give zero net phase sum per unit cell while
preserving the $1\times1$ translation periodicity~\cite{SM:OhgushiMurakamiNagaosa2000}.
For $\phi=0$, Eq.~\eqref{eq:G_flux_hamiltonian} reduces to
Eq.~\eqref{eq:A_kagome_h}.

For triangle phase $\phi=0.03$, the oriented bond phase is $0.01$
and the hexagon phase sum is $-0.06$. We keep $t=-1$, $\mu=0$, $U=-4.4$, and $V=1.1$ unchanged.
The imaginary part of each bond hopping has magnitude
$|t|\sin(0.01)\simeq0.01|t|$, while its total magnitude remains $|t|$.
We denote the pairing critical temperature of this full finite-flux
calculation by $T_c(\phi)$. At this temperature, the lowest eigenvalue of the
inverse pair propagator vanishes,
$\lambda_{\min}[\hat L_E^{-1}(\mathbf0,T_c(\phi);\phi)]=0$.
This gives $T_c(\phi)=0.7815|t|$, at which the microscopic projection yields
$\delta=3.7803\times10^{-4}|t|^{-1}$, with zero-momentum masses
$0$ and $2\delta$.

The microscopic projection above fixes the two zero-momentum masses at
$T_c(\phi)$. For the radial response below, we hold
$\delta=\delta(T_c(\phi),\phi)$ constant.
As a small-$\phi$ approximation, we use the unsplit values $a_E$ and $\gamma_{0,1}$
for the soft-mass temperature slope and gradient coefficients in
Eq.~\eqref{eq:G_split_kernel}.

\subsection{Exact radial integrals}

We can evaluate the radial integrals of the orbital magnetic susceptibility
in closed form for the kernel in Eq.~\eqref{eq:G_split_kernel}.
For convenience, using $y=p^2$ from section~\ref{app:kagome_response},
we define the following dimensionless quantities:
\begin{equation}
 x=\frac{\epsilon}{2\delta},\qquad
 z=\frac{\gamma_0y}{\delta},\qquad
 \beta_x=4x(x+1),\qquad
 c_r=1-r^2,\qquad \Upsilon=1+r^2\beta_x.
 \label{eq:G_dimensionless_defs}
\end{equation}
The ratio $x\simeq(T-T_c)/T_{\mathrm{split}}$ compares the distance
from the transition with the splitting scale
$T_{\mathrm{split}}=2\delta/a_E$.
The determinant $P$ of section~\ref{app:kagome_response} becomes
\begin{equation}
 P=(\epsilon+\gamma_0y)(\epsilon+2\delta+\gamma_0y)
       -\gamma_1^2y^2
   =\delta^2\widetilde P(z),\qquad
 \widetilde P(z)=c_rz^2+2(1+2x)z+\beta_x.
 \label{eq:G_determinant}
\end{equation}
Here $z$ measures the squared momentum in units of $\delta/\gamma_0$,
and $\beta_x$ is the constant term of the dimensionless function $\widetilde P(z)$.
Its discriminant is $4\Upsilon$, since
$(1+2x)^2-c_r\beta_x=\Upsilon>0$. This combination controls the root
separation and hence the logarithm in the radial integral.
Writing
$\mathcal A_T=e^{*2}T\gamma_0/(12\pi d_s)$, the two susceptibilities
are
\begin{equation}
 \chi_{zz}^{\mathrm K,\mathrm C}
 =\frac{3\mathcal A_T}{\delta}\int_0^\infty dz\,
       \mathcal I^{(z)}_{\mathrm K,\mathrm C}(z),
 \label{eq:G_scaled_integral}
\end{equation}
where $\mathcal I^{(z)}_{\mathrm K,\mathrm C}(z)
=(\delta^2/\gamma_0^2)\mathcal I_{\mathrm K,\mathrm C}(\delta z/\gamma_0)$ are the
dimensionless radial kernels:
\begin{equation}
 \mathcal I^{(z)}_{\mathrm K}
 =\frac23\left[
   \frac{2r^2-1}{\widetilde P}
   -\frac{2r^2c_rz^2}{\widetilde P^2}-\frac2{\widetilde P^2}\right],\qquad
 \mathcal I^{(z)}_{\mathrm C}
 =\frac{64r^2\Upsilon z^2}{3\widetilde P^3}.
 \label{eq:G_scaled_integrands}
\end{equation}
We can therefore factor
$\widetilde P(z)=c_r(z+z_+)(z+z_-)$, where
$z_\pm=(1+2x\pm\sqrt\Upsilon)/c_r>0$.
Integrating the resulting partial fractions gives the basic integral
\begin{equation}
 \mathcal J(x,r)=\int_0^\infty\frac{dz}{\widetilde P(z)}
 =\frac{1}{2\sqrt\Upsilon}
   \ln\frac{1+2x+\sqrt\Upsilon}{1+2x-\sqrt\Upsilon}.
 \label{eq:G_basic_integral}
\end{equation}
The remaining integrals can be obtained by differentiating the basic
integral. Temporarily treating the polynomial coefficients as independent,
write $\widetilde P(z)=c_2z^2+2c_1z+c_0$ and define
$\mathcal J_{\rm poly}(c_2,c_1,c_0)
=\int_0^\infty dz/(c_2z^2+2c_1z+c_0)$.
Differentiation with the other coefficients held fixed gives
\begin{equation}
 \int_0^\infty\frac{dz}{\widetilde P^2}
 =-\partial_{c_0}\mathcal J_{\rm poly},\quad
 \int_0^\infty\frac{z^2dz}{\widetilde P^2}
 =-\partial_{c_2}\mathcal J_{\rm poly},\quad
 \int_0^\infty\frac{z^2dz}{\widetilde P^3}
 =\tfrac12\partial_{c_0}\partial_{c_2}\mathcal J_{\rm poly}.
 \label{eq:G_moment_derivatives}
\end{equation}
The differentiation generates the moments needed in the two response
kernels without a separate partial-fraction expansion for each one.
Only after taking these derivatives do we impose
$(c_2,c_1,c_0)=(c_r,1+2x,\beta_x)$, obtaining
\begin{equation}
\begin{aligned}
 \int_0^\infty\frac{dz}{\widetilde P^2}
 &=\frac{(1+2x)/\beta_x-c_r\mathcal J}{2\Upsilon},\\
 \int_0^\infty\frac{z^2dz}{\widetilde P^2}
 &=\frac{(1+2x)/c_r-\beta_x\mathcal J}{2\Upsilon},\\
 \int_0^\infty\frac{z^2dz}{\widetilde P^3}
 &=\frac{[2(1+2x)^2+c_r\beta_x]\mathcal J-3(1+2x)}{8\Upsilon^2}.
\end{aligned}
 \label{eq:G_integral_identities}
\end{equation}
Substitution into Eq.~\eqref{eq:G_scaled_integral} yields the closed
forms
\begin{align}
 \chi_{zz}^{\mathrm K}
 &=\frac{2\mathcal A_T}{\delta}
       \left(r^2\mathcal J-\frac{1+2x}{\beta_x}\right),
 \label{eq:G_exact_H}\\
 \chi_{zz}^{\mathrm C}
 &=\frac{8\mathcal A_Tr^2}{\delta\Upsilon}
       \left[(2(1+2x)^2+c_r\beta_x)\mathcal J-3(1+2x)\right].
 \label{eq:G_exact_C}
\end{align}

\subsection{Degenerate and resolved regimes}

We take the $x\gg1$ and $x\ll1$ limits of
Eqs.~\eqref{eq:G_exact_H} and \eqref{eq:G_exact_C} at fixed $0\leq r<1$,
as for Eq.~(31) of the \textbf{Main Text}.
For $x\gg1$, the dominant fluctuations do not resolve the mass difference
between the branches. The basic integral becomes
$\mathcal J=\operatorname{atanh}r/[r(1+2x)]+O(x^{-3})$,
with $\operatorname{atanh}r/r\to1$ at $r=0$.
Using $\alpha=\epsilon+\delta$ in
Eqs.~\eqref{eq:G_exact_H} and \eqref{eq:G_exact_C} then gives
\begin{equation}
 \chi_{zz}^{\mathrm K} \simeq\frac{\mathcal A_T}{\alpha}\, 2\big(r\operatorname{atanh}r-1\big), \qquad \chi_{zz}^{\mathrm C} \simeq\frac{\mathcal A_T}{\alpha}\, 8\left[\left(\frac3r-r\right)\operatorname{atanh}r-3\right].
\label{eq:G_degenerate_scaling}
\end{equation}
For generic nonzero $r$, both sectors are of order $1/\alpha\simeq1/\epsilon$.
The $x\gg1$ result describes a preasymptotic window, or the limit in which
$\delta\to0$ is taken first.

In the resolved regime $x\ll1$, define
\begin{equation}
 \mathcal L=\ln\frac{1}{(1-r^2)x}
      =\ln\frac{2\delta}{(1-r^2)\epsilon}.
 \label{eq:G_resolved_log}
\end{equation}
The expansions $\mathcal J=\mathcal L/2+O(x\mathcal L)$ and
$2(1+2x)/\beta_x=(2x)^{-1}+1/2+O(x)$ give
\begin{align}
 \chi_{zz}^{\mathrm K}
 &=-\frac{\mathcal A_T}{\epsilon}
   +\frac{\mathcal A_T}{\delta}
       \left(r^2\mathcal L-\frac12\right)
   +O\!\left(\frac{\mathcal A_T\epsilon}{\delta^2}\mathcal L\right),
 \label{eq:G_resolved_H}\\
 \chi_{zz}^{\mathrm C}
 &=\frac{8\mathcal A_Tr^2}{\delta}(\mathcal L-3)
   +O\!\left(\frac{\mathcal A_T\epsilon}{\delta^2}\mathcal L\right).
 \label{eq:G_resolved_C}
\end{align}
The kinetic contribution retains the negative $1/\epsilon$ pole of one soft mode,
with a weaker positive logarithmic correction. For nonzero $r$, the
coherent contribution has a positive logarithmic divergence. In temperature
variables their leading behavior is
\begin{equation}
 \chi_{zz}^{\mathrm K}\propto-\frac1{T-T_c},\qquad
 \chi_{zz}^{\mathrm C}\propto
       \ln\frac{\delta}{a_E(T-T_c)}.
 \label{eq:G_temperature_scaling}
\end{equation}
Consequently the total minimally coupled response is
\begin{equation}
 \chi_{zz}^{\mathrm{fl}}
 =-\frac{\mathcal A_T}{\epsilon}
  +\frac{\mathcal A_T}{\delta}
       \left(9r^2\mathcal L-\frac12-24r^2\right)
  +O\!\left(\frac{\mathcal A_T\epsilon}{\delta^2}\mathcal L\right),
 \label{eq:G_resolved_total}
\end{equation}
so the response is diamagnetic in the critical limit.
For $r=0.8073$, the response is paramagnetic in the effectively degenerate
regime and diamagnetic sufficiently close to the transition. The crossover
occurs at $x_*$, defined by
$\chi_{zz}^{\mathrm K}(x_*)+\chi_{zz}^{\mathrm C}(x_*)=0$.
Solving this condition with Eqs.~\eqref{eq:G_exact_H} and
\eqref{eq:G_exact_C} gives $x_*=0.05852\ldots\simeq0.06$, as quoted in
the \textbf{Main Text}.

The different singularities can be traced to different momentum ranges.
At $p^2\sim\epsilon/\gamma_0$, the hard mass of
Eq.~\eqref{eq:G_split_kernel} remains finite, while the soft eigenvalue is
\begin{equation}
 \varepsilon_-(p)
 =\epsilon+\gamma_0p^2
       -\frac{\gamma_1^2p^4}{2\delta}+O(p^8).
 \label{eq:G_soft_dispersion}
\end{equation}
The leading integrand of the kinetic contribution is
$-\gamma_0^2/[3(\epsilon+\gamma_0y)^2]$, whose integral gives the
single-mode pole in Eq.~\eqref{eq:G_resolved_H}. The coherent kernel is
suppressed in the deep infrared and accumulates its logarithmic weight
over the intermediate range
$\epsilon/\gamma_0\ll y\ll\delta/\gamma_0$.
In this window, $P\simeq2\delta\gamma_0y$ and
\begin{equation}
 \mathcal I_{\mathrm C}(y)
 \simeq\frac{8\gamma_1^2}{3\delta\gamma_0}\frac1y.
 \label{eq:G_log_momentum}
\end{equation}
Integrating $1/y$ between these two momentum scales gives
$\ln(\delta/\epsilon)$ in Eq.~\eqref{eq:G_resolved_C}. Dividing Eqs.~\eqref{eq:G_resolved_H} and \eqref{eq:G_resolved_C} by $\chi_\epsilon=\mathcal A_T/\epsilon$ gives the normalized limits plotted in Fig.~3(b) of the \textbf{Main Text}. The kinetic contribution tends to $-1$, while the coherent contribution vanishes as $x\ln(1/x)$.

The weaker singularity also follows from the geometry at the soft-mode
minimum. For this quadratic texture,
$g_{ab}(\mathbf0)=\omega_{xy}(\mathbf0)=0$, so all terms in the
$1/\epsilon$ coefficient of the coherent response in
Eq.~\eqref{eq:E_soft_singular_parts} vanish. A general two-mode kernel
can instead have a finite hard mass together with nonzero
$\omega_{xy}(\mathbf0)$ and a coherent $1/\epsilon$ term.

\clearpage
\setcounter{NAT@ctr}{0}
\begingroup
\let\ArxivBibliographyLabel\label
\renewcommand{\label}[1]{\ArxivBibliographyLabel{SM:#1}}

\endgroup

\begin{thebibliography}{24}%
\makeatletter
\providecommand \@ifxundefined [1]{%
 \@ifx{#1\undefined}
}%
\providecommand \@ifnum [1]{%
 \ifnum #1\expandafter \@firstoftwo
 \else \expandafter \@secondoftwo
 \fi
}%
\providecommand \@ifx [1]{%
 \ifx #1\expandafter \@firstoftwo
 \else \expandafter \@secondoftwo
 \fi
}%
\providecommand \natexlab [1]{#1}%
\providecommand \enquote  [1]{``#1''}%
\providecommand \bibnamefont  [1]{#1}%
\providecommand \bibfnamefont [1]{#1}%
\providecommand \citenamefont [1]{#1}%
\providecommand \href@noop [0]{\@secondoftwo}%
\providecommand \href [0]{\begingroup \@sanitize@url \@href}%
\providecommand \@href[1]{\@@startlink{#1}\@@href}%
\providecommand \@@href[1]{\endgroup#1\@@endlink}%
\providecommand \@sanitize@url [0]{\catcode `\\12\catcode `\$12\catcode
  `\&12\catcode `\#12\catcode `\^12\catcode `\_12\catcode `\%12\relax}%
\providecommand \@@startlink[1]{}%
\providecommand \@@endlink[0]{}%
\providecommand \url  [0]{\begingroup\@sanitize@url \@url }%
\providecommand \@url [1]{\endgroup\@href {#1}{\urlprefix }}%
\providecommand \urlprefix  [0]{URL }%
\providecommand \Eprint [0]{\href }%
\providecommand \doibase [0]{https://doi.org/}%
\providecommand \selectlanguage [0]{\@gobble}%
\providecommand \bibinfo  [0]{\@secondoftwo}%
\providecommand \bibfield  [0]{\@secondoftwo}%
\providecommand \translation [1]{[#1]}%
\providecommand \BibitemOpen [0]{}%
\providecommand \bibitemStop [0]{}%
\providecommand \bibitemNoStop [0]{.\EOS\space}%
\providecommand \EOS [0]{\spacefactor3000\relax}%
\providecommand \BibitemShut  [1]{\csname bibitem#1\endcsname}%
\let\auto@bib@innerbib\@empty
%</preamble>
\bibitem [{\citenamefont {Nambu}(1960)}]{SM:Nambu1960quasi}%
  \BibitemOpen
  \bibfield  {author} {\bibinfo {author} {\bibfnamefont {Y.}~\bibnamefont
  {Nambu}},\ }\bibfield  {title} {\emph {\bibinfo {title} {Quasi-particles and
  gauge invariance in the theory of superconductivity}},\ }\href
  {https://doi.org/10.1103/physrev.117.648} {\bibfield  {journal} {\bibinfo
  {journal} {Phys. Rev.}\ }\textbf {\bibinfo {volume} {117}},\ \bibinfo {pages}
  {648} (\bibinfo {year} {1960})}\BibitemShut {NoStop}%
\bibitem [{\citenamefont {Ismail-Beigi}\ \emph {et~al.}(2001)\citenamefont
  {Ismail-Beigi}, \citenamefont {Chang},\ and\ \citenamefont
  {Louie}}]{SM:IsmailBeigiChangLouie2001}%
  \BibitemOpen
  \bibfield  {author} {\bibinfo {author} {\bibfnamefont {S.}~\bibnamefont
  {Ismail-Beigi}}, \bibinfo {author} {\bibfnamefont {E.~K.}\ \bibnamefont
  {Chang}},\ and\ \bibinfo {author} {\bibfnamefont {S.~G.}\ \bibnamefont
  {Louie}},\ }\bibfield  {title} {\emph {\bibinfo {title} {Coupling of nonlocal
  potentials to electromagnetic fields}},\ }\href
  {https://doi.org/10.1103/PhysRevLett.87.087402} {\bibfield  {journal}
  {\bibinfo  {journal} {Phys. Rev. Lett.}\ }\textbf {\bibinfo {volume} {87}},\
  \bibinfo {pages} {087402} (\bibinfo {year} {2001})}\BibitemShut {NoStop}%
\bibitem [{\citenamefont {M{\u a}ntoiu}\ and\ \citenamefont
  {Purice}(2004)}]{SM:MantoiuPurice2004}%
  \BibitemOpen
  \bibfield  {author} {\bibinfo {author} {\bibfnamefont {M.}~\bibnamefont {M{\u
  a}ntoiu}}\ and\ \bibinfo {author} {\bibfnamefont {R.}~\bibnamefont
  {Purice}},\ }\bibfield  {title} {\emph {\bibinfo {title} {The magnetic {Weyl}
  calculus}},\ }\href {https://doi.org/10.1063/1.1668334} {\bibfield  {journal}
  {\bibinfo  {journal} {J. Math. Phys.}\ }\textbf {\bibinfo {volume} {45}},\
  \bibinfo {pages} {1394} (\bibinfo {year} {2004})}\BibitemShut {NoStop}%
\bibitem [{\citenamefont {Provost}\ and\ \citenamefont
  {Vall{\'e}e}(1980)}]{SM:ProvostVallee1980}%
  \BibitemOpen
  \bibfield  {author} {\bibinfo {author} {\bibfnamefont {J.~P.}\ \bibnamefont
  {Provost}}\ and\ \bibinfo {author} {\bibfnamefont {G.}~\bibnamefont
  {Vall{\'e}e}},\ }\bibfield  {title} {\emph {\bibinfo {title} {Riemannian
  structure on manifolds of quantum states}},\ }\href
  {https://doi.org/10.1007/BF02193559} {\bibfield  {journal} {\bibinfo
  {journal} {Commun. Math. Phys.}\ }\textbf {\bibinfo {volume} {76}},\ \bibinfo
  {pages} {289} (\bibinfo {year} {1980})}\BibitemShut {NoStop}%
\bibitem [{\citenamefont {Berry}(1984)}]{SM:Berry1984}%
  \BibitemOpen
  \bibfield  {author} {\bibinfo {author} {\bibfnamefont {M.~V.}\ \bibnamefont
  {Berry}},\ }\bibfield  {title} {\emph {\bibinfo {title} {Quantal phase
  factors accompanying adiabatic changes}},\ }\href
  {https://doi.org/10.1098/rspa.1984.0023} {\bibfield  {journal} {\bibinfo
  {journal} {Proc. R. Soc. London A}\ }\textbf {\bibinfo {volume} {392}},\
  \bibinfo {pages} {45} (\bibinfo {year} {1984})}\BibitemShut {NoStop}%
\bibitem [{\citenamefont {Keskiner}\ and\ \citenamefont
  {Iskin}(2026)}]{SM:keskiner2026superconductivity}%
  \BibitemOpen
  \bibfield  {author} {\bibinfo {author} {\bibfnamefont {M.~A.}\ \bibnamefont
  {Keskiner}}\ and\ \bibinfo {author} {\bibfnamefont {M.}~\bibnamefont
  {Iskin}},\ }\emph {\bibinfo {title} {Superconductivity beyond band geometry:
  emergence of pair quantum geometry}},\ \Eprint
  {https://arxiv.org/abs/2606.06017} {arXiv:2606.06017}  (\bibinfo {year}
  {2026})\BibitemShut {NoStop}%
\bibitem [{\citenamefont {Kalkan}\ and\ \citenamefont
  {Iskin}(2026)}]{SM:KalkanIskin2026CollectivePairing}%
  \BibitemOpen
  \bibfield  {author} {\bibinfo {author} {\bibfnamefont {R.~N.}\ \bibnamefont
  {Kalkan}}\ and\ \bibinfo {author} {\bibfnamefont {M.}~\bibnamefont {Iskin}},\
  }\emph {\bibinfo {title} {Quantum geometry of collective pairing fluctuations
  in the superfluid weight of multiband superconductors}},\ \Eprint
  {https://arxiv.org/abs/2610.08336} {arXiv:2610.08336}  (\bibinfo {year}
  {2026})\BibitemShut {NoStop}%
\bibitem [{\citenamefont {B{\"o}ttcher}\ and\ \citenamefont
  {Wenzel}(2008)}]{SM:BottcherWenzel2008Commutator}%
  \BibitemOpen
  \bibfield  {author} {\bibinfo {author} {\bibfnamefont {A.}~\bibnamefont
  {B{\"o}ttcher}}\ and\ \bibinfo {author} {\bibfnamefont {D.}~\bibnamefont
  {Wenzel}},\ }\bibfield  {title} {\emph {\bibinfo {title} {The {Frobenius}
  norm and the commutator}},\ }\href
  {https://doi.org/10.1016/j.laa.2008.05.020} {\bibfield  {journal} {\bibinfo
  {journal} {Linear Algebra Appl.}\ }\textbf {\bibinfo {volume} {429}},\
  \bibinfo {pages} {1864} (\bibinfo {year} {2008})}\BibitemShut {NoStop}%
\bibitem [{\citenamefont {Fukuyama}(1971)}]{SM:Fukuyama1971}%
  \BibitemOpen
  \bibfield  {author} {\bibinfo {author} {\bibfnamefont {H.}~\bibnamefont
  {Fukuyama}},\ }\bibfield  {title} {\emph {\bibinfo {title} {Theory of orbital
  magnetism of {B}loch electrons: {C}oulomb interactions}},\ }\href
  {https://doi.org/10.1143/PTP.45.704} {\bibfield  {journal} {\bibinfo
  {journal} {Prog. Theor. Phys.}\ }\textbf {\bibinfo {volume} {45}},\ \bibinfo
  {pages} {704} (\bibinfo {year} {1971})}\BibitemShut {NoStop}%
\bibitem [{\citenamefont {Raoux}\ \emph {et~al.}(2015)\citenamefont {Raoux},
  \citenamefont {Pi{\'e}chon}, \citenamefont {Fuchs},\ and\ \citenamefont
  {Montambaux}}]{SM:RaouxEtAl2015}%
  \BibitemOpen
  \bibfield  {author} {\bibinfo {author} {\bibfnamefont {A.}~\bibnamefont
  {Raoux}}, \bibinfo {author} {\bibfnamefont {F.}~\bibnamefont {Pi{\'e}chon}},
  \bibinfo {author} {\bibfnamefont {J.-N.}\ \bibnamefont {Fuchs}},\ and\
  \bibinfo {author} {\bibfnamefont {G.}~\bibnamefont {Montambaux}},\ }\bibfield
   {title} {\emph {\bibinfo {title} {Orbital magnetism in coupled-bands
  models}},\ }\href {https://doi.org/10.1103/PhysRevB.91.085120} {\bibfield
  {journal} {\bibinfo  {journal} {Phys. Rev. B}\ }\textbf {\bibinfo {volume}
  {91}},\ \bibinfo {pages} {085120} (\bibinfo {year} {2015})}\BibitemShut
  {NoStop}%
\bibitem [{\citenamefont {Gao}\ \emph {et~al.}(2015)\citenamefont {Gao},
  \citenamefont {Yang},\ and\ \citenamefont {Niu}}]{SM:GaoYangNiu2015}%
  \BibitemOpen
  \bibfield  {author} {\bibinfo {author} {\bibfnamefont {Y.}~\bibnamefont
  {Gao}}, \bibinfo {author} {\bibfnamefont {S.~A.}\ \bibnamefont {Yang}},\ and\
  \bibinfo {author} {\bibfnamefont {Q.}~\bibnamefont {Niu}},\ }\bibfield
  {title} {\emph {\bibinfo {title} {Geometrical effects in orbital magnetic
  susceptibility}},\ }\href {https://doi.org/10.1103/PhysRevB.91.214405}
  {\bibfield  {journal} {\bibinfo  {journal} {Phys. Rev. B}\ }\textbf {\bibinfo
  {volume} {91}},\ \bibinfo {pages} {214405} (\bibinfo {year}
  {2015})}\BibitemShut {NoStop}%
\bibitem [{\citenamefont {Pi{\'e}chon}\ \emph {et~al.}(2016)\citenamefont
  {Pi{\'e}chon}, \citenamefont {Raoux}, \citenamefont {Fuchs},\ and\
  \citenamefont {Montambaux}}]{SM:PiechonEtAl2016}%
  \BibitemOpen
  \bibfield  {author} {\bibinfo {author} {\bibfnamefont {F.}~\bibnamefont
  {Pi{\'e}chon}}, \bibinfo {author} {\bibfnamefont {A.}~\bibnamefont {Raoux}},
  \bibinfo {author} {\bibfnamefont {J.-N.}\ \bibnamefont {Fuchs}},\ and\
  \bibinfo {author} {\bibfnamefont {G.}~\bibnamefont {Montambaux}},\ }\bibfield
   {title} {\emph {\bibinfo {title} {Geometric orbital susceptibility: Quantum
  metric without {B}erry curvature}},\ }\href
  {https://doi.org/10.1103/PhysRevB.94.134423} {\bibfield  {journal} {\bibinfo
  {journal} {Phys. Rev. B}\ }\textbf {\bibinfo {volume} {94}},\ \bibinfo
  {pages} {134423} (\bibinfo {year} {2016})}\BibitemShut {NoStop}%
\bibitem [{\citenamefont {Mishonov}\ \emph {et~al.}(2003)\citenamefont
  {Mishonov}, \citenamefont {Pachov}, \citenamefont {Genchev}, \citenamefont
  {Atanasova},\ and\ \citenamefont {Damianov}}]{SM:MishonovEtAl2003}%
  \BibitemOpen
  \bibfield  {author} {\bibinfo {author} {\bibfnamefont {T.~M.}\ \bibnamefont
  {Mishonov}}, \bibinfo {author} {\bibfnamefont {G.~V.}\ \bibnamefont
  {Pachov}}, \bibinfo {author} {\bibfnamefont {I.~N.}\ \bibnamefont {Genchev}},
  \bibinfo {author} {\bibfnamefont {L.~A.}\ \bibnamefont {Atanasova}},\ and\
  \bibinfo {author} {\bibfnamefont {D.~{\relax Ch}.}\ \bibnamefont
  {Damianov}},\ }\bibfield  {title} {\emph {\bibinfo {title} {Kinetics and
  {Boltzmann} kinetic equation for fluctuation {Cooper} pairs}},\ }\href
  {https://doi.org/10.1103/PhysRevB.68.054525} {\bibfield  {journal} {\bibinfo
  {journal} {Phys. Rev. B}\ }\textbf {\bibinfo {volume} {68}},\ \bibinfo
  {pages} {054525} (\bibinfo {year} {2003})}\BibitemShut {NoStop}%
\bibitem [{\citenamefont {Boev}\ and\ \citenamefont
  {Kovalev}(2024)}]{SM:BoevKovalev2024}%
  \BibitemOpen
  \bibfield  {author} {\bibinfo {author} {\bibfnamefont {M.~V.}\ \bibnamefont
  {Boev}}\ and\ \bibinfo {author} {\bibfnamefont {V.~M.}\ \bibnamefont
  {Kovalev}},\ }\bibfield  {title} {\emph {\bibinfo {title} {Photovoltaic
  {Hall} effect in two-dimensional fluctuating superconductors}},\ }\href
  {https://doi.org/10.1134/S002136402460294X} {\bibfield  {journal} {\bibinfo
  {journal} {JETP Letters}\ }\textbf {\bibinfo {volume} {120}},\ \bibinfo
  {pages} {494} (\bibinfo {year} {2024})}\BibitemShut {NoStop}%
\bibitem [{\citenamefont {Sun}\ \emph {et~al.}(2026)\citenamefont {Sun},
  \citenamefont {Xie},\ and\ \citenamefont {Nagaosa}}]{SM:sun2026quantum}%
  \BibitemOpen
  \bibfield  {author} {\bibinfo {author} {\bibfnamefont {Z.-T.}\ \bibnamefont
  {Sun}}, \bibinfo {author} {\bibfnamefont {Y.-M.}\ \bibnamefont {Xie}},\ and\
  \bibinfo {author} {\bibfnamefont {N.}~\bibnamefont {Nagaosa}},\ }\emph
  {\bibinfo {title} {Quantum information geometry of multicomponent
  superconducting fluctuation transport}},\ \Eprint
  {https://arxiv.org/abs/2606.15928} {arXiv:2606.15928}  (\bibinfo {year}
  {2026})\BibitemShut {NoStop}%
\bibitem [{\citenamefont {Kiesel}\ and\ \citenamefont
  {Thomale}(2012)}]{SM:KieselThomale2012}%
  \BibitemOpen
  \bibfield  {author} {\bibinfo {author} {\bibfnamefont {M.~L.}\ \bibnamefont
  {Kiesel}}\ and\ \bibinfo {author} {\bibfnamefont {R.}~\bibnamefont
  {Thomale}},\ }\bibfield  {title} {\emph {\bibinfo {title} {Sublattice
  interference in the kagome {Hubbard} model}},\ }\href
  {https://doi.org/10.1103/PhysRevB.86.121105} {\bibfield  {journal} {\bibinfo
  {journal} {Phys. Rev. B}\ }\textbf {\bibinfo {volume} {86}},\ \bibinfo
  {pages} {121105} (\bibinfo {year} {2012})}\BibitemShut {NoStop}%
\bibitem [{\citenamefont {Kiesel}\ \emph {et~al.}(2013)\citenamefont {Kiesel},
  \citenamefont {Platt},\ and\ \citenamefont
  {Thomale}}]{SM:KieselPlattThomale2013}%
  \BibitemOpen
  \bibfield  {author} {\bibinfo {author} {\bibfnamefont {M.~L.}\ \bibnamefont
  {Kiesel}}, \bibinfo {author} {\bibfnamefont {C.}~\bibnamefont {Platt}},\ and\
  \bibinfo {author} {\bibfnamefont {R.}~\bibnamefont {Thomale}},\ }\bibfield
  {title} {\emph {\bibinfo {title} {Unconventional {Fermi} surface
  instabilities in the {Kagome} {Hubbard} model}},\ }\href
  {https://doi.org/10.1103/PhysRevLett.110.126405} {\bibfield  {journal}
  {\bibinfo  {journal} {Phys. Rev. Lett.}\ }\textbf {\bibinfo {volume} {110}},\
  \bibinfo {pages} {126405} (\bibinfo {year} {2013})}\BibitemShut {NoStop}%
\bibitem [{\citenamefont {Wu}\ \emph {et~al.}(2021)\citenamefont {Wu},
  \citenamefont {Schwemmer}, \citenamefont {M{\"u}ller}, \citenamefont
  {Consiglio}, \citenamefont {Sangiovanni}, \citenamefont {{Di Sante}},
  \citenamefont {Iqbal}, \citenamefont {Hanke}, \citenamefont {Schnyder},
  \citenamefont {Denner}, \citenamefont {Fischer}, \citenamefont {Neupert},\
  and\ \citenamefont {Thomale}}]{SM:WuEtAl2021}%
  \BibitemOpen
  \bibfield  {author} {\bibinfo {author} {\bibfnamefont {X.}~\bibnamefont {Wu}}
  \emph {et~al.},\ }\bibfield  {title} {\emph {\bibinfo {title} {Nature of
  unconventional pairing in the {Kagome} superconductors {$A$V$_3$Sb$_5$
  ($A$=K,Rb,Cs)}}},\ }\href {https://doi.org/10.1103/PhysRevLett.127.177001}
  {\bibfield  {journal} {\bibinfo  {journal} {Phys. Rev. Lett.}\ }\textbf
  {\bibinfo {volume} {127}},\ \bibinfo {pages} {177001} (\bibinfo {year}
  {2021})}\BibitemShut {NoStop}%
\bibitem [{\citenamefont {Yu}\ and\ \citenamefont {Li}(2012)}]{SM:YuLi2012}%
  \BibitemOpen
  \bibfield  {author} {\bibinfo {author} {\bibfnamefont {S.-L.}\ \bibnamefont
  {Yu}}\ and\ \bibinfo {author} {\bibfnamefont {J.-X.}\ \bibnamefont {Li}},\
  }\bibfield  {title} {\emph {\bibinfo {title} {Chiral superconducting phase
  and chiral spin-density-wave phase in a {Hubbard} model on the kagome
  lattice}},\ }\href {https://doi.org/10.1103/PhysRevB.85.144402} {\bibfield
  {journal} {\bibinfo  {journal} {Phys. Rev. B}\ }\textbf {\bibinfo {volume}
  {85}},\ \bibinfo {pages} {144402} (\bibinfo {year} {2012})}\BibitemShut
  {NoStop}%
\bibitem [{\citenamefont {Larkin}\ and\ \citenamefont
  {Varlamov}(2005)}]{SM:LarkinVarlamov2005}%
  \BibitemOpen
  \bibfield  {author} {\bibinfo {author} {\bibfnamefont {A.}~\bibnamefont
  {Larkin}}\ and\ \bibinfo {author} {\bibfnamefont {A.}~\bibnamefont
  {Varlamov}},\ }\href
  {https://doi.org/10.1093/acprof:oso/9780198528159.001.0001} {\emph {\bibinfo
  {title} {Theory of Fluctuations in Superconductors}}}\ (\bibinfo  {publisher}
  {Oxford University Press},\ \bibinfo {address} {Oxford},\ \bibinfo {year}
  {2005})\BibitemShut {NoStop}%
\bibitem [{\citenamefont {Thouless}(1960)}]{SM:Thouless1960perturbation}%
  \BibitemOpen
  \bibfield  {author} {\bibinfo {author} {\bibfnamefont {D.~J.}\ \bibnamefont
  {Thouless}},\ }\bibfield  {title} {\emph {\bibinfo {title} {Perturbation
  theory in statistical mechanics and the theory of superconductivity}},\
  }\href {https://doi.org/10.1016/0003-4916(60)90122-6} {\bibfield  {journal}
  {\bibinfo  {journal} {Ann. Phys. (N.Y.)}\ }\textbf {\bibinfo {volume} {10}},\
  \bibinfo {pages} {553} (\bibinfo {year} {1960})}\BibitemShut {NoStop}%
\bibitem [{\citenamefont {Silaev}\ and\ \citenamefont
  {Babaev}(2012)}]{SM:SilaevBabaev2012}%
  \BibitemOpen
  \bibfield  {author} {\bibinfo {author} {\bibfnamefont {M.}~\bibnamefont
  {Silaev}}\ and\ \bibinfo {author} {\bibfnamefont {E.}~\bibnamefont
  {Babaev}},\ }\bibfield  {title} {\emph {\bibinfo {title} {Microscopic
  derivation of two-component {Ginzburg--Landau} model and conditions of its
  applicability in two-band systems}},\ }\href
  {https://doi.org/10.1103/PhysRevB.85.134514} {\bibfield  {journal} {\bibinfo
  {journal} {Phys. Rev. B}\ }\textbf {\bibinfo {volume} {85}},\ \bibinfo
  {pages} {134514} (\bibinfo {year} {2012})}\BibitemShut {NoStop}%
\bibitem [{\citenamefont {{\"O}rd}\ \emph {et~al.}(2012)\citenamefont
  {{\"O}rd}, \citenamefont {Vargunin},\ and\ \citenamefont
  {R{\"a}go}}]{SM:OrdVarguninRago2012}%
  \BibitemOpen
  \bibfield  {author} {\bibinfo {author} {\bibfnamefont {T.}~\bibnamefont
  {{\"O}rd}}, \bibinfo {author} {\bibfnamefont {A.}~\bibnamefont {Vargunin}},\
  and\ \bibinfo {author} {\bibfnamefont {K.}~\bibnamefont {R{\"a}go}},\ }\emph
  {\bibinfo {title} {Spatial correlation of fluctuations in multi-component
  superconducting systems}},\ \Eprint {https://arxiv.org/abs/1211.5287}
  {arXiv:1211.5287}  (\bibinfo {year} {2012})\BibitemShut {NoStop}%
\bibitem [{\citenamefont {Ohgushi}\ \emph {et~al.}(2000)\citenamefont
  {Ohgushi}, \citenamefont {Murakami},\ and\ \citenamefont
  {Nagaosa}}]{SM:OhgushiMurakamiNagaosa2000}%
  \BibitemOpen
  \bibfield  {author} {\bibinfo {author} {\bibfnamefont {K.}~\bibnamefont
  {Ohgushi}}, \bibinfo {author} {\bibfnamefont {S.}~\bibnamefont {Murakami}},\
  and\ \bibinfo {author} {\bibfnamefont {N.}~\bibnamefont {Nagaosa}},\
  }\bibfield  {title} {\emph {\bibinfo {title} {Spin anisotropy and quantum
  {Hall} effect in the kagom{\'e} lattice: Chiral spin state based on a
  ferromagnet}},\ }\href {https://doi.org/10.1103/PhysRevB.62.R6065} {\bibfield
   {journal} {\bibinfo  {journal} {Phys. Rev. B}\ }\textbf {\bibinfo {volume}
  {62}},\ \bibinfo {pages} {R6065} (\bibinfo {year} {2000})}\BibitemShut
  {NoStop}%
\end{thebibliography}

\begin{thebibliography}{58}%
\makeatletter
\providecommand \@ifxundefined [1]{%
 \@ifx{#1\undefined}
}%
\providecommand \@ifnum [1]{%
 \ifnum #1\expandafter \@firstoftwo
 \else \expandafter \@secondoftwo
 \fi
}%
\providecommand \@ifx [1]{%
 \ifx #1\expandafter \@firstoftwo
 \else \expandafter \@secondoftwo
 \fi
}%
\providecommand \natexlab [1]{#1}%
\providecommand \enquote  [1]{``#1''}%
\providecommand \bibnamefont  [1]{#1}%
\providecommand \bibfnamefont [1]{#1}%
\providecommand \citenamefont [1]{#1}%
\providecommand \href@noop [0]{\@secondoftwo}%
\providecommand \href [0]{\begingroup \@sanitize@url \@href}%
\providecommand \@href[1]{\@@startlink{#1}\@@href}%
\providecommand \@@href[1]{\endgroup#1\@@endlink}%
\providecommand \@sanitize@url [0]{\catcode `\\12\catcode `\$12\catcode
  `\&12\catcode `\#12\catcode `\^12\catcode `\_12\catcode `\%12\relax}%
\providecommand \@@startlink[1]{}%
\providecommand \@@endlink[0]{}%
\providecommand \url  [0]{\begingroup\@sanitize@url \@url }%
\providecommand \@url [1]{\endgroup\@href {#1}{\urlprefix }}%
\providecommand \urlprefix  [0]{URL }%
\providecommand \Eprint [0]{\href }%
\providecommand \doibase [0]{https://doi.org/}%
\providecommand \selectlanguage [0]{\@gobble}%
\providecommand \bibinfo  [0]{\@secondoftwo}%
\providecommand \bibfield  [0]{\@secondoftwo}%
\providecommand \translation [1]{[#1]}%
\providecommand \BibitemOpen [0]{}%
\providecommand \bibitemStop [0]{}%
\providecommand \bibitemNoStop [0]{.\EOS\space}%
\providecommand \EOS [0]{\spacefactor3000\relax}%
\providecommand \BibitemShut  [1]{\csname bibitem#1\endcsname}%
\let\auto@bib@innerbib\@empty
%</preamble>
\bibitem [{\citenamefont {Larkin}\ and\ \citenamefont
  {Varlamov}(2005)}]{LarkinVarlamov2005}%
  \BibitemOpen
  \bibfield  {author} {\bibinfo {author} {\bibfnamefont {A.}~\bibnamefont
  {Larkin}}\ and\ \bibinfo {author} {\bibfnamefont {A.}~\bibnamefont
  {Varlamov}},\ }\href
  {https://doi.org/10.1093/acprof:oso/9780198528159.001.0001} {\emph {\bibinfo
  {title} {Theory of Fluctuations in Superconductors}}}\ (\bibinfo  {publisher}
  {Oxford University Press},\ \bibinfo {address} {Oxford},\ \bibinfo {year}
  {2005})\BibitemShut {NoStop}%
\bibitem [{\citenamefont {Schmid}(1969)}]{schmid1969diamagnetic}%
  \BibitemOpen
  \bibfield  {author} {\bibinfo {author} {\bibfnamefont {A.}~\bibnamefont
  {Schmid}},\ }\bibfield  {title} {\emph {\bibinfo {title} {Diamagnetic
  susceptibility at the transition to the superconducting state}},\ }\href
  {https://doi.org/10.1103/PhysRev.180.527} {\bibfield  {journal} {\bibinfo
  {journal} {Phys. Rev.}\ }\textbf {\bibinfo {volume} {180}},\ \bibinfo {pages}
  {527} (\bibinfo {year} {1969})}\BibitemShut {NoStop}%
\bibitem [{\citenamefont {Lee}\ and\ \citenamefont
  {Payne}(1971)}]{LeePayne1971}%
  \BibitemOpen
  \bibfield  {author} {\bibinfo {author} {\bibfnamefont {P.~A.}\ \bibnamefont
  {Lee}}\ and\ \bibinfo {author} {\bibfnamefont {M.~G.}\ \bibnamefont
  {Payne}},\ }\bibfield  {title} {\emph {\bibinfo {title} {Theory of
  fluctuation-induced diamagnetism in superconductors}},\ }\href
  {https://doi.org/10.1103/PhysRevLett.26.1537} {\bibfield  {journal} {\bibinfo
   {journal} {Phys. Rev. Lett.}\ }\textbf {\bibinfo {volume} {26}},\ \bibinfo
  {pages} {1537} (\bibinfo {year} {1971})},\ \bibinfo {note} {erratum: Phys.
  Rev. Lett. 27, 170 (1971)}\BibitemShut {NoStop}%
\bibitem [{\citenamefont {Lee}\ and\ \citenamefont
  {Payne}(1972)}]{LeePayne1972}%
  \BibitemOpen
  \bibfield  {author} {\bibinfo {author} {\bibfnamefont {P.~A.}\ \bibnamefont
  {Lee}}\ and\ \bibinfo {author} {\bibfnamefont {M.~G.}\ \bibnamefont
  {Payne}},\ }\bibfield  {title} {\emph {\bibinfo {title} {Pair propagator
  approach to fluctuation-induced diamagnetism in superconductors---effects of
  impurities}},\ }\href {https://doi.org/10.1103/PhysRevB.5.923} {\bibfield
  {journal} {\bibinfo  {journal} {Phys. Rev. B}\ }\textbf {\bibinfo {volume}
  {5}},\ \bibinfo {pages} {923} (\bibinfo {year} {1972})}\BibitemShut {NoStop}%
\bibitem [{\citenamefont {Prange}(1970)}]{prange1970diamagnetic}%
  \BibitemOpen
  \bibfield  {author} {\bibinfo {author} {\bibfnamefont {R.~E.}\ \bibnamefont
  {Prange}},\ }\bibfield  {title} {\emph {\bibinfo {title} {Diamagnetic
  susceptibility at the transition to the superconducting state}},\ }\href
  {https://doi.org/10.1103/PhysRevB.1.2349} {\bibfield  {journal} {\bibinfo
  {journal} {Phys. Rev. B}\ }\textbf {\bibinfo {volume} {1}},\ \bibinfo {pages}
  {2349} (\bibinfo {year} {1970})}\BibitemShut {NoStop}%
\bibitem [{\citenamefont {Tinkham}(2004)}]{Tinkham2004}%
  \BibitemOpen
  \bibfield  {author} {\bibinfo {author} {\bibfnamefont {M.}~\bibnamefont
  {Tinkham}},\ }\href
  {https://store.doverpublications.com/products/9780486435039} {\emph {\bibinfo
  {title} {Introduction to Superconductivity}}},\ \bibinfo {edition} {2nd}\
  ed.\ (\bibinfo  {publisher} {Dover Publications},\ \bibinfo {year}
  {2004})\BibitemShut {NoStop}%
\bibitem [{\citenamefont {Gollub}\ \emph {et~al.}(1969)\citenamefont {Gollub},
  \citenamefont {Beasley}, \citenamefont {Newbower},\ and\ \citenamefont
  {Tinkham}}]{GollubEtAl1969Diamagnetism}%
  \BibitemOpen
  \bibfield  {author} {\bibinfo {author} {\bibfnamefont {J.~P.}\ \bibnamefont
  {Gollub}}, \bibinfo {author} {\bibfnamefont {M.~R.}\ \bibnamefont {Beasley}},
  \bibinfo {author} {\bibfnamefont {R.~S.}\ \bibnamefont {Newbower}},\ and\
  \bibinfo {author} {\bibfnamefont {M.}~\bibnamefont {Tinkham}},\ }\bibfield
  {title} {\emph {\bibinfo {title} {Observation of enhanced diamagnetism above
  {$T_c$} in indium due to thermodynamic fluctuations}},\ }\href
  {https://doi.org/10.1103/PhysRevLett.22.1288} {\bibfield  {journal} {\bibinfo
   {journal} {Phys. Rev. Lett.}\ }\textbf {\bibinfo {volume} {22}},\ \bibinfo
  {pages} {1288} (\bibinfo {year} {1969})}\BibitemShut {NoStop}%
\bibitem [{\citenamefont {Li}\ \emph {et~al.}(2010)\citenamefont {Li},
  \citenamefont {Wang}, \citenamefont {Komiya}, \citenamefont {Ono},
  \citenamefont {Ando}, \citenamefont {Gu},\ and\ \citenamefont
  {Ong}}]{LiEtAl2010Diamagnetism}%
  \BibitemOpen
  \bibfield  {author} {\bibinfo {author} {\bibfnamefont {L.}~\bibnamefont
  {Li}}, \bibinfo {author} {\bibfnamefont {Y.}~\bibnamefont {Wang}}, \bibinfo
  {author} {\bibfnamefont {S.}~\bibnamefont {Komiya}}, \bibinfo {author}
  {\bibfnamefont {S.}~\bibnamefont {Ono}}, \bibinfo {author} {\bibfnamefont
  {Y.}~\bibnamefont {Ando}}, \bibinfo {author} {\bibfnamefont {G.~D.}\
  \bibnamefont {Gu}},\ and\ \bibinfo {author} {\bibfnamefont {N.~P.}\
  \bibnamefont {Ong}},\ }\bibfield  {title} {\emph {\bibinfo {title}
  {Diamagnetism and {Cooper} pairing above {$T_c$} in cuprates}},\ }\href
  {https://doi.org/10.1103/PhysRevB.81.054510} {\bibfield  {journal} {\bibinfo
  {journal} {Phys. Rev. B}\ }\textbf {\bibinfo {volume} {81}},\ \bibinfo
  {pages} {054510} (\bibinfo {year} {2010})}\BibitemShut {NoStop}%
\bibitem [{\citenamefont {Adachi}\ and\ \citenamefont
  {Ikeda}(2016)}]{AdachiIkeda2016}%
  \BibitemOpen
  \bibfield  {author} {\bibinfo {author} {\bibfnamefont {K.}~\bibnamefont
  {Adachi}}\ and\ \bibinfo {author} {\bibfnamefont {R.}~\bibnamefont {Ikeda}},\
  }\bibfield  {title} {\emph {\bibinfo {title} {Fluctuation diamagnetism in
  two-band superconductors}},\ }\href
  {https://doi.org/10.1103/PhysRevB.93.134503} {\bibfield  {journal} {\bibinfo
  {journal} {Phys. Rev. B}\ }\textbf {\bibinfo {volume} {93}},\ \bibinfo
  {pages} {134503} (\bibinfo {year} {2016})}\BibitemShut {NoStop}%
\bibitem [{\citenamefont {Koshelev}\ \emph {et~al.}(2005)\citenamefont
  {Koshelev}, \citenamefont {Varlamov},\ and\ \citenamefont
  {Vinokur}}]{KoshelevVarlamovVinokur2005}%
  \BibitemOpen
  \bibfield  {author} {\bibinfo {author} {\bibfnamefont {A.~E.}\ \bibnamefont
  {Koshelev}}, \bibinfo {author} {\bibfnamefont {A.~A.}\ \bibnamefont
  {Varlamov}},\ and\ \bibinfo {author} {\bibfnamefont {V.~M.}\ \bibnamefont
  {Vinokur}},\ }\bibfield  {title} {\emph {\bibinfo {title} {Theory of
  fluctuations in a two-band superconductor: {MgB}$_2$}},\ }\href
  {https://doi.org/10.1103/PhysRevB.72.064523} {\bibfield  {journal} {\bibinfo
  {journal} {Phys. Rev. B}\ }\textbf {\bibinfo {volume} {72}},\ \bibinfo
  {pages} {064523} (\bibinfo {year} {2005})}\BibitemShut {NoStop}%
\bibitem [{\citenamefont {Sun}\ \emph {et~al.}(2026)\citenamefont {Sun},
  \citenamefont {Xie},\ and\ \citenamefont {Nagaosa}}]{sun2026quantum}%
  \BibitemOpen
  \bibfield  {author} {\bibinfo {author} {\bibfnamefont {Z.-T.}\ \bibnamefont
  {Sun}}, \bibinfo {author} {\bibfnamefont {Y.-M.}\ \bibnamefont {Xie}},\ and\
  \bibinfo {author} {\bibfnamefont {N.}~\bibnamefont {Nagaosa}},\ }\emph
  {\bibinfo {title} {Quantum information geometry of multicomponent
  superconducting fluctuation transport}},\ \Eprint
  {https://arxiv.org/abs/2606.15928} {arXiv:2606.15928}  (\bibinfo {year}
  {2026})\BibitemShut {NoStop}%
\bibitem [{\citenamefont {Keskiner}\ and\ \citenamefont
  {Iskin}(2026)}]{keskiner2026superconductivity}%
  \BibitemOpen
  \bibfield  {author} {\bibinfo {author} {\bibfnamefont {M.~A.}\ \bibnamefont
  {Keskiner}}\ and\ \bibinfo {author} {\bibfnamefont {M.}~\bibnamefont
  {Iskin}},\ }\emph {\bibinfo {title} {Superconductivity beyond band geometry:
  emergence of pair quantum geometry}},\ \Eprint
  {https://arxiv.org/abs/2606.06017} {arXiv:2606.06017}  (\bibinfo {year}
  {2026})\BibitemShut {NoStop}%
\bibitem [{\citenamefont {Kalkan}\ and\ \citenamefont
  {Iskin}(2026)}]{KalkanIskin2026CollectivePairing}%
  \BibitemOpen
  \bibfield  {author} {\bibinfo {author} {\bibfnamefont {R.~N.}\ \bibnamefont
  {Kalkan}}\ and\ \bibinfo {author} {\bibfnamefont {M.}~\bibnamefont {Iskin}},\
  }\emph {\bibinfo {title} {Quantum geometry of collective pairing fluctuations
  in the superfluid weight of multiband superconductors}},\ \Eprint
  {https://arxiv.org/abs/2610.08336} {arXiv:2610.08336}  (\bibinfo {year}
  {2026})\BibitemShut {NoStop}%
\bibitem [{\citenamefont {Nagashima}\ \emph {et~al.}(2024)\citenamefont
  {Nagashima}, \citenamefont {Tian}, \citenamefont {Haenel}, \citenamefont
  {Tsuji},\ and\ \citenamefont {Manske}}]{NagashimaEtAl2024}%
  \BibitemOpen
  \bibfield  {author} {\bibinfo {author} {\bibfnamefont {R.}~\bibnamefont
  {Nagashima}}, \bibinfo {author} {\bibfnamefont {S.}~\bibnamefont {Tian}},
  \bibinfo {author} {\bibfnamefont {R.}~\bibnamefont {Haenel}}, \bibinfo
  {author} {\bibfnamefont {N.}~\bibnamefont {Tsuji}},\ and\ \bibinfo {author}
  {\bibfnamefont {D.}~\bibnamefont {Manske}},\ }\bibfield  {title} {\emph
  {\bibinfo {title} {Classification of {Lifshitz} invariant in multiband
  superconductors: An application to {Leggett} modes in the linear response
  regime in {Kagome} lattice models}},\ }\href
  {https://doi.org/10.1103/PhysRevResearch.6.013120} {\bibfield  {journal}
  {\bibinfo  {journal} {Phys. Rev. Research}\ }\textbf {\bibinfo {volume}
  {6}},\ \bibinfo {pages} {013120} (\bibinfo {year} {2024})}\BibitemShut
  {NoStop}%
\bibitem [{\citenamefont {Holb{\ae}k}\ and\ \citenamefont
  {Fischer}(2025)}]{HolbaekFischer2025}%
  \BibitemOpen
  \bibfield  {author} {\bibinfo {author} {\bibfnamefont {S.~C.}\ \bibnamefont
  {Holb{\ae}k}}\ and\ \bibinfo {author} {\bibfnamefont {M.~H.}\ \bibnamefont
  {Fischer}},\ }\bibfield  {title} {\emph {\bibinfo {title} {Interplay of
  superconductivity and charge-density-wave order in kagome materials}},\
  }\href {https://doi.org/10.1103/PhysRevResearch.7.023129} {\bibfield
  {journal} {\bibinfo  {journal} {Phys. Rev. Research}\ }\textbf {\bibinfo
  {volume} {7}},\ \bibinfo {pages} {023129} (\bibinfo {year}
  {2025})}\BibitemShut {NoStop}%
\bibitem [{\citenamefont {Feng}\ \emph {et~al.}(2021)\citenamefont {Feng},
  \citenamefont {Jiang}, \citenamefont {Wang},\ and\ \citenamefont
  {Hu}}]{FengEtAl2021}%
  \BibitemOpen
  \bibfield  {author} {\bibinfo {author} {\bibfnamefont {X.}~\bibnamefont
  {Feng}}, \bibinfo {author} {\bibfnamefont {K.}~\bibnamefont {Jiang}},
  \bibinfo {author} {\bibfnamefont {Z.}~\bibnamefont {Wang}},\ and\ \bibinfo
  {author} {\bibfnamefont {J.}~\bibnamefont {Hu}},\ }\bibfield  {title} {\emph
  {\bibinfo {title} {Chiral flux phase in the {Kagome} superconductor
  {AV$_3$Sb$_5$}}},\ }\href {https://doi.org/10.1016/j.scib.2021.04.043}
  {\bibfield  {journal} {\bibinfo  {journal} {Sci. Bull.}\ }\textbf {\bibinfo
  {volume} {66}},\ \bibinfo {pages} {1384} (\bibinfo {year}
  {2021})}\BibitemShut {NoStop}%
\bibitem [{\citenamefont {Denner}\ \emph {et~al.}(2021)\citenamefont {Denner},
  \citenamefont {Thomale},\ and\ \citenamefont
  {Neupert}}]{DennerThomaleNeupert2021}%
  \BibitemOpen
  \bibfield  {author} {\bibinfo {author} {\bibfnamefont {M.~M.}\ \bibnamefont
  {Denner}}, \bibinfo {author} {\bibfnamefont {R.}~\bibnamefont {Thomale}},\
  and\ \bibinfo {author} {\bibfnamefont {T.}~\bibnamefont {Neupert}},\
  }\bibfield  {title} {\emph {\bibinfo {title} {Analysis of charge order in the
  kagome metal {AV$_3$Sb$_5$} {($A$=K,Rb,Cs)}}},\ }\href
  {https://doi.org/10.1103/PhysRevLett.127.217601} {\bibfield  {journal}
  {\bibinfo  {journal} {Phys. Rev. Lett.}\ }\textbf {\bibinfo {volume} {127}},\
  \bibinfo {pages} {217601} (\bibinfo {year} {2021})},\ \bibinfo {note}
  {erratum: Phys. Rev. Lett. 128, 099901 (2022)}\BibitemShut {NoStop}%
\bibitem [{\citenamefont {Wilson}\ and\ \citenamefont
  {Ortiz}(2024)}]{WilsonOrtiz2024Review}%
  \BibitemOpen
  \bibfield  {author} {\bibinfo {author} {\bibfnamefont {S.~D.}\ \bibnamefont
  {Wilson}}\ and\ \bibinfo {author} {\bibfnamefont {B.~R.}\ \bibnamefont
  {Ortiz}},\ }\bibfield  {title} {\emph {\bibinfo {title} {{AV$_3$Sb$_5$}
  kagome superconductors}},\ }\href
  {https://doi.org/10.1038/s41578-024-00677-y} {\bibfield  {journal} {\bibinfo
  {journal} {Nat. Rev. Mater.}\ }\textbf {\bibinfo {volume} {9}},\ \bibinfo
  {pages} {420} (\bibinfo {year} {2024})}\BibitemShut {NoStop}%
\bibitem [{\citenamefont {Neupert}\ \emph {et~al.}(2022)\citenamefont
  {Neupert}, \citenamefont {Denner}, \citenamefont {Yin}, \citenamefont
  {Thomale},\ and\ \citenamefont {Hasan}}]{NeupertEtAl2022}%
  \BibitemOpen
  \bibfield  {author} {\bibinfo {author} {\bibfnamefont {T.}~\bibnamefont
  {Neupert}}, \bibinfo {author} {\bibfnamefont {M.~M.}\ \bibnamefont {Denner}},
  \bibinfo {author} {\bibfnamefont {J.-X.}\ \bibnamefont {Yin}}, \bibinfo
  {author} {\bibfnamefont {R.}~\bibnamefont {Thomale}},\ and\ \bibinfo {author}
  {\bibfnamefont {M.~Z.}\ \bibnamefont {Hasan}},\ }\bibfield  {title} {\emph
  {\bibinfo {title} {Charge order and superconductivity in {K}agome
  materials}},\ }\href {https://doi.org/10.1038/s41567-021-01404-y} {\bibfield
  {journal} {\bibinfo  {journal} {Nat. Phys.}\ }\textbf {\bibinfo {volume}
  {18}},\ \bibinfo {pages} {137} (\bibinfo {year} {2022})}\BibitemShut
  {NoStop}%
\bibitem [{\citenamefont {Jiang}\ \emph {et~al.}(2023)\citenamefont {Jiang},
  \citenamefont {Wu}, \citenamefont {Yin}, \citenamefont {Wang}, \citenamefont
  {Hasan}, \citenamefont {Wilson}, \citenamefont {Chen},\ and\ \citenamefont
  {Hu}}]{JiangEtAl2023KagomeReview}%
  \BibitemOpen
  \bibfield  {author} {\bibinfo {author} {\bibfnamefont {K.}~\bibnamefont
  {Jiang}}, \bibinfo {author} {\bibfnamefont {T.}~\bibnamefont {Wu}}, \bibinfo
  {author} {\bibfnamefont {J.-X.}\ \bibnamefont {Yin}}, \bibinfo {author}
  {\bibfnamefont {Z.}~\bibnamefont {Wang}}, \bibinfo {author} {\bibfnamefont
  {M.~Z.}\ \bibnamefont {Hasan}}, \bibinfo {author} {\bibfnamefont {S.~D.}\
  \bibnamefont {Wilson}}, \bibinfo {author} {\bibfnamefont {X.}~\bibnamefont
  {Chen}},\ and\ \bibinfo {author} {\bibfnamefont {J.}~\bibnamefont {Hu}},\
  }\bibfield  {title} {\emph {\bibinfo {title} {Kagome superconductors
  {AV$_3$Sb$_5$} {($A$ = K, Rb, Cs)}}},\ }\href
  {https://doi.org/10.1093/nsr/nwac199} {\bibfield  {journal} {\bibinfo
  {journal} {Natl. Sci. Rev.}\ }\textbf {\bibinfo {volume} {10}},\ \bibinfo
  {pages} {nwac199} (\bibinfo {year} {2023})}\BibitemShut {NoStop}%
\bibitem [{\citenamefont {Yin}\ \emph {et~al.}(2022)\citenamefont {Yin},
  \citenamefont {Lian},\ and\ \citenamefont {Hasan}}]{YinLianHasan2022Review}%
  \BibitemOpen
  \bibfield  {author} {\bibinfo {author} {\bibfnamefont {J.-X.}\ \bibnamefont
  {Yin}}, \bibinfo {author} {\bibfnamefont {B.}~\bibnamefont {Lian}},\ and\
  \bibinfo {author} {\bibfnamefont {M.~Z.}\ \bibnamefont {Hasan}},\ }\bibfield
  {title} {\emph {\bibinfo {title} {Topological kagome magnets and
  superconductors}},\ }\href {https://doi.org/10.1038/s41586-022-05516-0}
  {\bibfield  {journal} {\bibinfo  {journal} {Nature (London)}\ }\textbf
  {\bibinfo {volume} {612}},\ \bibinfo {pages} {647} (\bibinfo {year}
  {2022})}\BibitemShut {NoStop}%
\bibitem [{\citenamefont {Yin}\ \emph {et~al.}(2026)\citenamefont {Yin},
  \citenamefont {Wu}, \citenamefont {Fischer}, \citenamefont {Yan},
  \citenamefont {Zhong},\ and\ \citenamefont {Okazaki}}]{YinEtAl2026}%
  \BibitemOpen
  \bibfield  {author} {\bibinfo {author} {\bibfnamefont {J.-X.}\ \bibnamefont
  {Yin}}, \bibinfo {author} {\bibfnamefont {X.}~\bibnamefont {Wu}}, \bibinfo
  {author} {\bibfnamefont {M.~H.}\ \bibnamefont {Fischer}}, \bibinfo {author}
  {\bibfnamefont {X.-Y.}\ \bibnamefont {Yan}}, \bibinfo {author} {\bibfnamefont
  {Y.}~\bibnamefont {Zhong}},\ and\ \bibinfo {author} {\bibfnamefont
  {K.}~\bibnamefont {Okazaki}},\ }\bibfield  {title} {\emph {\bibinfo {title}
  {Topological pair density waves in kagome superconductors}},\ }\href
  {https://doi.org/10.1038/s42254-026-00974-1} {\bibfield  {journal} {\bibinfo
  {journal} {Nat. Rev. Phys.}\ }\textbf {\bibinfo {volume} {8}},\ \bibinfo
  {pages} {667} (\bibinfo {year} {2026})}\BibitemShut {NoStop}%
\bibitem [{\citenamefont {{Di Sante}}\ \emph {et~al.}(2026)\citenamefont {{Di
  Sante}}, \citenamefont {Neupert}, \citenamefont {Sangiovanni}, \citenamefont
  {Thomale}, \citenamefont {Comin}, \citenamefont {Checkelsky}, \citenamefont
  {Zeljkovic},\ and\ \citenamefont {Wilson}}]{DiSanteEtAl2026KagomeMetals}%
  \BibitemOpen
  \bibfield  {author} {\bibinfo {author} {\bibfnamefont {D.}~\bibnamefont {{Di
  Sante}}}, \bibinfo {author} {\bibfnamefont {T.}~\bibnamefont {Neupert}},
  \bibinfo {author} {\bibfnamefont {G.}~\bibnamefont {Sangiovanni}}, \bibinfo
  {author} {\bibfnamefont {R.}~\bibnamefont {Thomale}}, \bibinfo {author}
  {\bibfnamefont {R.}~\bibnamefont {Comin}}, \bibinfo {author} {\bibfnamefont
  {J.~G.}\ \bibnamefont {Checkelsky}}, \bibinfo {author} {\bibfnamefont
  {I.}~\bibnamefont {Zeljkovic}},\ and\ \bibinfo {author} {\bibfnamefont
  {S.~D.}\ \bibnamefont {Wilson}},\ }\bibfield  {title} {\emph {\bibinfo
  {title} {Kagome metals}},\ }\href {https://doi.org/10.1103/1g9n-wm38}
  {\bibfield  {journal} {\bibinfo  {journal} {Rev. Mod. Phys.}\ }\textbf
  {\bibinfo {volume} {98}},\ \bibinfo {pages} {015002} (\bibinfo {year}
  {2026})}\BibitemShut {NoStop}%
\bibitem [{\citenamefont {Sigrist}\ and\ \citenamefont
  {Ueda}(1991)}]{SigristUeda1991}%
  \BibitemOpen
  \bibfield  {author} {\bibinfo {author} {\bibfnamefont {M.}~\bibnamefont
  {Sigrist}}\ and\ \bibinfo {author} {\bibfnamefont {K.}~\bibnamefont {Ueda}},\
  }\bibfield  {title} {\emph {\bibinfo {title} {Phenomenological theory of
  unconventional superconductivity}},\ }\href
  {https://doi.org/10.1103/RevModPhys.63.239} {\bibfield  {journal} {\bibinfo
  {journal} {Rev. Mod. Phys.}\ }\textbf {\bibinfo {volume} {63}},\ \bibinfo
  {pages} {239} (\bibinfo {year} {1991})}\BibitemShut {NoStop}%
\bibitem [{\citenamefont {Maiti}\ and\ \citenamefont
  {Hirschfeld}(2015)}]{MaitiHirschfeld2015}%
  \BibitemOpen
  \bibfield  {author} {\bibinfo {author} {\bibfnamefont {S.}~\bibnamefont
  {Maiti}}\ and\ \bibinfo {author} {\bibfnamefont {P.~J.}\ \bibnamefont
  {Hirschfeld}},\ }\bibfield  {title} {\emph {\bibinfo {title} {Collective
  modes in superconductors with competing $s$- and $d$-wave interactions}},\
  }\href {https://doi.org/10.1103/PhysRevB.92.094506} {\bibfield  {journal}
  {\bibinfo  {journal} {Phys. Rev. B}\ }\textbf {\bibinfo {volume} {92}},\
  \bibinfo {pages} {094506} (\bibinfo {year} {2015})}\BibitemShut {NoStop}%
\bibitem [{\citenamefont {Leggett}(1966)}]{Leggett1966}%
  \BibitemOpen
  \bibfield  {author} {\bibinfo {author} {\bibfnamefont {A.~J.}\ \bibnamefont
  {Leggett}},\ }\bibfield  {title} {\emph {\bibinfo {title} {Number-phase
  fluctuations in two-band superconductors}},\ }\href
  {https://doi.org/10.1143/PTP.36.901} {\bibfield  {journal} {\bibinfo
  {journal} {Prog. Theor. Phys.}\ }\textbf {\bibinfo {volume} {36}},\ \bibinfo
  {pages} {901} (\bibinfo {year} {1966})}\BibitemShut {NoStop}%
\bibitem [{\citenamefont {Altland}\ and\ \citenamefont
  {Simons}(2023)}]{Altland_Simons_2023}%
  \BibitemOpen
  \bibfield  {author} {\bibinfo {author} {\bibfnamefont {A.}~\bibnamefont
  {Altland}}\ and\ \bibinfo {author} {\bibfnamefont {B.}~\bibnamefont
  {Simons}},\ }\href {https://doi.org/10.1017/9781108781244} {\emph {\bibinfo
  {title} {Condensed Matter Field Theory}}},\ \bibinfo {edition} {3rd}\ ed.\
  (\bibinfo  {publisher} {Cambridge University Press},\ \bibinfo {year}
  {2023})\BibitemShut {NoStop}%
\bibitem [{\citenamefont {Gor'kov}(1959)}]{Gorkov1959GL}%
  \BibitemOpen
  \bibfield  {author} {\bibinfo {author} {\bibfnamefont {L.~P.}\ \bibnamefont
  {Gor'kov}},\ }\bibfield  {title} {\emph {\bibinfo {title} {Microscopic
  derivation of the {Ginzburg-Landau} equations in the theory of
  superconductivity}},\ }\href@noop {} {\bibfield  {journal} {\bibinfo
  {journal} {Sov. Phys. JETP}\ }\textbf {\bibinfo {volume} {9}},\ \bibinfo
  {pages} {1364} (\bibinfo {year} {1959})},\ \bibinfo {note} {{R}ussian
  original: Zh. Eksp. Teor. Fiz. 36, 1918--1923 (1959)}\BibitemShut {NoStop}%
\bibitem [{Sup()}]{SupplementaryMaterial}%
  \BibitemOpen
  \href@noop {} {\bibinfo {title} {The {Supplemental Material} for ``{Quantum}
  geometry drives orbital paramagnetism in superconducting fluctuations''}},\
  \bibinfo {note} {{which includes four parts: S1. Ward--Takahashi identities
  and gauge invariance; S2. Pair quantum geometry interpretation; S3.
  Microscopic model on kagome lattice; S4. Critical scaling with a chiral
  splitting.}}\BibitemShut {Stop}%
\bibitem [{\citenamefont {Bohm}(1949)}]{bohm1949note}%
  \BibitemOpen
  \bibfield  {author} {\bibinfo {author} {\bibfnamefont {D.}~\bibnamefont
  {Bohm}},\ }\bibfield  {title} {\emph {\bibinfo {title} {Note on a theorem of
  {Bloch} concerning possible causes of superconductivity}},\ }\href
  {https://doi.org/10.1103/PhysRev.75.502} {\bibfield  {journal} {\bibinfo
  {journal} {Phys. Rev.}\ }\textbf {\bibinfo {volume} {75}},\ \bibinfo {pages}
  {502} (\bibinfo {year} {1949})}\BibitemShut {NoStop}%
\bibitem [{\citenamefont {Liu}\ \emph {et~al.}(2015)\citenamefont {Liu},
  \citenamefont {Jing},\ and\ \citenamefont {Wang}}]{LiuJingWang2015RLD}%
  \BibitemOpen
  \bibfield  {author} {\bibinfo {author} {\bibfnamefont {J.}~\bibnamefont
  {Liu}}, \bibinfo {author} {\bibfnamefont {X.-X.}\ \bibnamefont {Jing}},\ and\
  \bibinfo {author} {\bibfnamefont {X.}~\bibnamefont {Wang}},\ }\bibfield
  {title} {\emph {\bibinfo {title} {Quantum metrology with unitary
  parametrization processes}},\ }\href {https://doi.org/10.1038/srep08565}
  {\bibfield  {journal} {\bibinfo  {journal} {Sci. Rep.}\ }\textbf {\bibinfo
  {volume} {5}},\ \bibinfo {pages} {8565} (\bibinfo {year} {2015})}\BibitemShut
  {NoStop}%
\bibitem [{\citenamefont {Nakahara}(2003)}]{Nakahara2003}%
  \BibitemOpen
  \bibfield  {author} {\bibinfo {author} {\bibfnamefont {M.}~\bibnamefont
  {Nakahara}},\ }\href@noop {} {\emph {\bibinfo {title} {Geometry, Topology and
  Physics}}},\ \bibinfo {edition} {2nd}\ ed.\ (\bibinfo  {publisher} {Institute
  of Physics Publishing},\ \bibinfo {address} {Bristol and Philadelphia},\
  \bibinfo {year} {2003})\BibitemShut {NoStop}%
\bibitem [{\citenamefont {B{\"o}ttcher}\ and\ \citenamefont
  {Wenzel}(2008)}]{BottcherWenzel2008Commutator}%
  \BibitemOpen
  \bibfield  {author} {\bibinfo {author} {\bibfnamefont {A.}~\bibnamefont
  {B{\"o}ttcher}}\ and\ \bibinfo {author} {\bibfnamefont {D.}~\bibnamefont
  {Wenzel}},\ }\bibfield  {title} {\emph {\bibinfo {title} {The {Frobenius}
  norm and the commutator}},\ }\href
  {https://doi.org/10.1016/j.laa.2008.05.020} {\bibfield  {journal} {\bibinfo
  {journal} {Linear Algebra Appl.}\ }\textbf {\bibinfo {volume} {429}},\
  \bibinfo {pages} {1864} (\bibinfo {year} {2008})}\BibitemShut {NoStop}%
\bibitem [{\citenamefont {Ussishkin}\ \emph {et~al.}(2002)\citenamefont
  {Ussishkin}, \citenamefont {Sondhi},\ and\ \citenamefont
  {Huse}}]{UssishkinSondhiHuse2002}%
  \BibitemOpen
  \bibfield  {author} {\bibinfo {author} {\bibfnamefont {I.}~\bibnamefont
  {Ussishkin}}, \bibinfo {author} {\bibfnamefont {S.~L.}\ \bibnamefont
  {Sondhi}},\ and\ \bibinfo {author} {\bibfnamefont {D.~A.}\ \bibnamefont
  {Huse}},\ }\bibfield  {title} {\emph {\bibinfo {title} {Gaussian
  superconducting fluctuations, thermal transport, and the {Nernst} effect}},\
  }\href {https://doi.org/10.1103/PhysRevLett.89.287001} {\bibfield  {journal}
  {\bibinfo  {journal} {Phys. Rev. Lett.}\ }\textbf {\bibinfo {volume} {89}},\
  \bibinfo {pages} {287001} (\bibinfo {year} {2002})}\BibitemShut {NoStop}%
\bibitem [{\citenamefont {Iskin}(2023)}]{Iskin2023GLGeometry}%
  \BibitemOpen
  \bibfield  {author} {\bibinfo {author} {\bibfnamefont {M.}~\bibnamefont
  {Iskin}},\ }\bibfield  {title} {\emph {\bibinfo {title} {Extracting
  quantum-geometric effects from {Ginzburg--Landau} theory in a multiband
  {Hubbard} model}},\ }\href {https://doi.org/10.1103/PhysRevB.107.224505}
  {\bibfield  {journal} {\bibinfo  {journal} {Phys. Rev. B}\ }\textbf {\bibinfo
  {volume} {107}},\ \bibinfo {pages} {224505} (\bibinfo {year}
  {2023})}\BibitemShut {NoStop}%
\bibitem [{\citenamefont {Chen}\ and\ \citenamefont
  {Law}(2024)}]{ChenLaw2024GLGeometry}%
  \BibitemOpen
  \bibfield  {author} {\bibinfo {author} {\bibfnamefont {S.~A.}\ \bibnamefont
  {Chen}}\ and\ \bibinfo {author} {\bibfnamefont {K.~T.}\ \bibnamefont {Law}},\
  }\bibfield  {title} {\emph {\bibinfo {title} {{Ginzburg--Landau} theory of
  flat-band superconductors with quantum metric}},\ }\href
  {https://doi.org/10.1103/PhysRevLett.132.026002} {\bibfield  {journal}
  {\bibinfo  {journal} {Phys. Rev. Lett.}\ }\textbf {\bibinfo {volume} {132}},\
  \bibinfo {pages} {026002} (\bibinfo {year} {2024})}\BibitemShut {NoStop}%
\bibitem [{\citenamefont {Iskin}(2024)}]{Iskin2024CoherenceGeometry}%
  \BibitemOpen
  \bibfield  {author} {\bibinfo {author} {\bibfnamefont {M.}~\bibnamefont
  {Iskin}},\ }\bibfield  {title} {\emph {\bibinfo {title} {Coherence length and
  quantum geometry in a dilute flat-band superconductor}},\ }\href
  {https://doi.org/10.1103/PhysRevB.110.144505} {\bibfield  {journal} {\bibinfo
   {journal} {Phys. Rev. B}\ }\textbf {\bibinfo {volume} {110}},\ \bibinfo
  {pages} {144505} (\bibinfo {year} {2024})}\BibitemShut {NoStop}%
\bibitem [{\citenamefont {Wu}\ \emph {et~al.}(2021)\citenamefont {Wu},
  \citenamefont {Schwemmer}, \citenamefont {M{\"u}ller}, \citenamefont
  {Consiglio}, \citenamefont {Sangiovanni}, \citenamefont {{Di Sante}},
  \citenamefont {Iqbal}, \citenamefont {Hanke}, \citenamefont {Schnyder},
  \citenamefont {Denner}, \citenamefont {Fischer}, \citenamefont {Neupert},\
  and\ \citenamefont {Thomale}}]{WuEtAl2021}%
  \BibitemOpen
  \bibfield  {author} {\bibinfo {author} {\bibfnamefont {X.}~\bibnamefont {Wu}}
  \emph {et~al.},\ }\bibfield  {title} {\emph {\bibinfo {title} {Nature of
  unconventional pairing in the {Kagome} superconductors {$A$V$_3$Sb$_5$
  ($A$=K,Rb,Cs)}}},\ }\href {https://doi.org/10.1103/PhysRevLett.127.177001}
  {\bibfield  {journal} {\bibinfo  {journal} {Phys. Rev. Lett.}\ }\textbf
  {\bibinfo {volume} {127}},\ \bibinfo {pages} {177001} (\bibinfo {year}
  {2021})}\BibitemShut {NoStop}%
\bibitem [{\citenamefont {Hossain}\ \emph {et~al.}(2025)\citenamefont
  {Hossain}, \citenamefont {Zhang}, \citenamefont {Choi}, \citenamefont
  {Ratkovski}, \citenamefont {L{\"u}scher}, \citenamefont {Li}, \citenamefont
  {Jiang}, \citenamefont {Litskevich}, \citenamefont {Cheng}, \citenamefont
  {Yin}, \citenamefont {Cochran}, \citenamefont {Casas}, \citenamefont {Kim},
  \citenamefont {Yang}, \citenamefont {Liu}, \citenamefont {Yao}, \citenamefont
  {Bangura}, \citenamefont {Wang}, \citenamefont {Fischer}, \citenamefont
  {Neupert}, \citenamefont {Balicas},\ and\ \citenamefont
  {Hasan}}]{HossainEtAl2025}%
  \BibitemOpen
  \bibfield  {author} {\bibinfo {author} {\bibfnamefont {M.~S.}\ \bibnamefont
  {Hossain}} \emph {et~al.},\ }\bibfield  {title} {\emph {\bibinfo {title}
  {Unconventional gapping behaviour in a kagome superconductor}},\ }\href
  {https://doi.org/10.1038/s41567-024-02770-z} {\bibfield  {journal} {\bibinfo
  {journal} {Nat. Phys.}\ }\textbf {\bibinfo {volume} {21}},\ \bibinfo {pages}
  {556} (\bibinfo {year} {2025})}\BibitemShut {NoStop}%
\bibitem [{\citenamefont {Wang}\ \emph {et~al.}(2026)\citenamefont {Wang},
  \citenamefont {Feng}, \citenamefont {Fang}, \citenamefont {Peng},
  \citenamefont {Sun}, \citenamefont {Yang}, \citenamefont {Xiong},
  \citenamefont {Liu}, \citenamefont {Zhao}, \citenamefont {Kang},
  \citenamefont {Wu}, \citenamefont {Zhang}, \citenamefont {Fu}, \citenamefont
  {Law}, \citenamefont {Lin},\ and\ \citenamefont {Yu}}]{WangEtAl2026TRSB}%
  \BibitemOpen
  \bibfield  {author} {\bibinfo {author} {\bibfnamefont {S.}~\bibnamefont
  {Wang}} \emph {et~al.},\ }\bibfield  {title} {\emph {\bibinfo {title}
  {Unconventional hysteresis due to time-reversal symmetry breaking
  superconductivity in {RbV$_3$Sb$_5$}}},\ }\href
  {https://doi.org/10.1038/s41467-025-68067-8} {\bibfield  {journal} {\bibinfo
  {journal} {Nat. Commun.}\ }\textbf {\bibinfo {volume} {17}},\ \bibinfo
  {pages} {1310} (\bibinfo {year} {2026})}\BibitemShut {NoStop}%
\bibitem [{\citenamefont {Mielke}\ \emph {et~al.}(2022)\citenamefont {Mielke},
  \citenamefont {Das}, \citenamefont {Yin}, \citenamefont {Liu}, \citenamefont
  {Gupta}, \citenamefont {Jiang}, \citenamefont {Medarde}, \citenamefont {Wu},
  \citenamefont {Lei}, \citenamefont {Chang}, \citenamefont {Dai},
  \citenamefont {Si}, \citenamefont {Miao}, \citenamefont {Thomale},
  \citenamefont {Neupert}, \citenamefont {Shi}, \citenamefont {Khasanov},
  \citenamefont {Hasan}, \citenamefont {Luetkens},\ and\ \citenamefont
  {Guguchia}}]{MielkeEtAl2022}%
  \BibitemOpen
  \bibfield  {author} {\bibinfo {author} {\bibfnamefont {C.}~\bibnamefont
  {Mielke}, \bibfnamefont {III}} \emph {et~al.},\ }\bibfield  {title} {\emph
  {\bibinfo {title} {Time-reversal symmetry-breaking charge order in a kagome
  superconductor}},\ }\href {https://doi.org/10.1038/s41586-021-04327-z}
  {\bibfield  {journal} {\bibinfo  {journal} {Nature (London)}\ }\textbf
  {\bibinfo {volume} {602}},\ \bibinfo {pages} {245} (\bibinfo {year}
  {2022})}\BibitemShut {NoStop}%
\bibitem [{\citenamefont {Cha}\ \emph {et~al.}(2026)\citenamefont {Cha},
  \citenamefont {Lee}, \citenamefont {Sim}, \citenamefont {Sur}, \citenamefont
  {Kim}, \citenamefont {Han}, \citenamefont {Kim}, \citenamefont {Lee},
  \citenamefont {Hyun}, \citenamefont {Lim}, \citenamefont {Ahn}, \citenamefont
  {Gim}, \citenamefont {Denlinger}, \citenamefont {Kim}, \citenamefont {Kim},
  \citenamefont {Lee}, \citenamefont {Han},\ and\ \citenamefont
  {Kim}}]{ChaEtAl2026}%
  \BibitemOpen
  \bibfield  {author} {\bibinfo {author} {\bibfnamefont {J.}~\bibnamefont
  {Cha}} \emph {et~al.},\ }\bibfield  {title} {\emph {\bibinfo {title}
  {Evidence of time-reversal symmetry breaking above the charge density wave
  order in a kagome metal}},\ }\href
  {https://doi.org/10.1038/s41567-026-03331-2} {\bibfield  {journal} {\bibinfo
  {journal} {Nat. Phys.}\ }\textbf {\bibinfo {volume} {22}},\ \bibinfo {pages}
  {1245} (\bibinfo {year} {2026})}\BibitemShut {NoStop}%
\bibitem [{\citenamefont {Kang}\ \emph {et~al.}(2022)\citenamefont {Kang},
  \citenamefont {Fang}, \citenamefont {Kim}, \citenamefont {Ortiz},
  \citenamefont {Ryu}, \citenamefont {Kim}, \citenamefont {Yoo}, \citenamefont
  {Sangiovanni}, \citenamefont {Di~Sante}, \citenamefont {Park}, \citenamefont
  {Jozwiak}, \citenamefont {Bostwick}, \citenamefont {Rotenberg}, \citenamefont
  {Kaxiras}, \citenamefont {Wilson}, \citenamefont {Park},\ and\ \citenamefont
  {Comin}}]{KangEtAl2022}%
  \BibitemOpen
  \bibfield  {author} {\bibinfo {author} {\bibfnamefont {M.}~\bibnamefont
  {Kang}} \emph {et~al.},\ }\bibfield  {title} {\emph {\bibinfo {title}
  {Twofold van {Hove} singularity and origin of charge order in topological
  kagome superconductor {CsV$_3$Sb$_5$}}},\ }\href
  {https://doi.org/10.1038/s41567-021-01451-5} {\bibfield  {journal} {\bibinfo
  {journal} {Nat. Phys.}\ }\textbf {\bibinfo {volume} {18}},\ \bibinfo {pages}
  {301} (\bibinfo {year} {2022})}\BibitemShut {NoStop}%
\bibitem [{\citenamefont {R{\o}mer}\ \emph {et~al.}(2022)\citenamefont
  {R{\o}mer}, \citenamefont {Bhattacharyya}, \citenamefont {Valent{\'i}},
  \citenamefont {Christensen},\ and\ \citenamefont {Andersen}}]{RomerEtAl2022}%
  \BibitemOpen
  \bibfield  {author} {\bibinfo {author} {\bibfnamefont {A.~T.}\ \bibnamefont
  {R{\o}mer}}, \bibinfo {author} {\bibfnamefont {S.}~\bibnamefont
  {Bhattacharyya}}, \bibinfo {author} {\bibfnamefont {R.}~\bibnamefont
  {Valent{\'i}}}, \bibinfo {author} {\bibfnamefont {M.~H.}\ \bibnamefont
  {Christensen}},\ and\ \bibinfo {author} {\bibfnamefont {B.~M.}\ \bibnamefont
  {Andersen}},\ }\bibfield  {title} {\emph {\bibinfo {title} {Superconductivity
  from repulsive interactions on the kagome lattice}},\ }\href
  {https://doi.org/10.1103/PhysRevB.106.174514} {\bibfield  {journal} {\bibinfo
   {journal} {Phys. Rev. B}\ }\textbf {\bibinfo {volume} {106}},\ \bibinfo
  {pages} {174514} (\bibinfo {year} {2022})}\BibitemShut {NoStop}%
\bibitem [{\citenamefont {Yu}\ and\ \citenamefont {Li}(2012)}]{YuLi2012}%
  \BibitemOpen
  \bibfield  {author} {\bibinfo {author} {\bibfnamefont {S.-L.}\ \bibnamefont
  {Yu}}\ and\ \bibinfo {author} {\bibfnamefont {J.-X.}\ \bibnamefont {Li}},\
  }\bibfield  {title} {\emph {\bibinfo {title} {Chiral superconducting phase
  and chiral spin-density-wave phase in a {Hubbard} model on the kagome
  lattice}},\ }\href {https://doi.org/10.1103/PhysRevB.85.144402} {\bibfield
  {journal} {\bibinfo  {journal} {Phys. Rev. B}\ }\textbf {\bibinfo {volume}
  {85}},\ \bibinfo {pages} {144402} (\bibinfo {year} {2012})}\BibitemShut
  {NoStop}%
\bibitem [{\citenamefont {Wang}\ \emph {et~al.}(2013)\citenamefont {Wang},
  \citenamefont {Li}, \citenamefont {Xiang},\ and\ \citenamefont
  {Wang}}]{WangEtAl2013}%
  \BibitemOpen
  \bibfield  {author} {\bibinfo {author} {\bibfnamefont {W.-S.}\ \bibnamefont
  {Wang}}, \bibinfo {author} {\bibfnamefont {Z.-Z.}\ \bibnamefont {Li}},
  \bibinfo {author} {\bibfnamefont {Y.-Y.}\ \bibnamefont {Xiang}},\ and\
  \bibinfo {author} {\bibfnamefont {Q.-H.}\ \bibnamefont {Wang}},\ }\bibfield
  {title} {\emph {\bibinfo {title} {Competing electronic orders on kagome
  lattices at van {Hove} filling}},\ }\href
  {https://doi.org/10.1103/PhysRevB.87.115135} {\bibfield  {journal} {\bibinfo
  {journal} {Phys. Rev. B}\ }\textbf {\bibinfo {volume} {87}},\ \bibinfo
  {pages} {115135} (\bibinfo {year} {2013})}\BibitemShut {NoStop}%
\bibitem [{\citenamefont {Ohgushi}\ \emph {et~al.}(2000)\citenamefont
  {Ohgushi}, \citenamefont {Murakami},\ and\ \citenamefont
  {Nagaosa}}]{OhgushiMurakamiNagaosa2000}%
  \BibitemOpen
  \bibfield  {author} {\bibinfo {author} {\bibfnamefont {K.}~\bibnamefont
  {Ohgushi}}, \bibinfo {author} {\bibfnamefont {S.}~\bibnamefont {Murakami}},\
  and\ \bibinfo {author} {\bibfnamefont {N.}~\bibnamefont {Nagaosa}},\
  }\bibfield  {title} {\emph {\bibinfo {title} {Spin anisotropy and quantum
  {Hall} effect in the kagom{\'e} lattice: Chiral spin state based on a
  ferromagnet}},\ }\href {https://doi.org/10.1103/PhysRevB.62.R6065} {\bibfield
   {journal} {\bibinfo  {journal} {Phys. Rev. B}\ }\textbf {\bibinfo {volume}
  {62}},\ \bibinfo {pages} {R6065} (\bibinfo {year} {2000})}\BibitemShut
  {NoStop}%
\bibitem [{\citenamefont {Tazai}\ \emph {et~al.}(2026)\citenamefont
  {Tazai}, \citenamefont {Yamakawa},\ and\ \citenamefont
  {Kontani}}]{TazaiYamakawaKontani2026}%
  \BibitemOpen
  \bibfield {author} {\bibinfo {author} {\bibfnamefont {R.}~\bibnamefont
  {Tazai}}, \bibinfo {author} {\bibfnamefont {Y.}~\bibnamefont {Yamakawa}},\
  and\ \bibinfo {author} {\bibfnamefont {H.}~\bibnamefont {Kontani}},\ }
  \bibfield {title} {\emph {\bibinfo {title} {Nematic and chiral
  superconductivity emerging within the loop-current phase in kagome metals}},\ }
  \href {https://doi.org/10.1038/s41467-026-75259-3} {\bibfield {journal}
  {\bibinfo {journal} {Nat. Commun.}\ }\textbf {\bibinfo {volume} {17}},\
  \bibinfo {pages} {9114} (\bibinfo {year} {2026})}\BibitemShut {NoStop}%
\bibitem [{\citenamefont {Jiang}\ \emph {et~al.}(2014)\citenamefont {Jiang},
  \citenamefont {Li},\ and\ \citenamefont
  {Rosenstein}}]{JiangLiRosenstein2014}%
  \BibitemOpen
  \bibfield  {author} {\bibinfo {author} {\bibfnamefont {X.}~\bibnamefont
  {Jiang}}, \bibinfo {author} {\bibfnamefont {D.}~\bibnamefont {Li}},\ and\
  \bibinfo {author} {\bibfnamefont {B.}~\bibnamefont {Rosenstein}},\ }\bibfield
   {title} {\emph {\bibinfo {title} {Strong thermal fluctuations in cuprate
  superconductors in magnetic fields above {$T_c$}}},\ }\href
  {https://doi.org/10.1103/PhysRevB.89.064507} {\bibfield  {journal} {\bibinfo
  {journal} {Phys. Rev. B}\ }\textbf {\bibinfo {volume} {89}},\ \bibinfo
  {pages} {064507} (\bibinfo {year} {2014})}\BibitemShut {NoStop}%
\bibitem [{\citenamefont {Xiao}\ \emph {et~al.}(2005)\citenamefont {Xiao},
  \citenamefont {Shi},\ and\ \citenamefont {Niu}}]{XiaoShiNiu2005}%
  \BibitemOpen
  \bibfield  {author} {\bibinfo {author} {\bibfnamefont {D.}~\bibnamefont
  {Xiao}}, \bibinfo {author} {\bibfnamefont {J.}~\bibnamefont {Shi}},\ and\
  \bibinfo {author} {\bibfnamefont {Q.}~\bibnamefont {Niu}},\ }\bibfield
  {title} {\emph {\bibinfo {title} {{Berry} phase correction to electron
  density of states in solids}},\ }\href
  {https://doi.org/10.1103/PhysRevLett.95.137204} {\bibfield  {journal}
  {\bibinfo  {journal} {Phys. Rev. Lett.}\ }\textbf {\bibinfo {volume} {95}},\
  \bibinfo {pages} {137204} (\bibinfo {year} {2005})}\BibitemShut {NoStop}%
\bibitem [{\citenamefont {Shi}\ \emph {et~al.}(2007)\citenamefont {Shi},
  \citenamefont {Vignale}, \citenamefont {Xiao},\ and\ \citenamefont
  {Niu}}]{ShiVignaleXiaoNiu2007}%
  \BibitemOpen
  \bibfield  {author} {\bibinfo {author} {\bibfnamefont {J.}~\bibnamefont
  {Shi}}, \bibinfo {author} {\bibfnamefont {G.}~\bibnamefont {Vignale}},
  \bibinfo {author} {\bibfnamefont {D.}~\bibnamefont {Xiao}},\ and\ \bibinfo
  {author} {\bibfnamefont {Q.}~\bibnamefont {Niu}},\ }\bibfield  {title} {\emph
  {\bibinfo {title} {Quantum theory of orbital magnetization and its
  generalization to interacting systems}},\ }\href
  {https://doi.org/10.1103/PhysRevLett.99.197202} {\bibfield  {journal}
  {\bibinfo  {journal} {Phys. Rev. Lett.}\ }\textbf {\bibinfo {volume} {99}},\
  \bibinfo {pages} {197202} (\bibinfo {year} {2007})}\BibitemShut {NoStop}%
\bibitem [{\citenamefont {Xiao}\ \emph {et~al.}(2010)\citenamefont {Xiao},
  \citenamefont {Chang},\ and\ \citenamefont {Niu}}]{XiaoChangNiu2010}%
  \BibitemOpen
  \bibfield  {author} {\bibinfo {author} {\bibfnamefont {D.}~\bibnamefont
  {Xiao}}, \bibinfo {author} {\bibfnamefont {M.-C.}\ \bibnamefont {Chang}},\
  and\ \bibinfo {author} {\bibfnamefont {Q.}~\bibnamefont {Niu}},\ }\bibfield
  {title} {\emph {\bibinfo {title} {{Berry} phase effects on electronic
  properties}},\ }\href {https://doi.org/10.1103/RevModPhys.82.1959} {\bibfield
   {journal} {\bibinfo  {journal} {Rev. Mod. Phys.}\ }\textbf {\bibinfo
  {volume} {82}},\ \bibinfo {pages} {1959} (\bibinfo {year}
  {2010})}\BibitemShut {NoStop}%
\bibitem [{\citenamefont {Fukuyama}(1971)}]{Fukuyama1971}%
  \BibitemOpen
  \bibfield  {author} {\bibinfo {author} {\bibfnamefont {H.}~\bibnamefont
  {Fukuyama}},\ }\bibfield  {title} {\emph {\bibinfo {title} {Theory of orbital
  magnetism of {B}loch electrons: {C}oulomb interactions}},\ }\href
  {https://doi.org/10.1143/PTP.45.704} {\bibfield  {journal} {\bibinfo
  {journal} {Prog. Theor. Phys.}\ }\textbf {\bibinfo {volume} {45}},\ \bibinfo
  {pages} {704} (\bibinfo {year} {1971})}\BibitemShut {NoStop}%
\bibitem [{\citenamefont {Ogata}\ and\ \citenamefont
  {Fukuyama}(2015)}]{OgataFukuyama2015}%
  \BibitemOpen
  \bibfield  {author} {\bibinfo {author} {\bibfnamefont {M.}~\bibnamefont
  {Ogata}}\ and\ \bibinfo {author} {\bibfnamefont {H.}~\bibnamefont
  {Fukuyama}},\ }\bibfield  {title} {\emph {\bibinfo {title} {Orbital magnetism
  of {Bloch} electrons {I}. {G}eneral formula}},\ }\href
  {https://doi.org/10.7566/JPSJ.84.124708} {\bibfield  {journal} {\bibinfo
  {journal} {J. Phys. Soc. Jpn.}\ }\textbf {\bibinfo {volume} {84}},\ \bibinfo
  {pages} {124708} (\bibinfo {year} {2015})}\BibitemShut {NoStop}%
\bibitem [{\citenamefont {Gao}\ \emph {et~al.}(2015)\citenamefont {Gao},
  \citenamefont {Yang},\ and\ \citenamefont {Niu}}]{GaoYangNiu2015}%
  \BibitemOpen
  \bibfield  {author} {\bibinfo {author} {\bibfnamefont {Y.}~\bibnamefont
  {Gao}}, \bibinfo {author} {\bibfnamefont {S.~A.}\ \bibnamefont {Yang}},\ and\
  \bibinfo {author} {\bibfnamefont {Q.}~\bibnamefont {Niu}},\ }\bibfield
  {title} {\emph {\bibinfo {title} {Geometrical effects in orbital magnetic
  susceptibility}},\ }\href {https://doi.org/10.1103/PhysRevB.91.214405}
  {\bibfield  {journal} {\bibinfo  {journal} {Phys. Rev. B}\ }\textbf {\bibinfo
  {volume} {91}},\ \bibinfo {pages} {214405} (\bibinfo {year}
  {2015})}\BibitemShut {NoStop}%
\bibitem [{\citenamefont {Pi{\'e}chon}\ \emph {et~al.}(2016)\citenamefont
  {Pi{\'e}chon}, \citenamefont {Raoux}, \citenamefont {Fuchs},\ and\
  \citenamefont {Montambaux}}]{PiechonEtAl2016}%
  \BibitemOpen
  \bibfield  {author} {\bibinfo {author} {\bibfnamefont {F.}~\bibnamefont
  {Pi{\'e}chon}}, \bibinfo {author} {\bibfnamefont {A.}~\bibnamefont {Raoux}},
  \bibinfo {author} {\bibfnamefont {J.-N.}\ \bibnamefont {Fuchs}},\ and\
  \bibinfo {author} {\bibfnamefont {G.}~\bibnamefont {Montambaux}},\ }\bibfield
   {title} {\emph {\bibinfo {title} {Geometric orbital susceptibility: Quantum
  metric without {B}erry curvature}},\ }\href
  {https://doi.org/10.1103/PhysRevB.94.134423} {\bibfield  {journal} {\bibinfo
  {journal} {Phys. Rev. B}\ }\textbf {\bibinfo {volume} {94}},\ \bibinfo
  {pages} {134423} (\bibinfo {year} {2016})}\BibitemShut {NoStop}%
\bibitem [{\citenamefont {Lee}\ and\ \citenamefont
  {Shenoy}(1972)}]{LeeShenoy1972}%
  \BibitemOpen
  \bibfield  {author} {\bibinfo {author} {\bibfnamefont {P.~A.}\ \bibnamefont
  {Lee}}\ and\ \bibinfo {author} {\bibfnamefont {S.~R.}\ \bibnamefont
  {Shenoy}},\ }\bibfield  {title} {\emph {\bibinfo {title} {Effective
  dimensionality change of fluctuations in superconductors in a magnetic
  field}},\ }\href {https://doi.org/10.1103/PhysRevLett.28.1025} {\bibfield
  {journal} {\bibinfo  {journal} {Phys. Rev. Lett.}\ }\textbf {\bibinfo
  {volume} {28}},\ \bibinfo {pages} {1025} (\bibinfo {year}
  {1972})}\BibitemShut {NoStop}%
\bibitem [{\citenamefont {Sumiyoshi}\ and\ \citenamefont
  {Fujimoto}(2014)}]{SumiyoshiFujimoto2014}%
  \BibitemOpen
  \bibfield  {author} {\bibinfo {author} {\bibfnamefont {H.}~\bibnamefont
  {Sumiyoshi}}\ and\ \bibinfo {author} {\bibfnamefont {S.}~\bibnamefont
  {Fujimoto}},\ }\bibfield  {title} {\emph {\bibinfo {title} {Giant {Nernst}
  and {Hall} effects due to chiral superconducting fluctuations}},\ }\href
  {https://doi.org/10.1103/PhysRevB.90.184518} {\bibfield  {journal} {\bibinfo
  {journal} {Phys. Rev. B}\ }\textbf {\bibinfo {volume} {90}},\ \bibinfo
  {pages} {184518} (\bibinfo {year} {2014})}\BibitemShut {NoStop}%
\end{thebibliography}
\end{document}